\documentclass[reprint,aps]{revtex4-1}
\usepackage{amsfonts}
\usepackage{amssymb}
\usepackage{amsmath}
\usepackage{bbm}
\usepackage{bm}
\usepackage{relsize}
\usepackage{graphicx}
\usepackage{xcolor}
\usepackage{float}
\graphicspath{{../plotfiles/}}

\newcommand{\sumint}[1]{\mathrel{\sum_{#1}\!\!\!\!\!\!\!\!\int}}

\DeclareMathOperator{\sinc}{sinc}

\DeclareMathAlphabet\mathbfcal{OMS}{cmsy}{b}{n}

\newcommand{\be}{\begin{equation}}
\newcommand{\ee}{\end{equation}}
\newcommand{\bea}{\begin{eqnarray}}
\newcommand{\eea}{\end{eqnarray}}
\newcommand{\Eq}[1]{Eq.\,(\ref{#1})}
\newcommand{\Eqs}[2]{Eqs.\,(\ref{#1}) and (\ref{#2})}
\newcommand{\Eqsss}[3]{Eqs.\,(\ref{#1}), (\ref{#2}), and (\ref{#3})}

\newcommand{\Figs}[2]{Figs.\,\ref{#1} and \ref{#2}}
\newcommand{\Fig}[1]{Fig.\,\ref{#1}}
\newcommand{\Sec}[1]{Sec.\,\ref{#1}}
\newcommand{\Secs}[2]{Secs.\,\ref{#1} and \ref{#2}}
\newcommand{\App}[1]{Appendix\,\ref{#1}}

\newcommand\one{\hat{\mathbf{1}}}

\newcommand{\eps}{\varepsilon}

\newcommand{\E}{\textbf{E}}
\newcommand{\F}{\textbf{F}}

\renewcommand{\H}{\textbf{H}}

\newcommand{\Fc}{\mathbfcal{F}}
\newcommand{\Ec}{\mathbfcal{E}}
\newcommand{\Hc}{\mathbfcal{H}}

\newcommand{\GF}{\hat{\mathcal{G}}}

\newcommand{\hG}{\hat{G}}
\newcommand{\hL}{\hat{L}}
\newcommand{\hP}{\hat{P}}
\newcommand{\hD}{\hat{D}}
\newcommand{\hO}{\hat{O}}

\newcommand{\hS}{\hat{U}}

\newcommand{\hcL}{\hat{\cal L}}
\newcommand{\hcD}{\hat{\cal D}}

\newcommand{\p}{\partial}

\date{\today}
\begin{document}

\title{Resonant-state expansion for planar photonic-crystal structures}
\author{Sam Neale}
\affiliation{School of Physics and Astronomy, Cardiff University, Cardiff CF24 3AA, United Kingdom}
\author{Egor A. Muljarov}
\affiliation{School of Physics and Astronomy, Cardiff University, Cardiff CF24 3AA, United Kingdom}

	\begin{abstract}
		We present a new paradigm in the field of photonic crystals and metamaterials, applying the resonant-state expansion (RSE) to planar photonic-crystal structures. The RSE allows us to understand and quantify optical resonances in photonic-crystal structures in terms of the analytic resonant states of a homogeneous planar waveguide.
The RSE provides an efficient and reliable tool for accurate calculation of a complete set of the resonant states of a photonic-crystal slab, which is required for the correct description and a better  understanding of its optical spectra. For the proof of principle, numerical verification of the RSE, and demonstration of its unprecedented accuracy and convergence, an infinite planar photonic crystal slab periodic in one dimension is taken as an example. To illustrate the power of the present approach, we consider the mode evolution with the amplitude of the periodic modulation, revealing the role of the guided modes in the formation of bound states in the continuum.	

\end{abstract}
	\maketitle
	
	\section{Introduction}

	Photonic crystal structures exhibit a number of fundamental optical properties, such as strong confinement and Bragg scattering of light, which can be used e.g. for light propagation control in grating couplers \cite{LiuAPL10}, photonic integrated circuits \cite{McnabOS03,McGurnPRB00}, and beam splitters \cite{BayindirAPL00}. The band structure of an idealized photonic crystal (PC), infinitely extended in all directions, is already very complicated~\cite{YablonovitchPRL91}.  Planar PC systems provide an opportunity for the light trapped within an optical waveguide (WG) to couple to the photonic continuum outside the system~\cite{WhittakerPRB99,TikhodeevPRB02,FanPRB02,ZhouPQE14}, which makes finding the light eigenmodes of a PC slab an even more challenging task.

	The complex transmission spectrum of a PC slab can be intuitively understood as a superposition of a large number of resonances of different linewidth and lineshape~\cite{TikhodeevPRB02}, which can be rigorously described by
the resonant states (RSs). Being introduced in quantum mechanics nearly a century ago~\cite{GamowZP28,SiegertPR39}, the RSs in electro-magnetics are the discrete eigenmodes of an optical system -- solutions to Maxwell's wave equation with outgoing boundary conditions (BCs)~\cite{Weinstein}. Physically, RSs describe a culmination of various constructive and destructive interferences of waves due to multiple reflections within the optical system. The RS eigenfrequency is generally complex, with the quality factor (Q-factor)  being the half of the ratio of its real to imaginary part. In planar systems, the RSs include, as a special case, modes with purely real frequency, such as WG modes formed in planar systems as a result of the total internal reflection.
Additionally, a typical optical spectrum of a PC slab contains non-resonant features known as  Rayleigh-Wood anomalies~\cite{WoodPM02}, which are caused by the opening of new diffraction orders into free space. Mathematically, they correspond to the branch cuts in the complex frequency plane caused by the square root in the light dispersion which can be represented as a continuum of modes that lie along the cuts~\cite{LobanovPRA17} and that are similar to the RSs. These cuts can be observed, e.g. in a form of isolated resonances -- cut modes, in finite-difference time-domain calculations with perfectly matched layers~\cite{GrasOL19}.  In some cases, these continua of modes can even be entirely eliminated from the spectral representation~\cite{Akimov11,ArmitagePRA14}.

The positions and linewidths of the RSs can be modified e.g. by introducing imperfections or changes to the permittivity. This makes RSs of particular importance for sensing applications, ranging from sensing
the refractive index and chirality of a medium~\cite{GovorovNL10,WeissPRL16,WeissPRB17} to biosensing of individual molecules and atoms~\cite{VollmerNMe08,VollmerAPL02,RosenblitPRA04}. RSs have found their application also in miniature lasers \cite{FrateschiAPL95} and low-loss guiding of light in photonic crystal fibers~\cite{McnabOS03}, to name a few. The concept of RSs (also known as ``quasi-normal modes'') is widely used the literature as a natural tool for understanding the optical properties of micro- and nano-resonators, see e.g. a recent review~\cite{LalanneLPR18}.

The resonant-state expansion (RSE) is a novel rigorous approach developed in electrodynamics~\cite{MuljarovEPL10} for calculating the RSs of an optical system.  Using a complete set of the RSs of a simpler system as a basis, the RSE performs a mapping of Maxwell's wave equation onto a linear eigenvalues problem, which determines the RSs of the complex system of interest. In addition to a higher numerical efficiency compared to other computational methods, such as finite-difference in time-domain, finite-element, and Fourier modal method, as demonstrated in~\cite{DoostPRA14,LobanovPRA17}, the RSE provides an intuitive physical picture of resonant phenomena, capable of explaining features observed in optical spectra. Also, unlike other approaches, the RSE guarantees the completeness of the set of the RSs found within the selected spectral range, provided that the basis set used as input for the RSE is also complete. The latter is easy to achieve and to verify when choosing the basis system to be analytically solvable. Other approaches, in turn, are not able to guarantee that all relevant modes are found and that there are no spurious solutions. The effect of other modes, not included in the optical spectrum, is commonly treated as a background contribution adjusted with a number of fit parameters~\cite{SauvanPRL13,FloessPRX17}.

So far the RSE has been applied to finite open optical systems of different geometry and dimensionality~\cite{DoostPRA12,DoostPRA13,DoostPRA14}, as well as to homogeneous~\cite{ArmitagePRA14,ArmitagePRA18} and inhomogeneous planar waveguides~\cite{LobanovPRA17}.
Recently, the RSE was generalized to systems with frequency dispersion of the permittivity~\cite{MuljarovPRB16} and later on to magnetic, chiral and bi-aniso\-tropic optical systems~\cite{MuljarovOL18}, enabling its further application to metamaterials. The RSE has been also used in first order of the perturbation theory for PC structures to describe the refractive index sensing~\cite{WeissPRL16},
and a rigorous analytic mode normalization in PC structures has been presented in~\cite{WeissPRL16,WeissPRB17}.
However, the RSE has never been applied in full to PC systems.

In this paper, we develop a photonic-crystal RSE (PC-RSE), a new rigorous approach for the accurate calculation of RSs in planar PC structures.  The key idea of the PC-RSE is to use the analytically solvable homogeneous slab as basis system and to treat a PC structure as a periodic modulation on top of the slab. This idea echoes back to the famous nearly-free electron model in the solid state theory~\cite{Ashcroft}. However, unlike states of a free electron, the basis RSs of an open optical system are generally leaky (having finite Q-factors), which makes the implementation of the same idea in optics entirely different.

The idea of using the eigenmodes of a homogeneous slab has been already implemented as a guided mode expansion method~\cite{AndreaniPRB06}. Among different methods available in the literature, this approach has been considered as the most efficient way of calculating the eigenmodes of PC structures. It provides, in particular, a powerful tool for optimizing parameters of photonic crystal cavities~\cite{MinkovSR14}. This method, however, has a significant disadvantage: A set of the guided modes it uses as a basis is incomplete. This breach has been patched by introducing an additional procedure, similar to Fermi's golden rule, for an approximate treatment of light leakage from the system resulting in finite Q-factors of modes~\cite{AndreaniPRB06}. Our approach instead adds to this incomplete set of guided modes all the missing RSs having finite Q-factors, as well as the cut modes responsible for the Rayleigh-Wood anomalies observed in optical spectra. Having a complete set of modes, which is generated by the RSE, allows us to quantify precisely any optical observable~\cite{YanPRB18,LobanovPRA18,WeissPRB18}, such as the transmission, reflection, scattering, and extinction of light.

The periodicity of a PC structure mixes all possible Bragg harmonics. Therefore, the basis RSs have to be taken with different in-plane wave numbers. As a result, the dyadic Green's function of the set of Maxwell's equations has branch cuts in the complex frequency plane which have to be taken into account in the PC-RSE along with the RSs. This presents the major complication of the PC-RSE which we have dealt with by splitting the cuts into series of discrete, artificial cut modes added for completeness to the basis RSs, as it was done e.g. in~\cite{DoostPRA13,LobanovPRA17}.

A significant technical advantage of treating periodic modulations of a homogeneous slab as perturbations is that all the diagonal elements of the perturbation matrix are vanishing due to periodicity (without homogeneous perturbation). This guarantees a low level of numerical errors even for small basis sizes and strong periodic modulations, as we show in this paper. Using for illustration, a dielectric slab in a vacuum periodically modulated in one dimension (1D), we demonstrate the accuracy and efficiency of the PC-RSE in finding the RSs of PC structures.

For verification of the PC-RSE, we compare it with the scattering matrix method (SMM)~\cite{WhittakerPRB99,TikhodeevPRB02},  also known in the literature as Fourier modal method~\cite{LalanneOL00,SilbersteinJOSAA01,LiJOA03,WeissJOA09}, which has been considered in the literature as the most accurate and reliable way of calculating the optical spectra of infinitely extended periodic open systems. In fact, the SMM is asymptotically exact, having the total numbers of the Bragg diffractions channels taken into account as the only parameter of the method.
However, when used for finding the RSs of the system, the SMM is limited by rather low numbers of Bragg channels, struggling to find all modes in a given frequency range and often returning spurious modes. Recently,
a so called ``mode expansion method'' has been introduced in~\cite{GaoSR16} for an accurate calculation and qualitative study of bound states in the continuum (BICs) in PC slabs. We note, however, that this method, which appeared under a new name, is nothing else than the well-known SMM.

The PC-RSE presented here is not only a simple numerical tool that allows us to accurately calculate a complete set of the RSs of a planar PC slab with a complicated structure. It is a new paradigm in the field of PC systems revealing the origin and the properties of their RSs. To demonstrate this, we study BICs of a dielectric PC slab with 1D grating. These modes, having infinite Q-factors while residing inside the continuum, were predicted in non-relativistic quantum systems almost a century ago~\cite{NeumannPZ29} but only recently have become a subject of particular interest~\cite{MarinicaPRL08,BulgakovPRB08,MoiseyevPRL09} and have been observed in optics~\cite{PlotnikPRL11}. We show in particular how BICs are formed and how different types of basis RSs of a homogeneous slab contribute to them as compared to other resonances of the PC slab, such as quasi-guided~\cite{TikhodeevPRB02}
and Fabry-P\'erot modes.

	\section{Formalism of the PC-RSE}
\label{Sec:Formalism}

Consider a PC slab occupying the region $|z|\leqslant a$, where $z$ is the coordinate in the growth direction. Assuming the permittivity and the permeability are isotropic everywhere,
the electric field $\E$, magnetic field $\H$, and the frequency $\omega$ of a given RS of the PC slab satisfy the following Maxwell equations (the speed of light $c=1$)
\bea
\nabla\times\E&=&\omega(\mu+\Delta\mu) i\H\,,
\label{ME1}
\\
\nabla\times i\H&=&\omega(\eps+\Delta\eps) \E\,,
\label{ME2}
\eea
which have to be solved together with outgoing wave BCs outside the PC slab.
Here we have explicitly separated the total permittivity [permeability],
periodic in $x$-direction with the period $d$,  into a homogeneous part $\eps(z)$ [$\mu(z)$] and a periodic part $\Delta\eps(x,z)$ [$\Delta\mu(x,z)$], obeying
\be
\Delta\eps(x+d,z)=\Delta\eps(x,z)\,, \ \ \  \Delta\mu(x+d,z)=\Delta\mu(x,z)\,.
\ee
For the purpose of a clearer illustration of our approach, we consider the case of the transverse electric (TE) and transverse magnetic (TM) polarizations not coupled to each other, which is achieved by assuming that the $y$-component of the in-plane momentum is zero.
We note, however, that generalization to the case of a non-zero $y$-component of the momentum and to 2D periodicity of the PC slab is straightforward, and the whole formalism remains essentially the same as presented here.

Since the TE and TM polarizations are not coupled, each polarization can be treated separately. However, due to the symmetry of Maxwell's equations (\ref{ME1})--(\ref{ME2}) with respect to a simultaneous exchange of $\E\leftrightarrow i\H$, $\eps\leftrightarrow\mu$, and $\Delta\eps\leftrightarrow\Delta\mu$, it is sufficient to treat only one of the two polarizations, for example, the TE polarization, while retaining the permeability in all results, even if all the constituent materials are non-magnetic.

For the TE polarization, Maxwell's equations (\ref{ME1})--(\ref{ME2}) reduce to
\be
(\hL+\Delta\hL)\F=0\,,
\label{ME}
\ee
where the vector field $\F$ is formed from three non-vanishing components of the electric and magnetic fields,
\be
\F= \left(\begin{array}{c}
E_y\\
iH_x\\
iH_z
\end{array}\right),
\label{F}
\ee
and
\bea
\hL(x,z;\omega)&=&\omega \hP(z)-\hD(x,z)\,,\\
\Delta\hL(x,z;\omega)&=&\omega \Delta\hP(x,z)
\eea
are linear operators, consisting of the generalized permittivity~\cite{MuljarovOL18}
\be
\hP(z)=
\left(\begin{array}{ccc}
\eps(z)&0&0\\
0&\mu(z)&0\\
0&0&\mu(z)
\end{array}\right)\,,
\label{P}
\ee
the curl operator
\be
\hD(x,z)=
\left(\begin{array}{ccc}
0&\p_z&-\p_x\\
-\p_z&0&0\\
\p_x&0&0
\end{array}\right)\,,
\label{D}
\ee
and the perturbation
\be
\Delta\hP(x,z)=
\left(\begin{array}{ccc}
\Delta\eps(x,z)&0&0\\
0&\Delta\mu(x,z)&0\\
0&0&\Delta\mu(x,z)
\end{array}\right)\,.
\ee

Owing to the periodicity in the $x$-direction, the wave function $\F$ obeys the Bloch theorem,
\be
\F(x+d,z)=\F(x,z) e^{ipd}\,,
\ee
determining the quasi-momentum $p$ in the $x$-direction.
We therefore solve the Maxwell equations (\ref{ME}) for the given $p$, using a {\em periodic} dyadic Green's function (GF) $\hG_p(x;z,z')$ of the homogeneous slab:
\be
\F(x,z)\!=\!-\frac{\omega}{d}\!\int_d\!\! dx'\! \!\int \!\!dz' \hG_p(x-x';z,z')\Delta\hP(x',z')\F(x',z'),
\label{F-int}
\ee
where $\int_d dx$ implies integration over any period interval.
This GF has the same value of $p$ and satisfies Maxwell's equations with a periodic array of sources:
\be
\hL(x,z;\omega) \hG_p=\one \delta(z-z') d\sum_m e^{ipmd}\delta(x-x'-md)\,,
\ee
where $m$ is an integer and $\one$ is the $3\times3$ unit matrix. Using Bloch's theorem again, the periodic GF can be written as
\be
\hG_p(x;z,z')=\sum_g \GF_{p+g}(z,z')e^{i(p+g)x}\,,
\label{GFsum}
\ee
where
\be
g=\frac{2\pi m}{d}\,,
\ee
and $\GF_{p}(z,z')$ is another, $x$-independent GF of the homogeneous slab
satisfying an equation
\be
\hcL_p(z;\omega) \GF_{p}(z,z')=\one \delta(z-z') \,,
\label{GF-equ}
\ee
with a modified operator
\be
\hcL_p(z;\omega)=\omega\hP(z)-\hcD_p(z)\,,
\ee
which consists of the homogeneous generalized permittivity $\hP(z)$ given by \Eq{P} and the curl operator $\hcD_p(z)$ given by \Eq{D} with $\p_x$ replaced by $ip$: $\hcD_p(z)=\hD_{\p_x\to ip}$.

The homogeneous GF $\GF_{p}$ can be written, using the Mittag-Leffler (ML) theorem~\cite{Arfken01}, in terms of the RSs of the homogeneous slab,
\be
\GF_{p}(z,z')=\sumint{n}\frac{\Fc_n(z;p)\otimes\Fc_n(z';-p)}{\omega-\omega_n}
\label{GF-ML}
\ee
where $\otimes$ denotes the dyadic vector product, $\Fc_n(z;p)$ is the vectorial wave function of the RS $n$ of the homogeneous slab, satisfying Maxwell's equation
\be
\hcL_p(z;\omega_n) \Fc_n(z;p)=0
\ee
and outgoing BCs, and $\omega_n$ is the RS eigenfrequency. Equation~(\ref{GF-ML}) is valid if the RSs are normalized according to a general normalization condition~\cite{WeissPRB17,MuljarovOL18} applied to the homogeneous planar system,
\be
1=\int_{z_1}^{z_2} \Fc^\dagger_n \cdot \hP \Fc_n dz
+\left.\left(  \Ec^\dagger_n \times i\Hc'_n -\Ec'_n \times i\Hc^\dagger_n\right)\cdot{\bf e}_z\right|_{z_1}^{z_2},
\label{gen-norm}
\ee
where ${\bf e}_z$ is the unit vector in the $z$-direction, $z_1$ and $z_2$ are two arbitrary coordinates outside of the system, such that $z_1\leqslant -a$ and $z_2\geqslant a$,
and $\Fc^\dagger_n(z;p)=\Fc_n(z;-p)$ is the adjoint field. $\Ec_n$  and $\Hc_n$ in \Eq{gen-norm} are, respectively, the electric and magnetic fields of the RS $n$,  combined together into the vector $\Fc_n=\{\Ec_n, i\Hc_n\}$ having in general six components which are reduced to only three for TE or TM polarization, in accordance with \Eq{F};  $\Ec'_n$  and $\Hc'_n$ are the frequency derivatives of the analytic continuation  of the fields $\Ec_n$  and $\Hc_n$ into the complex frequency plane (for more details see \cite{DoostPRA14,MuljarovOL18}). Note that the ML series \Eq{GF-ML} includes also {\em cut modes}, in addition to all the RSs lying on the ``physical'' Riemann sheet of complex frequency -- for details and derivation, see Appendices \ref{App:GF} and \ref{App:ML}.

Substituting \Eq{GFsum} into \Eq{F-int} and using the ML expansion \Eq{GF-ML}, we obtain, owing to the completeness of the basis and in agreement with Bloch's theorem,
an expansion of the wave function $\F$ of the RS of the PC slab into the RSs of the homogeneous slab,
\be
\F(x,z)=\sum_g\sumint{n} c_n^g \Fc_n(z;p+g) e^{i(p+g)x}\,,
\label{F-exp}
\ee
where the expansion coefficients are given by
\bea
c_n^g&=&-\frac{\omega}{\omega-\omega_n^g}\frac{1}{d} \int_d dx e^{-i(p+g)x} \nonumber\\
&&\times \int dz \Fc_n(z;-p-g)\cdot \Delta\hP(x,z)\F(x,z)\,.
\label{coef}
\eea
Then, substituting the expansion \Eq{F-exp} into \Eq{coef}, we arrive at the
the key equation of the PC-RSE:
\be
\omega\sum_{g'}\sumint{n' }\left( \delta_{nn'}\delta_{gg'}+ V_{nn'}^{gg'}\right)c_{n'}^{g'}=\omega_n^g c_n^g\,,
\label{RSE}
\ee
in which $\delta_{ij}$ is the Kronecker delta, and the matrix elements of the perturbation are defined as
\be
V_{nn'}^{gg'}= \int \Fc_n(z;-p-g)\cdot \Delta\hP_{g-g'}(z)\Fc_{n'}(z;p+g')dz
\label{V}
\ee
with $ \Delta\hP_{g}(z)$ being the Fourier coefficients of the generalized permittivity perturbation,
\be
\Delta\hP_{g}(z) =\frac{1}{d}\int_d \Delta\hP(x,z) e^{-igx}dx\,.
\ee
Note that in \Eqs{coef}{RSE}, we have added index $g$ to $\omega_n$ in order emphasize the dependence of the basis RS frequencies on the Bragg channel number $g$.

Equation~(\ref{RSE}) presents a matrix eigenvalue problem, linear in $\omega$ (the eigenfrequency of a perturbed RS of the PC slab) and can be solved simply by diagonalizing a complex symmetric matrix. This equation is very similar to the RSE equation for a finite open optical system~\cite{MuljarovEPL10}. However, the main difference between the two is that \Eq{RSE} contains a summation over all Bragg channels, labeled by the index $g$. Also, the contribution of the cuts, denoted by the integral, is included in \Eq{RSE}, in the same way as it was done in the RSE applied to 2D open optical systems~\cite{DoostPRA13} or to inhomogeneous waveguides~\cite{LobanovPRA17}.

Note that the RSE has been recently formulated for PC systems~\cite{WeissPRL16,WeissPRB17}, in a form of a perturbation theory treating some modifications of the already existing periodic structure, i.e. using a PC slab as a basis system. It has been applied so far to either weak perturbations, limiting the RSE basis to a single mode~\cite{WeissPRL16}, or to moderate perturbations of quasi-degenerate modes, limiting the basis to such a pair of modes~\cite{WeissPRB17}. In the present work we consider instead a homogeneous basis containing up to several thousands of modes of the homogeneous basis. This choice of the basis presents a significant advantage in implementation of the RSE. For example, all different Bragg channels are fully isolated in the homogeneous basis, whereas a PC basis has all these channels already coupled together. This has, in particular, a dramatic consequence on inclusion of the branch cuts in the basis, which is impossible to do in practice with Bragg channels mixed up as in the PC basis.

\section{Results}

\subsection{PC-RSE for permittivity perturbations}
\label{Sec:permittivity}

We now use the PC-RSE derived in \Sec{Sec:Formalism} for finding the RSs of a nonmagnetic PC slab with a 
periodic modulation of the permittivity. In this case the perturbation matrix in the TE polarization simplifies to
\be
V_{nn'}^{gg'}= \int E_n^g(z)\Delta\eps_{g-g'}(z)E_{n'}^{g'}(z)dz
\label{Veps}
\ee
where
\be
\Delta\eps_{g}(z) =\frac{1}{d}\int_d \Delta\eps(x,z) e^{-igx}dx
\label{epsG}
\ee
and $E_n^g(z)$ is the electric field (directed along $y$) of the homogeneous-slab RS with index $n$ and momentum $p+g$ along $x$. In general, this RS is a solution of the Maxwell wave equation (\ref{E-hom}) with the outgoing boundary condition \Eq{BC} and the normalization given by \Eq{RS-norm}, see \App{App:ML}. Here, we have added index $g$ to the electric field $E_n^g$ and the eigenfrequency $\omega_n^g$, in order to distinguish different Bragg channels contributing to the PC-RSE.

Note that in order to treat a permittivity perturbation in the TM polarization, one should instead set $\Delta\eps=0$ in \Eq{V} and use $\Delta\mu$ for the modulation of the permittivity, along with replacements $iH_x\to E_x$ and  $iH_z\to E_z$
in the unperturbed wave function $\Fc_n(z;p)$.

As a basis system, we choose a homogeneous dielectric slab in vacuum, of thickness $2a$, permittivity $\epsilon>1$, and permeability $\mu=1$. The full permittivity profile of the slab system is given by \Eq{eps-hom}, and the basis RSs and cut densities are provided in \App{App:HS}. In practical use of \Eq{RSE} we apply a cut discretization, described in detail in \App{App:RSE-w}, which modifies the PC-RSE equation to
\be
\omega\sum_{\bar{n}'g' }\left( \delta_{\bar{n}\bar{n}'}\delta_{gg'}+ V_{\bar{n}\bar{n}'}^{gg'}\right)c_{\bar{n}'}^{g'}=\omega_{\bar{n}}^g c_{\bar{n}}^g
\label{RSEdis}
\ee
with index $\bar{n}$ labeling both the RS and the cut modes, see \Eq{nbar}. Within the slab, $|z|\leqslant a$, the electric fields of the RSs and cut modes are described by the same functions
\be
E^g_{\bar{n}}(z)=B^g_{\bar{n}}(e^{iq^g_{\bar{n}} z}+(-1)^{\bar{n}} e^{-iq^g_{\bar{n}}z})
\ee
with the normalization constants for the RSs and cut modes given by \Eqs{Bn}{Bnu}, respectively.  The eigenfrequencies $\omega_{\bar{n}}^g$ are determined by the secular equation (\ref{secular}) for the RSs  and by \Eq{omega-nu} for the cut modes. Furthermore, the link between the mode frequency $\omega^g_{\bar{n}}$ and the wave number $k^g_{\bar{n}}$ in vacuum and  $q^g_{\bar{n}}$ in the medium is provided by the following light dispersion relations:
\bea
(\omega^g_{\bar{n}})^2&=&(k^g_{\bar{n}})^2+(p+g)^2\,,\\
\epsilon(\omega^g_{\bar{n}})^2&=&(q^g_{\bar{n}})^2+(p+g)^2\,.
\eea
	\begin{figure}
		\includegraphics*[clip,width=0.35\textwidth]{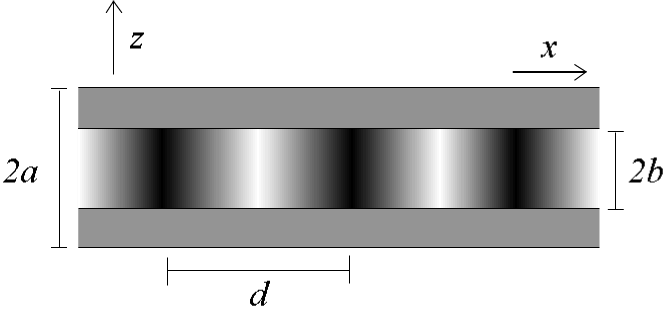}
		\caption{Schematic of the perturbed system -- photonic crystal slab of the total thickness $2a$, periodically modulated in $x$ direction, with period $d$, within the layer of thickness $2b$ at the center of the slab.}
\label{system}
	\end{figure}

For illustration purposes and also for the ease of comparison with the SMM, the perturbation of the homogeneous slab is taken in the most simple harmonic form:
\be
\Delta\eps(x,z)= \left (\alpha+\beta\cos{\frac{2\pi x}{d}}\right) \Theta(b-|z|)\,,
\label{pert}
\ee
where $\Theta(z)$ is the Heaviside step function, $b\leqslant a$, and $\alpha$ and $\beta$ are some parameters, see \Fig{system}.
We note, however, that the RSE can equally deal with any other shape of the periodic perturbation, not requiring the separation of variables which the perturbation in the form of \Eq{pert} possesses. The SMM in turn, requires this separation. In fact,  the transfer matrices that form the scattering matrix are calculated layer-by-layer through the system, see~\cite{TikhodeevPRB02}. Therefore any system changing smoothly in the growth direction will be approximated by a stack of slices homogeneous in $z$,  which are ideally infinitely thin.

For the perturbation given by \Eq{pert}, the matrix elements take the following explicit form:
\be
V_{\bar{n}\bar{n}'}^{gg'}=B^g_{\bar{n}}B^{g'}_{\bar{n}'}\left(2\alpha \delta_{gg'}+
\beta X_{g-g'} \right)b Z^{gg'}_{\bar{n}\bar{n}'}\,,
\label{Vnm}
\ee
where
\be
X_{g}=\delta_{g,g_1}+\delta_{g,g_{-1}}
\ee
with $g_{\pm1}=\pm 2\pi/d$ and
\bea			
Z^{gg'}_{\bar{n}\bar{n}'}&=&
\left(1+(-1)^{\bar{n}+\bar{n}'}\right)\sinc\frac{q^g_{\bar{n}}+q^{g'}_{\bar{n}'}}{\pi}b\\
			&&+\left((-1)^{\bar{n}}+(-1)^{\bar{n}'}\right)\sinc\frac{q^g_{\bar{n}}-q^{g'}_{\bar{n}'}}{\pi}b
\eea
with $\sinc z=\sin z/z$.

For homogeneous perturbations, used in \App{App:RSE for slab} for the RSE verification and comparison of the $k$- and $\omega$-representations, we use $\alpha\neq 0$ and $\beta=0$.
For all periodic perturbations we instead take $\alpha=0$ and $\beta\neq 0$, so that all the matrix elements \Eq{Veps} within the same channel ($g=g'$) are vanishing, since $\Delta\eps_{0}(z)=0$, according to \Eq{epsG}.
Using this property results in a general quick convergence of the PC-RSE. In fact, since the diagonal elements of the perturbation matrix are all zeros, the first-order contribution of the PC-RSE is vanishing~\cite{DoostPRA14}.  Then the lowest-order non-vanishing contribution of the perturbation can only be quadratic in $V_{nn'}^{gg'}$, making its overall effect quantitatively small and the PC-RSE converging quickly to the exact solution.

\subsection{Basis for the PC-RSE}
\label{Sec:basis}
	\begin{figure}
	 	\includegraphics*[clip,width=0.48\textwidth]{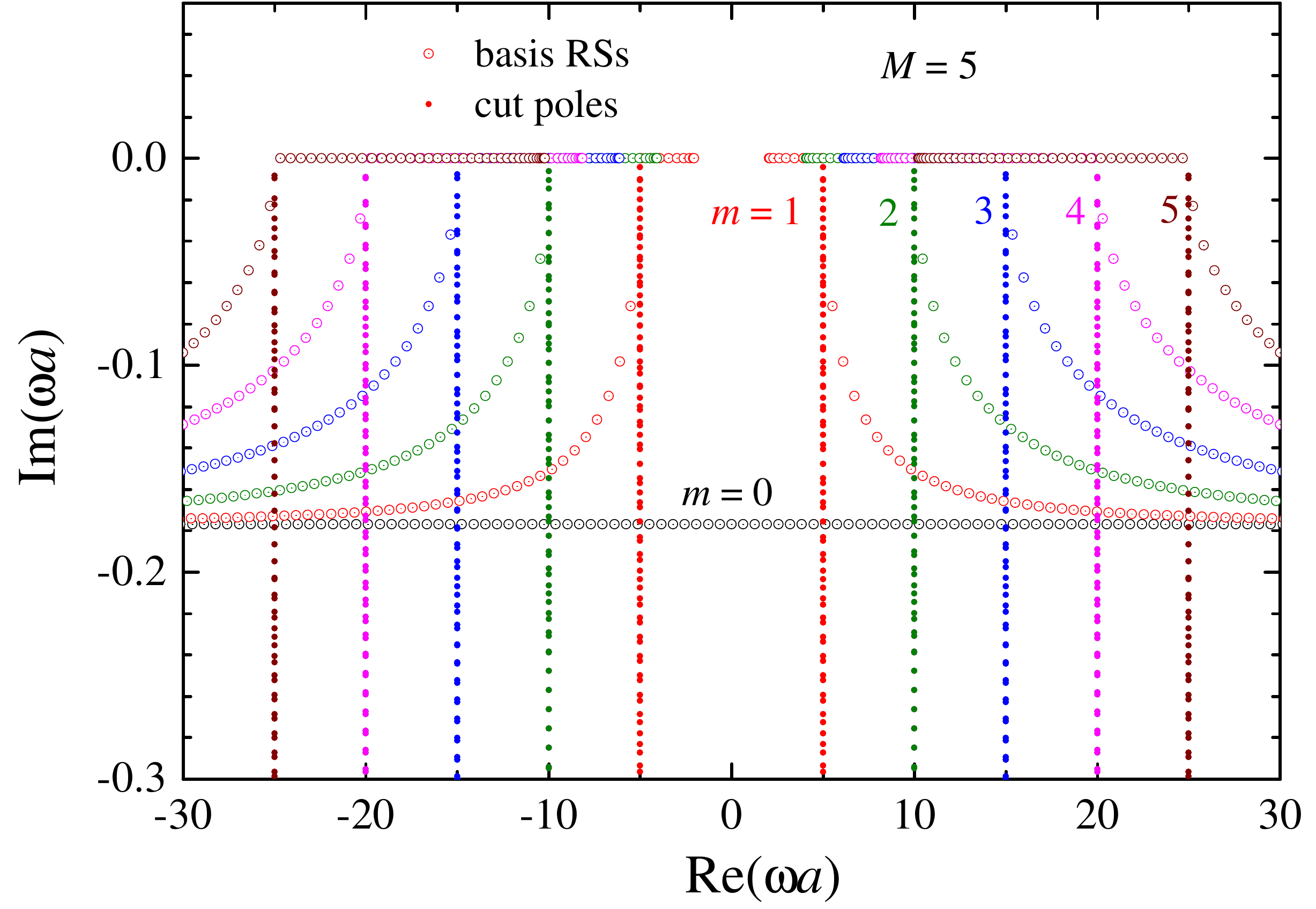}
	 	\caption{Basis RSs and cut poles (used for comparison of the PC-RSE with SMM) for $\epsilon=6$, $\alpha=\beta=0$, and $M=5$ leading to $11$ Bragg channels in the basis, each channel labelled with index $m$. The frequencies of the RS and cut modes of channels $m$ and $-m$ are identical. }
	 	\label{crystal basis}
	\end{figure}
	
The full basis for the PC-RSE consists of an infinite number of RSs and cut modes, taken for all Bragg channels $g=2\pi m/d$, where $m=0,\pm1,\pm2...$. Periodic perturbations, such as the one given by \Eq{pert}, introduce coupling between the basis states belonging to different Bragg channels, so that for obtaining the exact result one needs to take into account all of them simultaneously. In practice, we introduce a truncation, limiting the number of RSs and cut modes for each Bragg channel as well the number of Bragg channels themselves. We do both truncations by introducing a single real parameter $\omega_{\rm max}$ which defines a circle $|\omega|<\omega_{\rm max}$ in the complex frequency plane containing the basis RSs and cut modes included in the PC-RSE. The set of included RSs, defined in this way, typically contains a large number of WG modes. In fact, only the $m=0$ channel consists of equidistant FP modes which we call in the following {\it leaky modes}. All other channels contain WG modes having real frequencies $\omega^g_n$ within the intervals $(p+g)/\sqrt{\epsilon}<|\omega^g_n|<(p+g)$, the number of WG modes within each channel is growing linearly with $g$, with the total number increasing quadratically with $g$. The WG modes are separated from FP modes by series of cut poles of the GF which are positioned vertically down below the branch points at $\omega=\pm(p+g)$. Above $\omega= p+g$ and below $\omega=-(p+g)$ there are two infinite series of FP modes for each Bragg channel. All this means, in particular, that for a given radius $\omega_{\rm max}$, the basis includes $2M+1\approx \omega_{\rm max}\sqrt{\epsilon}d/\pi$ channels, most of which consist of only WG modes.

For the purpose of verification of the PC-RSE by comparing it with the SMM, which is presented in \Sec{Sec:verification} below, we use however a different criterion: We limit the number of Bragg channels to $|m|\leqslant M$, where $M$ is a fixed number, and truncate the RSs and cut modes independent of $M$, i.e. using the same number of modes for each selected channel. This is done in order to avoid a computationally expensive root searching within the SMM related to the increase of the S-matrix  size with $M$. Clearly, for adequate comparison, it is essential to keep the truncation number $M$ the same for both PC-RSE and SMM. However, the necessity to keep $M$ low demonstrates the major weakness of the SMM.

The PC-RSE basis used for the comparison with the SMM is illustrated in \Fig{crystal basis}, for $p=0$ and $M=5$, showing the eigenfrequencies of both the RSs and cut modes for all selected 11 Bragg channels.
Clearly, for $p=0$,  the positive- and negative-$m$ channels are degenerate (giving the same RS eigenfrequencies), and both degenerate channels must be included in the basis. Additionally, there are now $2M$ cuts with discretized cut modes added to the basis. These cut modes are also degenerate for the same reason as the RSs.

\subsection{Verification of the PC-RSE}
\label{Sec:verification}

Before applying the PC-RSE we first consider a homogeneous dielectric slab in vacuum. Taking $\epsilon=6$ and $pa=5$ as an example, we demonstrate numerically in \App{App:HS} a quick convergence with the basis size $N$ of the ML series for the GF to its exact values, given by the analytic formula \Eq{GFanal}. We show in particular that both $k$- and $\omega$-representations of the GF (the first without and the second with cuts in the basis) converge in the same way, with the absolute error scaling as $1/N^2$. The contribution of the cuts to the ML expansion in the $\omega$-representation is taken into account in this case in a form of a numerical integration.

We then  apply in \App{App:RSE for slab} the RSE to a uniform perturbation of the homogeneous slab for $p\neq0$, demonstrating for both $k$- and $\omega$-representations a quick convergence of the RSE to the analytic solution available for the core-shell geometry used, with the relative error for the wave numbers scaling as $1/N^3$. While the RSE in the $k$-representation essentially reproduces the results of~\cite{ArmitagePRA14}, the RSE in the $\omega$-representation is applied here to homogeneous systems for the first time. In this representation, the cut contribution is taken into account in a form of a subset of artificial cut modes obtained by a numerical discretization of the cuts and added to the basis. The procedure of the cut discretization is described in detail in \App{App:RSE-w}. These cut modes are then used in the PC-RSE.

\begin{figure}
		\includegraphics*[clip,width=0.48\textwidth]{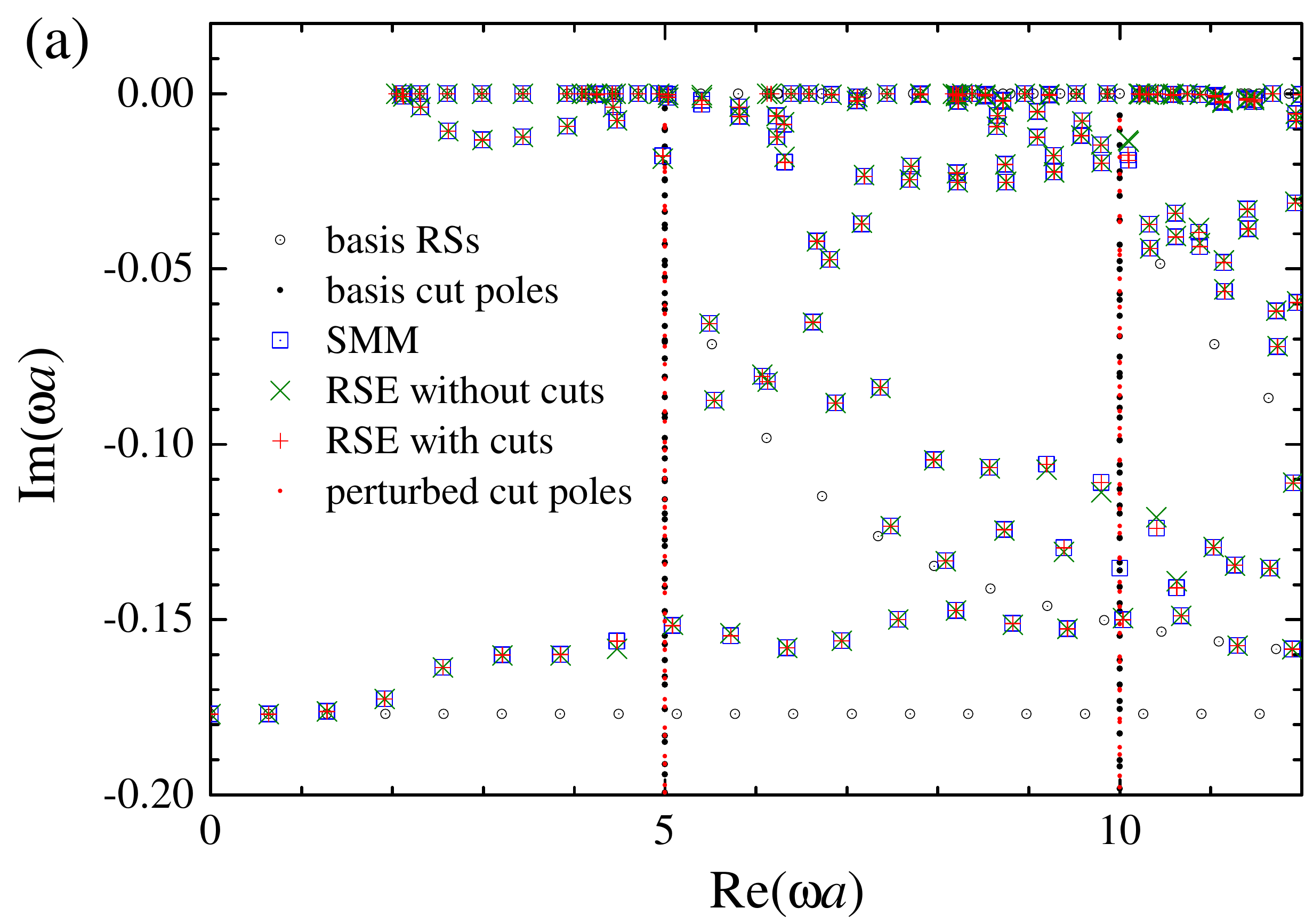}
		\includegraphics*[clip,width=0.48\textwidth]{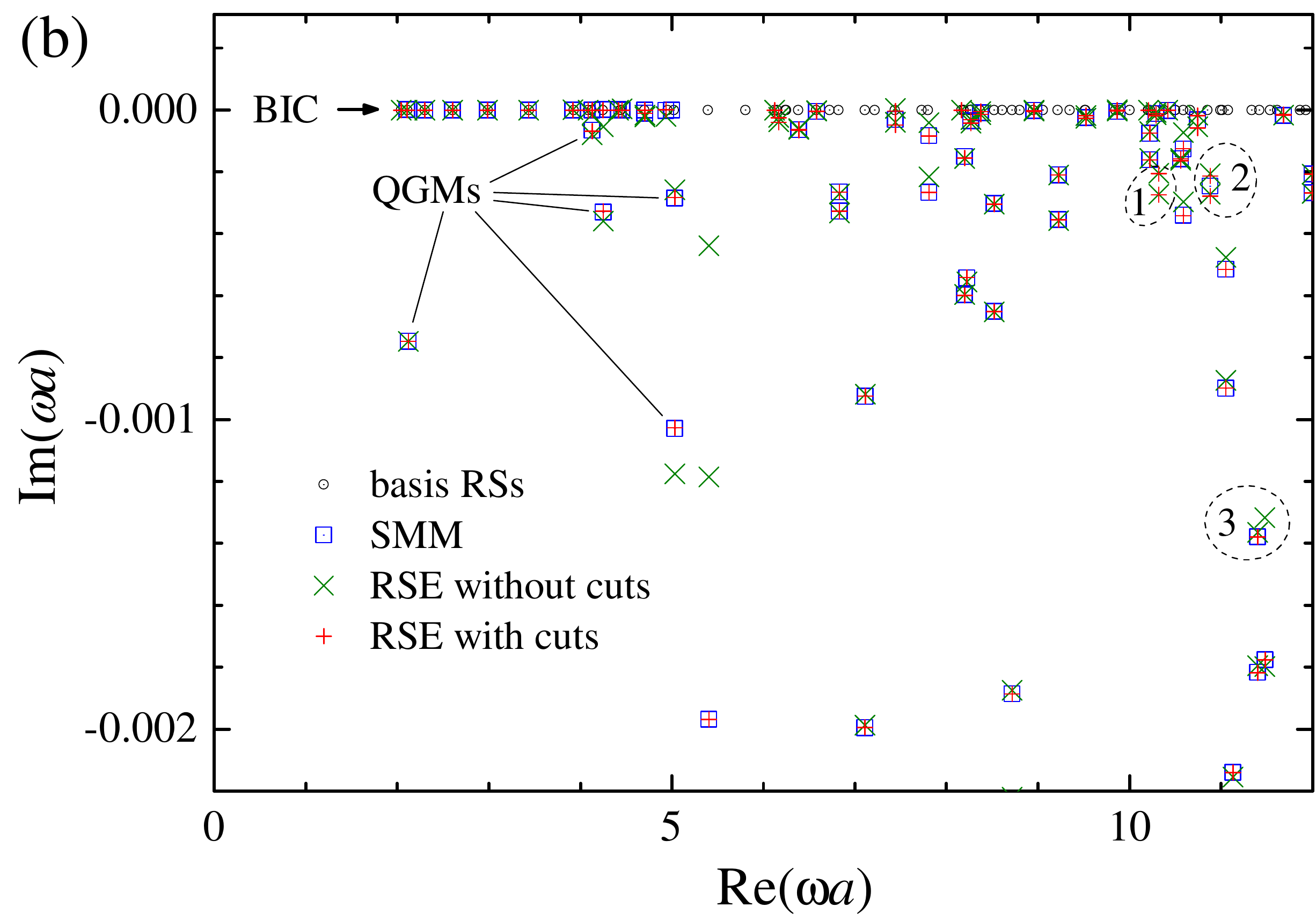}
		\caption{(a) RS frequencies of a PC slab with $\epsilon=6$, $\alpha=0$, $\beta=1$, $b=a/2$, and $d=2\pi/5$, calculated for $M=5$ using the PC-RSE ($N=1998$, $F=1$) with and without cut modes (red and green crosses) and the SMM (blue squared with dots). Unperturbed RSs and both unperturbed and perturbed cut modes are also shown (black circles with dots, black and blue dots, respectively). (b) Zoom of (a) showing RSs near and on the real axis.
}
		\label{crystal rse}
	\end{figure}

Now, in order to verify the PC-RSE, we consider the periodic perturbation \Eq{pert} with $\alpha=0$, $\beta=1$, $d=2\pi/5$, and $b=a/2$. The unperturbed system is a homogeneous slab in vacuum of thickness $2a$ and permittivity $\epsilon=6$. Its RSs and cut modes for TE polarization and $p=0$ are shown in \Fig{crystal basis}. Perturbed RSs of the PC slab, calculated via the PC-RSE (with and without cut modes) and the SMM, are shown in \Fig{crystal rse} along with all the unperturbed RSs and cut modes present in the displayed spectral range. As already mentioned in \Sec{Sec:basis}, the same truncation of the Bragg   channels with $M=5$ was used for both the PC-RSE and SMM.

While the periodic perturbation is not small ($\beta=1$), leading to a considerable modification of the RSs, one can see in \Fig{crystal rse} a very good visual agreement between the SMM and the PC-RSE, even when no cut modes are included in the basis (green crosses $\times$). In fact, in this case, there is only a slight discrepancy between the two calculations seen for some RSs close to the cuts, see e.g. region 3 in \Fig{crystal rse}(b). These discrepancies are fully removed when cut modes are included in the PC-RSE (red crosses $+$). Interestingly, the PC-RSE also returns cut modes of the perturbed system, positioned along the same cut lines at ${\rm Re}\, (\omega a)=5$ and 10,  but shifted vertically with respect to their unperturbed positions (compare red and black points).

\begin{figure}
		\includegraphics*[clip,width=0.48\textwidth]{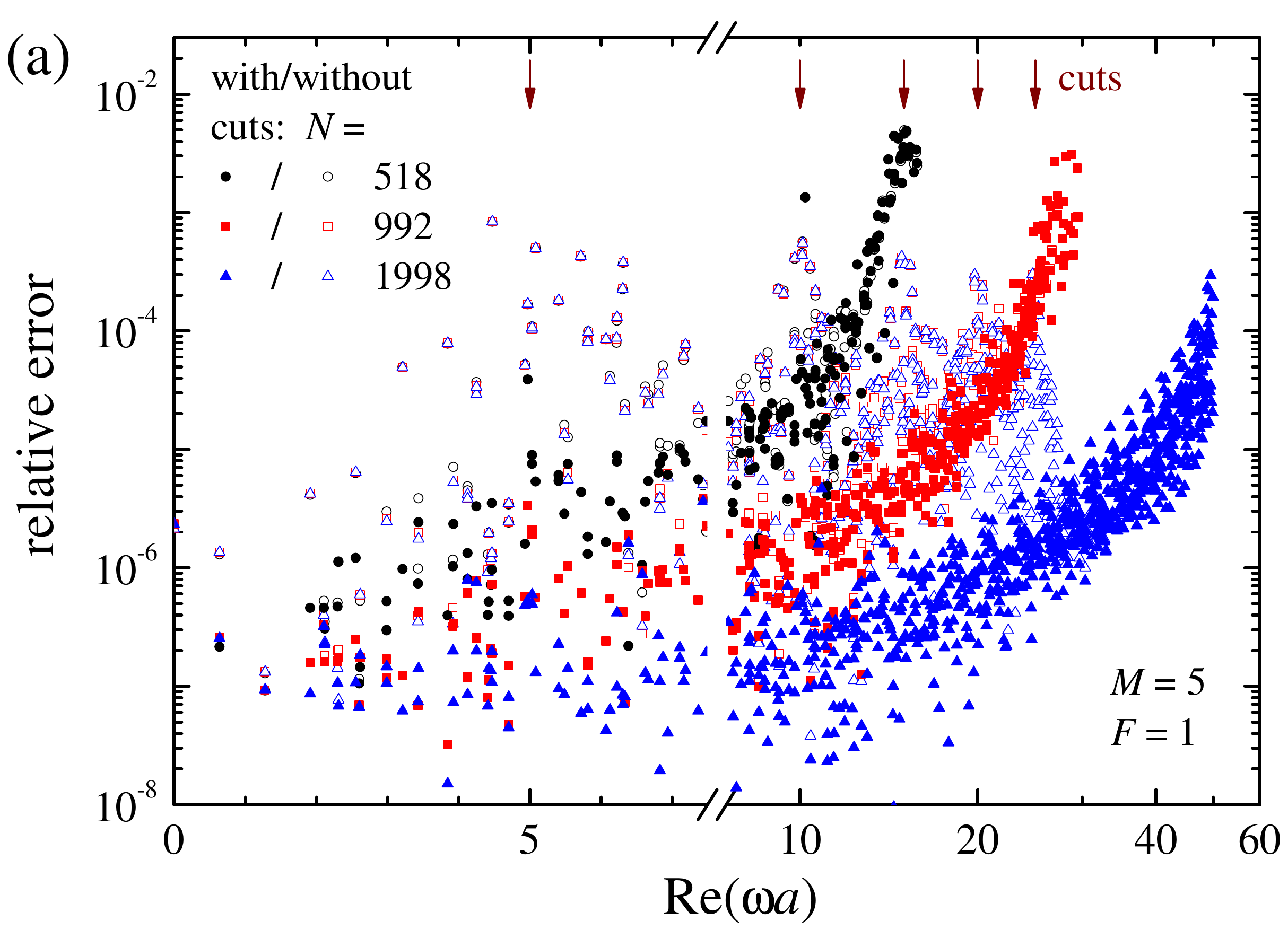}
		\includegraphics*[clip,width=0.48\textwidth]{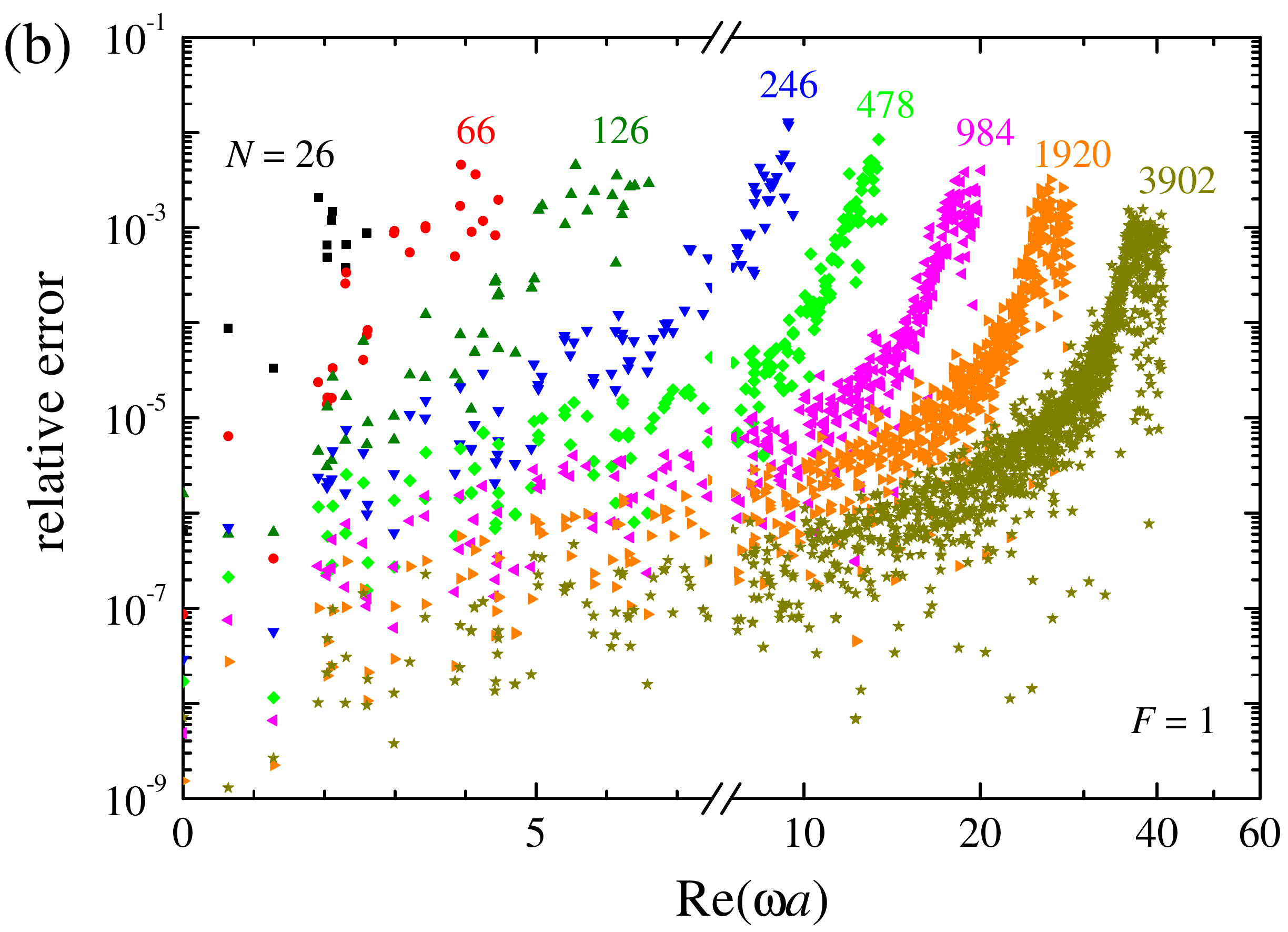}
		\caption{(a) Relative error of the PC-RSE compared to the SMM result, taking the latter as ``exact'',  calculated with and without cut modes, for different basis sized as labeled, $F=1$,  and parameters of the PC slab as in \Fig{crystal rse}. The basis consists of the closest to the origin modes, with the same number of modes for each Bragg channel, for $M=5$.
(b) Relative error of the PC-RSE using the $N\approx 6000$ result as ``exact'', for the same PC slab and the basis size as labeled. The basis consists of all RSs and cut modes within the circle $|\omega|<\omega_{\rm max}$ in the complex frequency plane.}
		\label{crystal rse error}
	\end{figure}

To quantify the agreement between the PC-RSE and SMM we study the relative error for the RS frequencies, which is shown in \Fig{crystal rse error}(a). Increasing the basis size $N$, the error is not improving for some RSs near the cuts, if the cut modes are not included in the basis. Including the cuts, the relative error gradually decreases for all RSs, as the basis size grows. Interestingly, the cut modes do not contribute to all the RSs evenly, and some RSs close to the cuts  show rather small errors, which are not improving much when the cut modes are included.

We also show in \Fig{crystal rse error}(b)	the relative error of the PC-RSE with respect to itself for a larger basis size, using the single truncation parameter $\omega_{\rm max}$, as described in \Sec{Sec:basis}.  We take in particular the eigenfrequencies calculated for the total basis size of $N_{\rm tot}\approx 12000$ as the ``exact'' values in finding the errors shown in \Fig{crystal rse error}(b) for different basis sizes. Here $N_{\rm tot}=N+FN$, where $N$ and $FN$ are, respectively, the number of RSs and cut modes in the basis. We study the dependence on $F$ of the relative error in \App{App:RSE-w} (see \Fig{w-RSE-homo}(c)) and show that $F\approx1$ presents an optimal value for
determining the RSs within a rather wide spectral range. This value of $F$ is thus used in the calculations presented in this section.
	
	\begin{figure*}
		\includegraphics*[clip,width=0.8\textwidth]{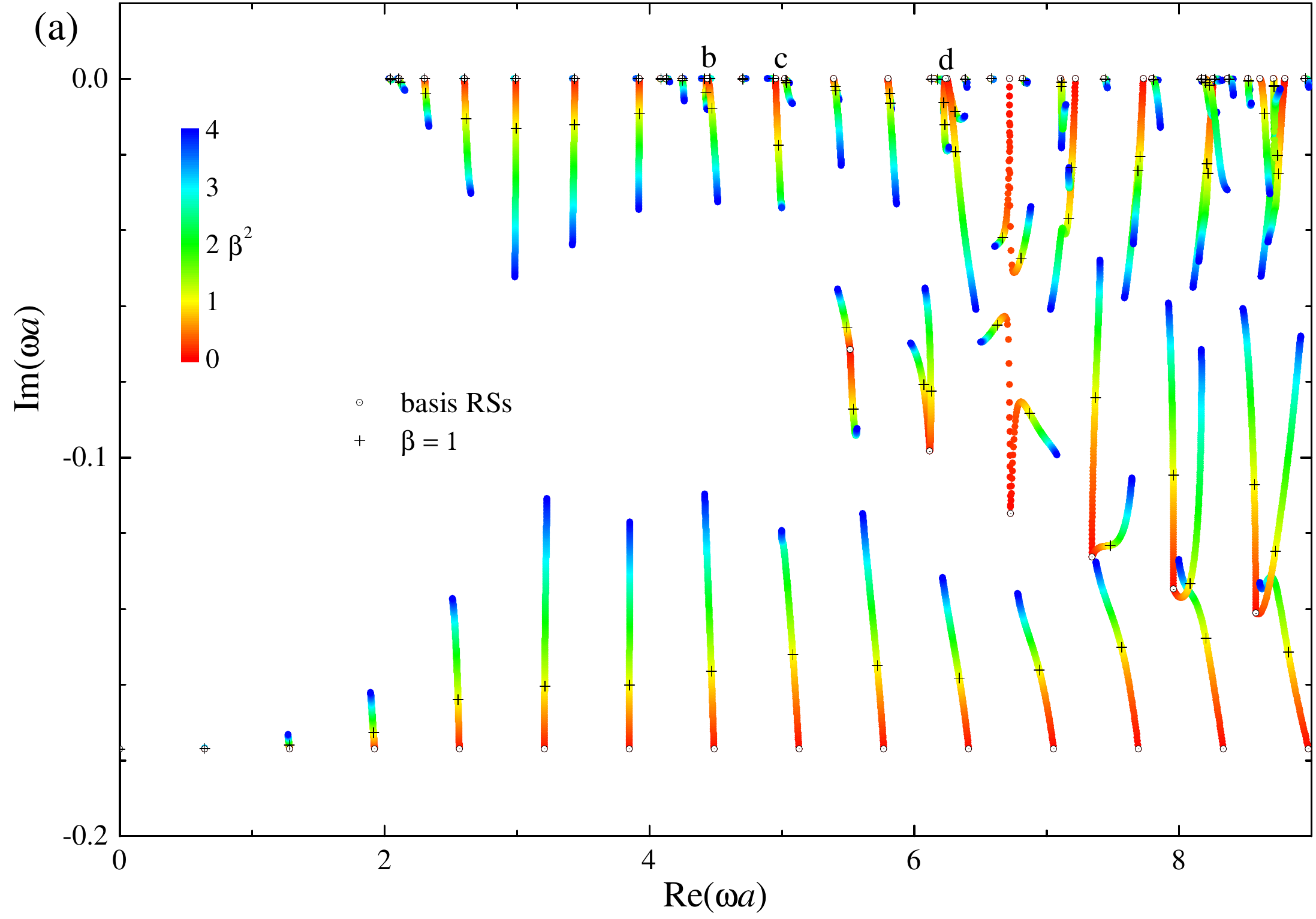}
\vskip2mm
		\includegraphics*[clip,width=0.495\textwidth]{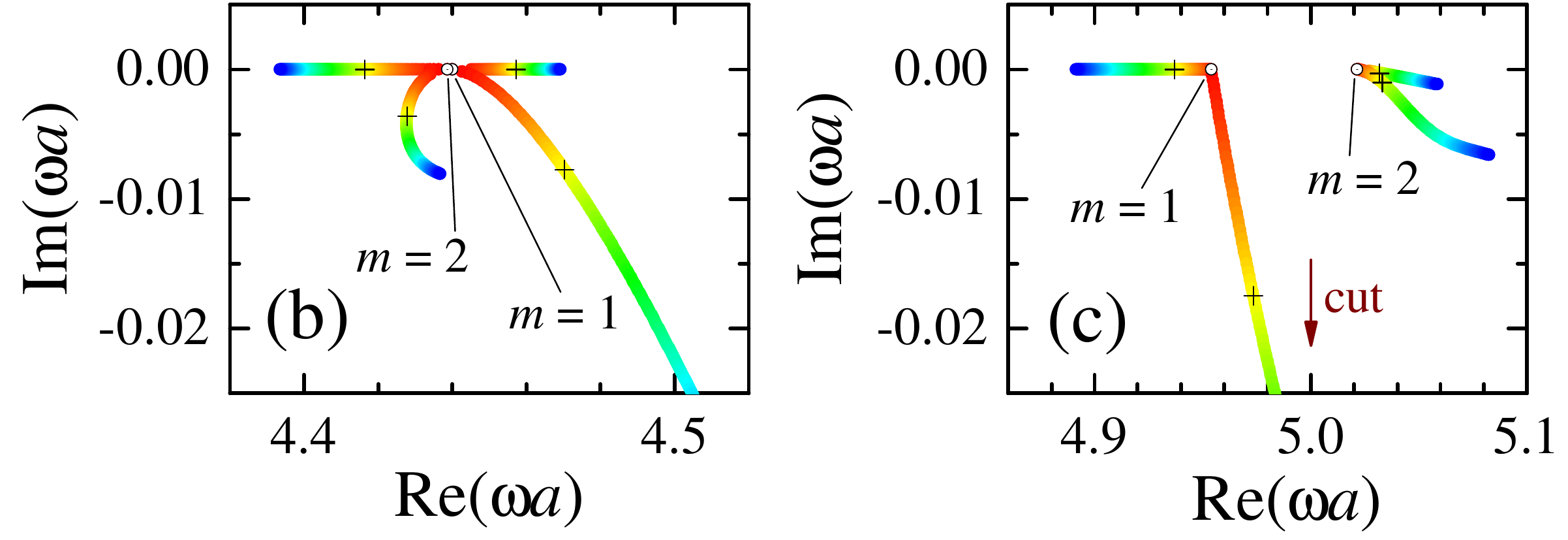}
		\includegraphics*[clip,width=0.485\textwidth]{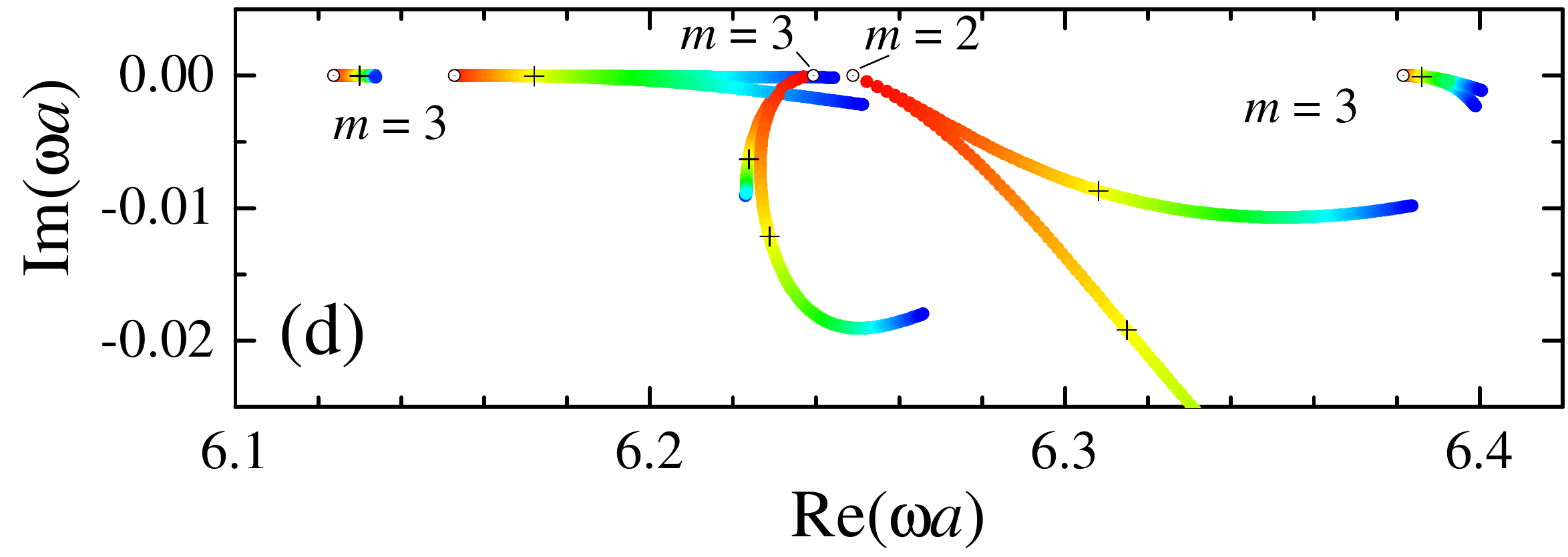}
		\caption{Evolution of the RS eigenfrequencies of a PC slab for the amplitude of the period modulation changing from $\beta=0$ to $\beta=2$ as given by the color code. Black dotted circles represent the unperturbed modes ($\beta=0$) and the black crosses the modes shown in Fig. \ref{crystal rse} ($\beta=1$). (b-d) Zoom-in of particular regions in the complex frequency plane close to the real axis, with the values of $m$ indicating the Bragg order of the unperturbed modes.}
		\label{Mode Shifts}	
	\end{figure*}

Looking at the dependence of the error on the basis size, presented in \Fig{crystal rse error}(b), we see that
the error decreases by roughly an order as the basis size doubles, which is close to the $1/N^3$ dependence observed for effective 1D systems treated by the RSE~\cite{MuljarovEPL10,DoostPRA12,DoostPRA14}. This demonstrates a high efficiency of the PC-RSE. In fact, its slowest element is matrix diagonalization for which the compute time scales as $N^3$. More important is however the overall level of errors: Even for 26 RSs and no cut modes in the basis, the perturbed RSs of the PC slab are calculated with the accuracy of about or less than $10^{-3}$. The same level of errors is seen in \Fig{crystal rse error}(a) when cut modes are not taken into account,  for a large number of RSs found within a much wider spectral range. This can be understood by the already mentioned fact that all diagonal elements $V_{\bar{n}\bar{n}}^{gg}=0$, leading to the effect of the perturbation vanishing in first-order, and therefore to a rather low level of corrections and errors.

A large number of the RSs shown in \Fig{crystal rse}(a) seem to have the imaginary part of the eigenfrequency close to zero. To see this more clearly, we zoom in the view of the imaginary part in \Fig{crystal rse}(b) by a factor of 100. This allows us to distinguish two types of modes. 
The first type is known as bound states in the continuum (BICs). These modes, much like WG modes in a planar waveguide, have strictly zero imaginary part and therefore infinite Q-factor or lifetime. However, unlike the WG modes, BICs lie in the spectral range where they could (but in reality do not) communicate with the photonic continuum outside of the system. The second type we call quasi-guided modes (QGMs)~\cite{TikhodeevPRB02} which usually have a very small but non-zero imaginary part of the eigenfrequency (high Q-factor), as compared to e.g. FP modes. This is again due to the dominant role of the WG modes in their formation, like for the BIC, while the small imaginary part is caused by the coupling between the WG and leaky modes, see \Sec{Sec:Mode evolution} below for a more detailed analysis of both types of RSs.

We also see in \Fig{crystal rse}(b) some failures of the SMM, showcasing the superiority of the RSE method. The SMM ultimately relies on the Newton-Raphson method of finding the poles of the S-matrix~\cite{TikhodeevPRB02,BykovJLT13}. This means that it uses a small but finite tolerance playing the role of the parameter. If the mode splitting is below the tolerance level, the SMM is unable to resolve them, such as in region $1$ of \Fig{crystal rse}(b). Reducing the tolerance can fix this issue, however, with a potential to generate at the same time spurious solution at another place, such as in region $2$ of \Fig{crystal rse}(b). The RSE in turn does not require a tolerance and returns the correct number of RSs in a selected region, neither missing any modes nor producing any spurious solutions.
This is an important and unique property of the RSE, following from the completeness of the basis used. Owing to its linearity in $\omega$, the RSE always returns as output a set of perturbed modes which is also complete. Furthermore, the number of perturbed modes is always equal to the number of basis modes used.
	

\subsection{Origin and evolution of the RSs in a PC slab}
\label{Sec:Mode evolution}

Here we use the advantage of the PC-RSE being an efficient and accurate tool for finding the complete set of the RSs of a PC structure and study the origin of the RSs in a PC slab, their formation and further evolution with change of the system parameters. Here we change the most important parameter: the amplitude of the periodic modulation $\beta$. \App{App:other} also presents results for varying the thickness of the modulation layer $2b$ and its period $d$.

Figure~\ref{Mode Shifts} shows the evolution of the RSs eigenfrequencies with increase of the amplitude $\beta$ of the periodic modulation, which we call the perturbation strength. In the $p=0$ case treated here, all unperturbed RSs (and cut modes) except $m=0$ channel are doubly degenerate, as discussed in \Sec{Sec:basis}. The periodic modulation lifts this degeneracy, which is well seen in the figure. Furthemore, leaky and FP modes shift upwards, increasing their Q-factors, while the majority of the RSs originating from WG modes are moving down, away from the real axis, in this way having their Q-factors reduced with $\beta$. Overall, this picture demonstrates a complicated mixing of basis modes with both infinite and finite Q-factors.

	\begin{figure}
		\includegraphics*[clip,width=0.48\textwidth]{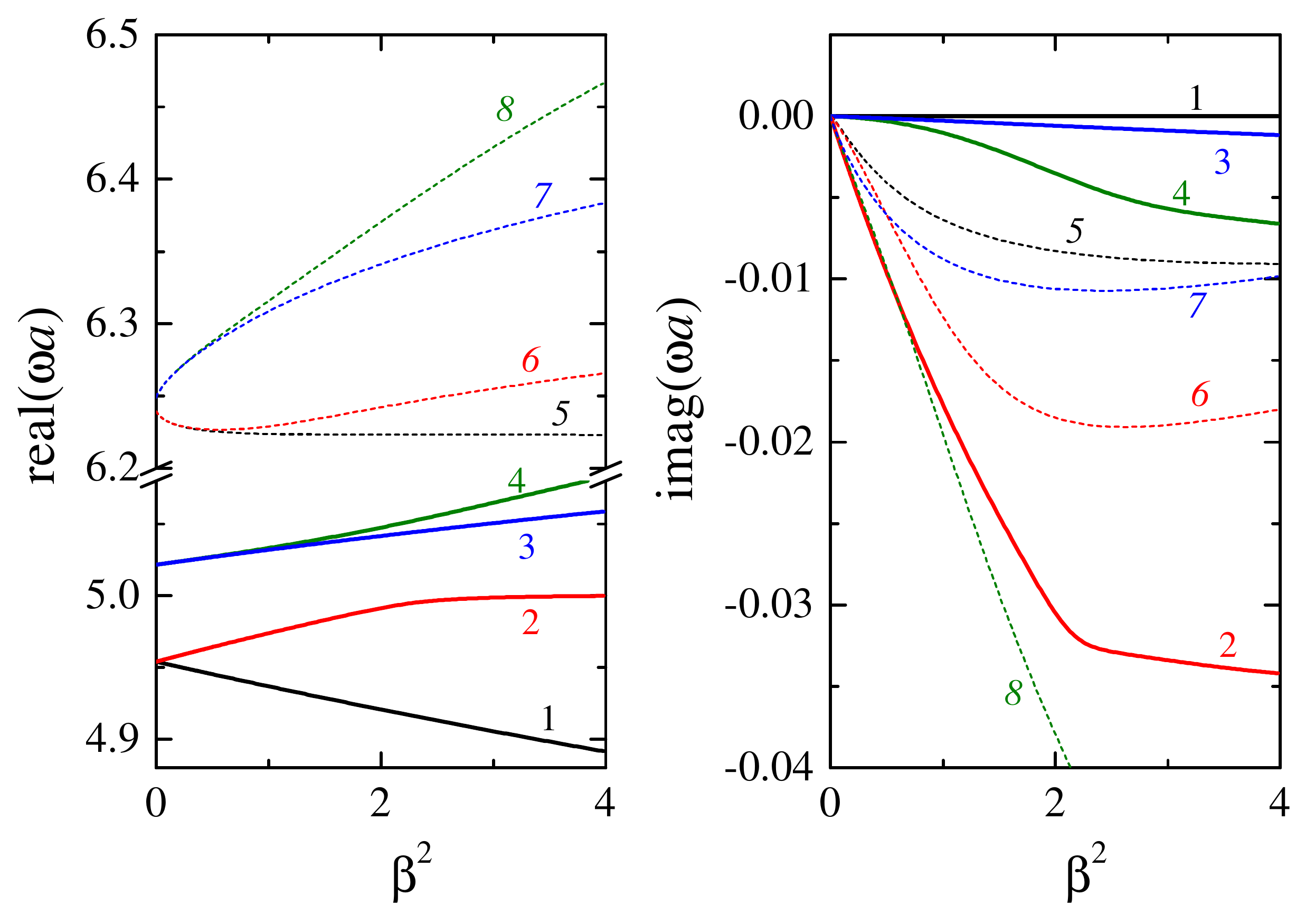}
		\caption{The real and imaginary parts of the modes shown in Fig \ref{Mode Shifts}(c) and four central modes in Fig \ref{Mode Shifts}(d)  as function of the permittivity modulation amplitude.}
		\label{Mode Shifts against beta}	
	\end{figure}

	\begin{figure}
		\includegraphics*[clip,width=0.48\textwidth]{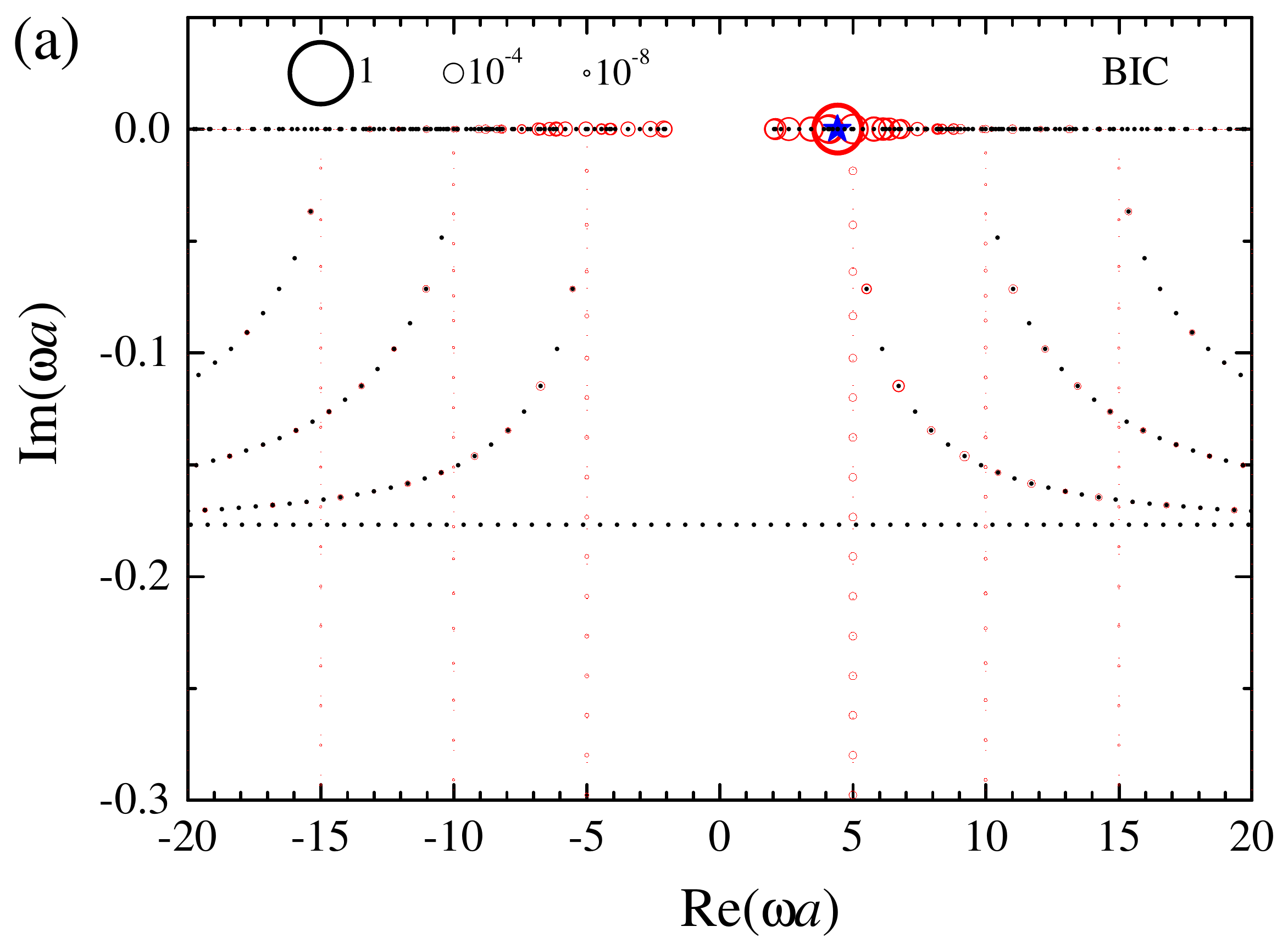}
		\includegraphics*[clip,width=0.48\textwidth]{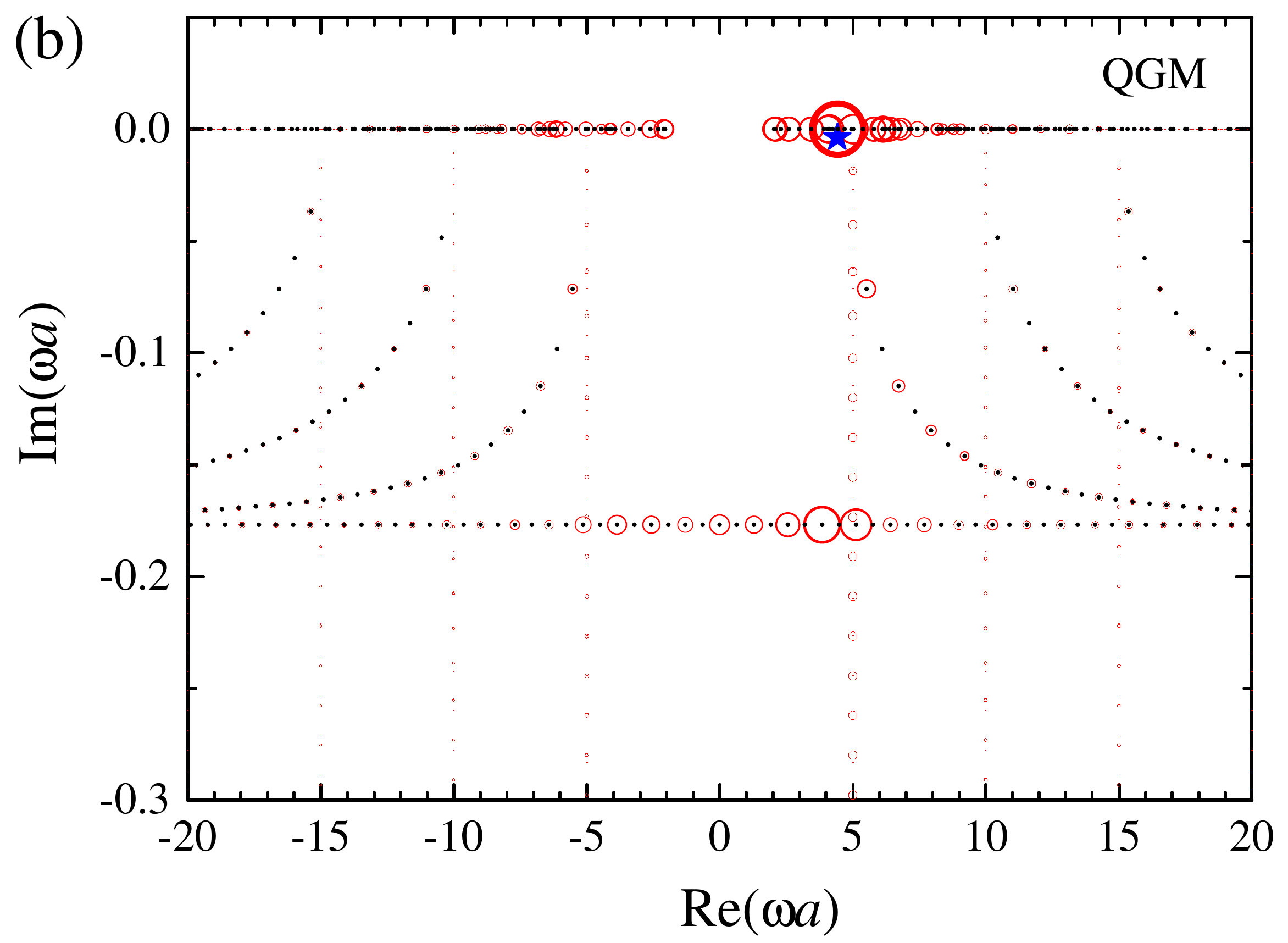}
		\caption{Basis mode contribution to (a) bound state in the continuum and (b) quasi-guided mode, a pair of modes originating from $m=1$ WG mode and shown by black crosses  in \Fig{Mode Shifts}(c). The area of each circle is proportional to $\sqrt{|c_n^g|}$. A key showing the relationship between the circle area and $|{c_n^g}|^2$ is given as black circles.
}
		\label{Mode Contributions}	
	\end{figure}

Of particular interest are RSs found close to the real axis of $\omega$. As mentioned in \Sec{Sec:verification} above, perturbing the WG modes results in two types of RSs, BICs and QGMs. A closer look provided in \Fig{Mode Shifts}(b-d) shows however that BICs are formed only within the region $-2\pi/d<{\rm Re}\,\omega<2\pi/d$, bound by the cuts of the $m=\pm1$ Bragg channels.
Figure~\ref{Mode Shifts}(b) demonstrates that each pair of WG modes within this region produces one BIC and one QGM, the latter loosing its quality very quickly with the perturbation strength. Outside this region we see instead that the RSs originating from the WG modes of the homogeneous slab have either Q-factors also quickly decreasing with $\beta$ or very high Q-factors weakly depending on the strength of periodic modulation. The latter can be called quasi-BICs, the term which has recently become widely used in the literature for such modes~\cite{TaghizadehAPL17,BulgakovPRA19}.  It is also interesting to see a formation of some of the RSs as a result of a rather strong coupling between WG modes which are close in frequency but belong to  different Bragg channels, see e.g in \Fig{Mode Shifts}(b) (\Fig{Mode Shifts}(d)) the mode repulsion due to the coupling between WG modes of $m=2$ and $m=1$ ($m=3$ and $m=2$) channels.

As already discussed in \Secs{Sec:permittivity}{Sec:verification}, the perturbation does not contribute in first-order  (for $\alpha=0$), and thus the RSs eigenfrequencies change  $\propto\beta^2$ for small $\beta$, in accordance with Eq.\,(38) of~\cite{DoostPRA14}, see  \Fig{Mode Shifts against beta}. However, in the case of the above mentioned strong coupling between the channels, this linear in $\beta^2$ regime takes place only at very low values of $\beta$. Another interesting feature well seen in  \Fig{Mode Shifts against beta} is that the degenerate pair of basis WG modes producing a BIC-QGM pair shows a linear in $\beta^2$ splitting, while any other pair of states, not containing BICs, remains degenerate in this order.  This makes BICs even more peculiar states.

To understand this and some other properties of BICs, we look at the basis mode contribution to different RSs.  We show in \Fig{Mode Contributions} the mode contribution to the BIC-QGM pair selected close to the cut of the 1st Bragg channel. Other types of modes -- FP, leaky and cut modes -- are considered in \App{App:cng}. The size of each circle represents how much that mode contributes, with the circle area proportional to  $\sqrt{|c_n^g|}$, chosen (instead of the natural $|c_n^g|^2$) in order to demonstrate more clearly the role of different basis modes. We see that all the basis RSs and cut modes of the given parity  contribute to these states, even though the relevant WG mode has the dominant and indeed very large contribution to both states. It becomes also clear that the main difference between the BIC and the QGM in the studied pair is that the leaky modes of the 0th Bragg channel do not contribute to BIC. This confirms that the BICs found in this system are symmetry-protected~\cite{BulgakovPRA14,ZhenPRB14,HsuNat16,BykovPRA19}.

In fact, owing to the symmetry of the system, this BIC has a wave function which is odd in the $x$ direction and thus does not couple to the $m=0$ channel containing only even states. In other words, the $m=0$ channel is not present in the subgroup of odd states, and therefore this BIC is not even falling into the continuum:  For this subgroup, the continuum starts at the cut positions of the $m=\pm1$ channels. Furthermore, all doubly degenerate basis states contribute to the QGM (BIC) with the same (opposite) amplitude, $c_n^g=\pm c_n^{-g}$, also reflecting selection by symmetry.

Note that the other mirror symmetry of the system, which is in the $z$-direction, also results in the formation of the basis and perturbed RSs of even and odd parity. It is clear, for example, that every other FP mode does not contribute to the states in \Fig{Mode Contributions}. Indeed, modes of the opposite parity in $z$ do not couple to each other, and both the QGM and BIC shown in \Fig{Mode Contributions} are of even parity. This implies in particular that the basis size can be halved for this kind of perturbations.

It is also clear from \Fig{Mode Contributions} that the cut modes contribute very little to the perturbed BIC and QGM but are nonetheless required for accuracy. We have seen, in particular that including even one cut mode, representing the full cut, is far better than not including any cut modes, despite the cut being very badly approximated.

Finally, the observed splitting $\propto\beta^2$ within the BIC-QGM pair can be understood as a result of the leaky modes of the $m=0$ Bragg channel affecting (not affecting) the QGM (BIC). If it contributes, the effect of the leaky modes appears already in the 2nd order in the perturbation $\beta$, and thus causes the splitting of the BIC and QGM in this order.

	\section{Conclusions}

The present paper offers a novel concept in the theory of photonic-crystal structures,  revealing the nature of resonances in the optical spectra and quantifying them precisely. This concept is based on using the analytical  resonant states of a homogeneous slab, along with its cut modes responsible for Rayleigh-Wood anomalies, as the most natural and simple basis for expanding the resonant states of a photonic-crystal slab. We present photonic-crystal resonant-state expansion (PC-RSE), capable of doing this accurately and efficiently, and illustrate it on examples of TE-polarized modes in a photonic crystal slab with a 1D harmonic modulation. These examples provide a proof of concept and also a verification of the PC-RSE by comparing it with the accurate scattering-matrix method.

We present a general formalism of the PC-RSE and its application to dielectric photonic-crystal structures. We demonstrate that the PC-RSE is an asymptotically exact approach, which  (i) depends on a single parameter -- truncation  frequency $\omega_{\rm max}$, determining the basis size;
(ii) guarantees completeness, i.e. has no missing or spurious modes, so that any observable can be represented as a superposition of the modes found;
(iii) technically reduces solving Maxwell's wave equation to a matrix diagonalization, thus making the application of the PC-RSE a fully automated and straightforward procedure, not requiring any supervision.
The PC-RSE provides an accurate and efficient tool for calculating all physically relevant eigenstates of the PC system within the selected spectral range. In particular, it can be effectively used repeatedly many times or in parallel, in order to investigate an arbitrarily large space of the physical parameters characterizing the system, for revealing and optimizing its fundamental properties. One of the immediate important applications of the PC-RSE is optimization of photonic-crystal cavities, with a correct account of the radiative losses of high-quality modes.

We have also demonstrated how the PC-RSE can be used to study the origin and physical properties of the optical modes. For illustration, we have traced the evolution of the optical modes in a photonic crystal slab increasing the amplitude of the periodic modulation of the permittivity, starting from a homogeneous slab with no modulation. This allowed us, in particular, to reveal the dominant role of the waveguide modes in the formation of bound states in the continuum and quasi-guided modes. Furthermore, the PC-RSE allows us to quantify precisely the contribution of each basis state to the optical mode of interest, which we have also demonstrated in the present work.

\acknowledgments S.N. acknowledges support by the EPSRC under the DTA scheme.

	\appendix
\section{Dyadic Green's function of a homogeneous slab}
\label{App:GF}

Let us consider arbitrary dependencies of the permittivity $\eps(z)$ and permeability $\mu(z)$ within a homogeneous slab occupying the region $|z|\leqslant a$ and surrounded by vacuum, so that outside the slab $\eps=\mu=1$. Denoting the components of the dyadic GF as $(\GF_p)_{ij}=G_{ij}(z,z')$, \Eq{GF-equ} becomes
\be
\left(\begin{array}{ccc}
\omega\eps&\p_z&-ip\\
-\p_z&\omega\mu&0\\
ip&0&\omega\mu
\end{array}\right)
\left(\begin{array}{ccc}
G_{11}&G_{12}&G_{13}\\
G_{21}&G_{22}&G_{23}\\
G_{31}&G_{32}&G_{33}
\end{array}\right)
=\one\delta(z-z')\,.
\label{GFij-equ}
\ee
Owing to the reciprocity of the optical system, the GF has the following property:
\be
G_{ij}(z,z')=G^\dagger_{ji}(z',z)\,,
\label{reciprocity}
\ee
where the adjoint $\dagger$ means replacing $p\to-p$.

From \Eq{GFij-equ} we obtain for the first column of $\GF_p$:
\bea
\hS(\mu;\omega)G_{11}(z,z';\omega)&=&\omega\mu(z)\delta(z-z')\,,
\label{G11}
\\
G_{21}(z,z')&=&\frac{1}{\omega\mu(z)}\p_zG_{11}(z,z')\,,
\label{G21}
\\
G_{31}(z,z')&=&-\frac{ip}{\omega\mu(z)}G_{11}(z,z')\,,
\eea
where the operator $\hS$ is defined as
\be
\hS(\zeta;\omega)=\zeta(z)\p_z\frac{1}{\zeta(z)}\p_z +\omega^2\eps(z)\mu(z)-p^2
\label{U}
\ee
with $\zeta(z)$ being a weight function.
For the second column of $\GF_p$, it follows from \Eq{GFij-equ} that
\bea
G_{12}(z,z')&=&-\frac{1}{\omega\chi(z)}\p_zG_{22}(z,z')\,,\\
\hS(\chi;\omega)G_{22}(z,z')&=&\omega\chi(z)\delta(z-z')\,,
\label{G22}
\\
G_{32}(z,z')&=&-\frac{ip}{\omega\mu(z)}G_{12}(z,z')\,,
\eea
where
\be
\chi(z)=\eps(z)-\frac{p^2}{\omega^2\mu(z)}\,.
\ee
Note that \Eq{G22} is essentially the same as \Eq{G11}, provided that $\mu(z)$ is replaced with $\chi(z)$.
Finally, for the third column of $\GF_p$ we obtain
\bea
\hS(\mu;\omega)G_{13}(z,z')&=&ip\delta(z-z')\,,
\label{G13}
\\
G_{23}(z,z')&=&\frac{1}{\omega\mu(z)}\p_zG_{13}(z,z')\,,
\label{G23}
\\
G_{33}(z,z')&=&-\frac{ip}{\omega\mu(z)}G_{13}(z,z')+\frac{\delta(z-z')}{\omega\mu(z)}\,,\nonumber\\
\eea
\vskip-3mm
\noindent demonstrating in particular that the longitudinal component $G_{33}(z,z')$ is divergent at $z=z'$, due to the $\delta$ function in the last term. Also, \Eq{G13} contains exactly the same operator as in \Eq{G11}, and therefore
\be
G_{13}(z,z')= \frac{ip}{\omega\mu(z')} G_{11}(z,z')\,.
\label{G13G11}
\ee

Using the reciprocity \Eq{reciprocity} and \Eqsss{G21}{G23}{G13G11}, we further obtain
\bea
G_{12}(z,z')&=&G_{21}(z',z)=\frac{1}{\omega\mu(z')}\p_{z'}G_{11}(z,z')\,,\\
G_{32}(z,z')&=&G_{23}^\dagger(z',z)=-\frac{ip}{\omega^2\mu(z)\mu(z')}\p_{z'}G_{11}(z,z')\,,
\nonumber\\
\eea
\vskip-3mm
\noindent and therefore
\bea
G_{22}(z,z')&=&\frac{1}{\omega\mu(z)}\p_zG_{12}(z,z')+\frac{\delta(z-z')}{\omega\mu(z)}\nonumber\\
&=&\frac{1}{\omega^2\mu(z)\mu(z')}\p_z\p_{z'}G_{11}(z,z')+\frac{\delta(z-z')}{\omega\mu(z)}\,.\nonumber\\
\eea
\vskip-3mm
\noindent
Finally,
\be
G_{13}(z,z')=G_{31}^\dagger(z',z)=\frac{ip}{\omega\mu(z')}G_{11}(z,z')\,,
\ee
in agreement with \Eq{G13G11}.
Collecting all this information about the GF components, we find a compact expression for the full dyadic GF:
\bea
\GF_p(z,z')&=&\hO_p(z;\omega)\otimes\hO_{-p}(z';\omega)G_{11}(z,z')\nonumber\\
&&+\frac{\one_2+\one_3}{\omega\mu(z)}\delta(z-z')\,,
\label{GFgen}
\eea
where  $\one_j$ are $3\times3$ matrices with elements $(\one_j)_{ii'}=\delta_{ii'}\delta_{ij}$
and $\hO_p$ is a vector operator defined as
\be
\hO_p(z;\omega)=
\left(\begin{array}{c}
1\\
\frac{1}{\omega\mu(z)}\p_z\\
-\frac{ip}{\omega\mu(z)}
\end{array}\right)\,.
\label{O}
\ee

Solving \Eq{G11} with outgoing boundary conditions and studying the pole structure and the cuts of the GF in the complex $\omega$-plane,  as done in \App{App:ML} for a general planar system, allows us to find the ML expansion for $G_{11}$:
\be
G_{11}(z,z)=\sumint{n}\frac{E_n(z)E_n(z')}{\omega-\omega_n}\,,
\label{G11-ML}
\ee
and also for
\be
g(z,z')= \frac{G_{11}(z,z)}{\omega}=\sumint{n}\frac{E_n(z)E_n(z')}{\omega_n(\omega-\omega_n)}\,,
\label{g-ML}
\ee
where $E_n(z)$ is the only non-vanishing component (along $y$) of the electric field of the RS $n$, satisfying an equation
\be
\hS(\mu;\omega_n)E_n(z)=0\,,
\ee
or $ E_n(z)$  is the wave function of the cut pole, see \App{App:ML} for details.
Obviously, $ E_n(z)$ depends on $p^2$, as so does the operator $\hS$, and thus $ E_n(z)$ is not sensitive to a change of sign of $p$, so that $ E_n^\dagger(z)=E_n(z)$.

Substituting \Eqs{G11-ML}{g-ML} into \Eq{G11}, we obtain a closure relation
\be
\eps(z)\sumint{n}E_n(z)E_n(z')=\delta(z-z')
\label{closure}
\ee
and a sum rule
\be
\sumint{n}\frac{E_n(z)E_n(z')}{\omega_n}=0\,.
\label{sum}
\ee
Using \Eqsss{G11-ML}{closure}{sum}, we obtain from \Eq{GFgen} the ML expansion \Eq{GF-ML} of the full dyadic GF, in which
\be
\Fc_n(z;p)=\hO_p(z;\omega_n)E_n(z)\,.
\label{Fc}
\ee
Note that in general, one also needs to include in \Eq{GF-ML} for completeness longitudinal static modes, in order to take into account the effect of the $\omega=0$ pole of the GF.

\section{Mittag-Leffler expansion of the GF of a general homogeneous planar system}
\label{App:ML}

Let us now consider the scalar GF $g(z,z')$ of a general homogeneous system, described by arbitrary functions $\eps(z)$ and $\mu(z)$, and derive its ML expansion given by \Eq{g-ML}. The GF satisfies an equation
\be
\hS_{k^2}(z) g(z,z')= \mu(z)\delta(z-z')\,,
\ee
where the operator $\hS_{k^2} (z)$ is given by
\be
\hS_{k^2} (z) =\mu(z)\p_z\frac{1}{\mu(z)}\p_z +k^2\eps(z)\mu(z)+p^2[\eps(z)\mu(z)-1]\,,
\label{g-equ}
\ee
in accordance with it definition \Eq{U}, now written in terms of $k^2=\omega^2-p^2$.

To simplify the derivation, we assume that $\eps(z)=\mu(z)=1$  outside the system ($|z|>a$). In this case \Eq{g-equ} outside the system takes the form
\be
(\p_z^2 +k^2)g(z,z')= \delta(z-z')
\ee
with $k$ being the normal component of the wave number in vacuum. Applying the outgoing wave boundary conditions, we find that
\be
g(z,z')=g(\pm a,z') e^{\pm ikz} \ \ \ {\rm for}\ \ |z|>a\ \ {\rm and}\ \ |z'|<a\,,
\ee
where $+$ ($-$) refers to $z>a$ ($z<-a$). The outgoing wave boundary conditions for solving \Eq{g-equ} can therefore be written in the following way
\be
\left(\frac{1}{\mu(z)}\p_z \mp ik\right)\!\!\left.g(z,z')\right|_{z=\pm a}=0\,,
\label{g-BC}
\ee
explicitly showing that $g(z,z')$ is an {\em analytic} function of $k$. Having a countable number of simple poles in the complex $k$-plane, which are at the RS wave numbers, $k_n=\sqrt{\omega_n^2-p^2}$, and being vanishing at $k\to\infty$, the GF $g(z,z')$ can be written as
\be
g(z,z')=\sum_n\frac{R_n(z,z')}{k-k_n}\,,
\label{ML10}
\ee
using the Mittag-Leffler theorem~\cite{Arfken01}.
To find an explicit form of the residue $R_n(z,z')$ we use  Maxwell's wave equation without sources
\be
\hS_{k^2_n}(z) E_n(z)=0\,,
\label{E-hom}
\ee
determining  the RS wave functions $E_n(z)$, as well as the one with a source term~\cite{DoostPRA14,MuljarovOL18},
\be
\hS_{k^2}(z) E(z;k)=\mu(z)(k-k_n)\sigma_n(z)\,,
\label{E-inhom}
\ee
determining its analytic continuation $E(z;k)$ in the complex $k$-plane about the point $k_n$, such that  $E(z;k_n)=E_n(z)$. The source $\sigma_n(z)$ can be any function  vanishing outside the system and normalized in such a way that
\be
\int_{-a}^a E_n(z)\sigma_n(z) dz= k_n\,.
\label{norm-cond}
\ee
In optical systems with degenerate RSs (e.g due to symmetry), such that $k_n=k_m$ for $m \ne n$, $\sigma_n(z)$ is chosen in such a way that $\int_{-a}^a E_m(z)\sigma_n(z)dz =\delta_{nm}$.

Solving \Eq{E-inhom} with the help of the GF $g(z,z')$ and using its ML expansion \Eq{ML10}, we find
\be
E(z;k)=\sum_{n'}\frac{k-k_n}{k-k_{n'}} \int_{-a}^a R_n(z,z')\sigma_n(z') dz'\,.
\label{E-sol}
\ee
Then taking the limit $k\to k_n$, \Eq{E-sol} becomes
\be
E_n(z)=\int_{-a}^a R_n(z,z')\sigma_n(z') dz'\,,
\ee
which can be written, after combining it with \Eq{norm-cond}, as
\be
\int_{-a}^a \left[\frac{E_n(z)E_n(z')}{k_n}-R_n(z,z')\right]\sigma_n(z') dz'=0\,.
\ee
The last equation must be satisfied for any normalized $\sigma_n(z)$,  suited for generating the analytic continuation. Clearly, such a function is not unique, therefore the integrand in the last equation should be vanishing, which gives
$R_n(z,z')=E_n(z)E_n(z')/k_n$ and results in the following series for the GF:
\be
g(z,z')=\sum_n\frac{E_n(z)E_n(z')}{k_n(k-k_n)}\,.
\label{ML11}
\ee

We now find the normalization of $E_n(z)$ which is determined by the ML form \Eq{ML11}, which in turn follows from the normalization of the source \Eq{norm-cond}. We therefore use it again, substituting $\sigma_n(z)$ from \Eq{E-inhom} into \Eq{norm-cond} and subtracting a similar integral vanishing due to \Eq{E-hom}:
\bea
&&(k-k_n)k_n=\int_{-a}^a \frac{1}{\mu}E_n\hS_{k^2} E dz-
\int_{-a}^a \frac{1}{\mu}E\hS_{k^2_n} E_n dz\nonumber\\
&&= \left[\frac{E_n}{\mu}\p_z E-\frac{E}{\mu}\p_z E_n\right]_{-a}^a+(k^2-k_n^2)\int_{-a}^a E_n E\eps dz\,,
\nonumber
\eea
where the first term  in the second line is obtained integrating by parts. Finally, using the outgoing wave boundary conditions for $E$ and $E_n$,
\bea
\left(\frac{1}{\mu(z)}\p_z \mp ik\right)\!\!\left.E(z;k)\right|_{z=\pm a}&=&0\,, \\
\left(\frac{1}{\mu(z)}\p_z \mp ik_n\right)\!\!\left.E_n(z)\right|_{z=\pm a}&=&0\,,
\label{BC}
\eea
similar to \Eq{g-BC}, we arrive, after taking again the limit $k\to k_n$, at the normalization condition for the RS wave function:
\be
2\int_{-a}^a E^2_n(z) \eps(z) dz - \frac{E^2_n(a)+E^2_n(-a)}{ik_n}=1\,,
\label{RS-norm}
\ee
which is the same as the one provided in \cite{ArmitagePRA14} without proof.

Note that \Eq{RS-norm} is equivalent to the general normalization \Eq{gen-norm} used for the TE polarization. In fact,
using \Eq{gen-norm} for $z_1=-a$ and $z_2=a$ and the fields replaced by their analytic continuations for the purpose of taking the frequency derivatives, we find
\be
1=\int_{-a}^{a} \left[ \eps E^2+\mu(H_x^2-H_z^2)\right] dz
-\left[E\p_\omega H_x-H_x\p_\omega E\right]_{-a}^{a},
\label{norm3}
\ee
see \Eqsss{F}{O}{Fc}. Then using
\be
H_x=-\frac{1}{\omega\mu}\p_z E\,,\ \ \ H_z=\frac{ip}{\omega\mu} E\,,
\ee
and
\be
\left(\mu\p_z\frac{1}{\mu}\p_z +\omega^2\eps\mu-p^2\right)E=0\,,
\label{E-equ}
\ee
valid for $k=k_n$, we  find, integrating by parts:
\bea
&&\int_{-a}^{a} \mu(H_x^2-H_z^2)dz = \frac{1}{\omega^2}\int_{-a}^{a} \frac{1}{\mu}[(\p_zE)^2+p^2E^2]dz \nonumber\\
&&=\frac{[E\p_z E]_{-a}^a}{\omega^2}+\frac{1}{\omega^2}\int_{-a}^{a} \frac{1}{\mu}\left[-E\mu\p_z\frac{1}{\mu}\p_z E+p^2E^2\right]dz
\nonumber\\
&&=\frac{[E\p_z E]_{-a}^a}{\omega^2}+\int_{-a}^{a} \eps E^2dz\,.
\eea
We then use the analytic form of the fields outside the system:
\be
E(z;k)=A_\pm e^{\pm ikz}\,,\ \ \ H(z;k)=\mp\frac{ik}{\omega} E(z;k)\,,
\label{Eout}
\ee
where, again, $+$ ($-$) corresponds to $z\geqslant a$ ($z\leqslant-a$), and the amplitudes $A_\pm$ are also functions of $\omega$ or $k$. However, their frequency dependence does not contribute to the normalization, since
\be
E\p_\omega H_x-H_x\p_\omega E=\mp E^2\p_\omega\frac{ik}{\omega}=\mp \frac{p^2}{ik\omega^2}E^2\,.
\ee
Collecting all the ``surface'' terms and differentiating the field outside the system, using the explicit form of $E$ given by \Eq{Eout}, we obtain
\bea
&&-\left[E\p_\omega H_x-H_x\p_\omega E\right]_{-a}^{a} +\frac{[E\p_z E]_{-a}^a}{\omega^2}
\nonumber\\
&&=\left(\frac{ip^2}{k\omega^2}+\frac{ik}{\omega^2}\right)\left[E^2(a)+E^2(-a)\right]
\nonumber\\
&&=-\frac{E^2(a)+E^2(-a)}{ik}\,,
\eea
which proves that \Eq{gen-norm} results in the normalization given by \Eq{RS-norm}.
	\begin{figure}
		\includegraphics[width=0.48\textwidth]{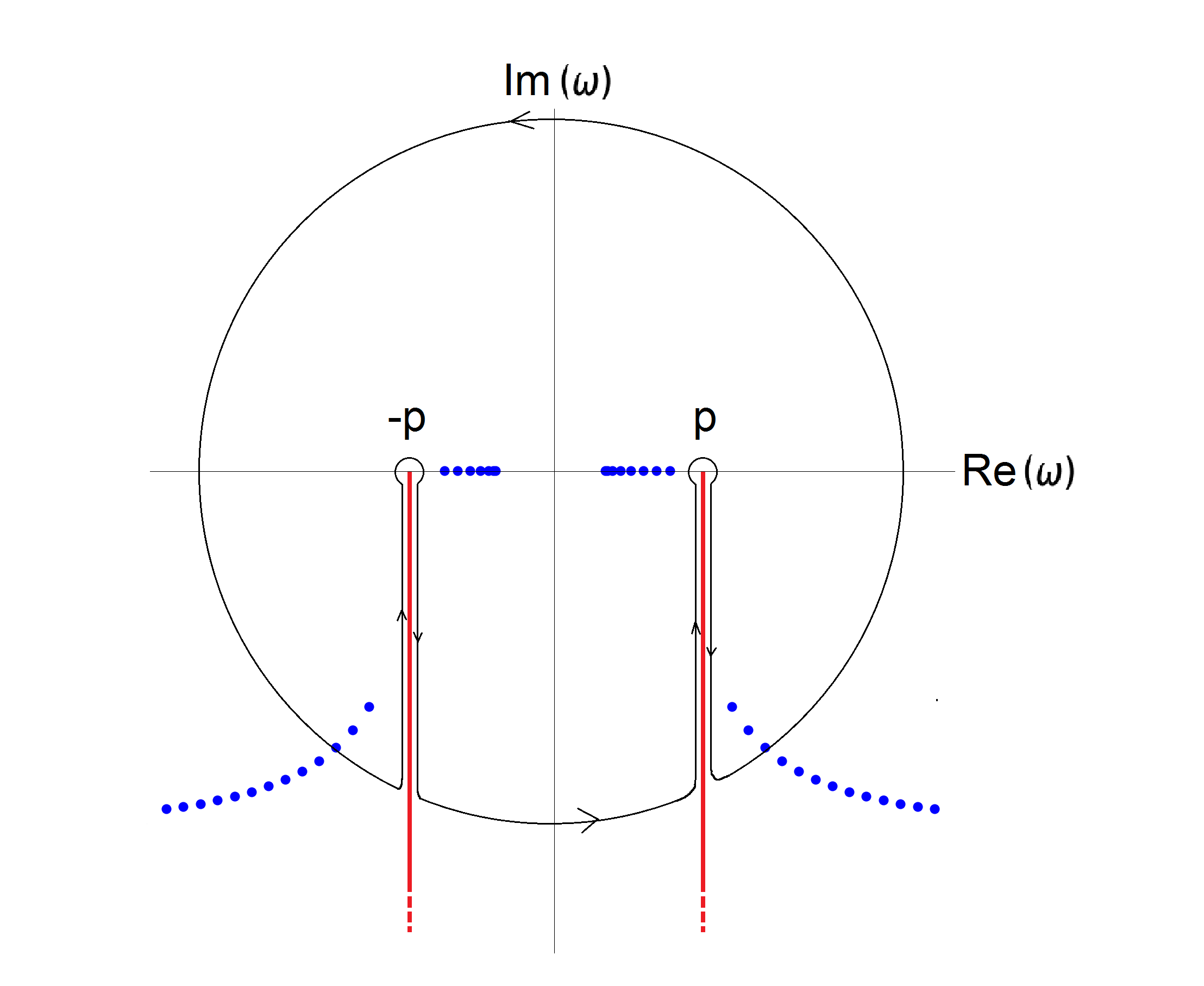}
		\caption{Poles (blue dots) and cuts (red lines) of the GF in the complex $\omega$ plane, as well the contour of integration leading to the ML expansion \Eq{ML21}, which includes the cut contributions displayed.}
\label{Fig:complex-w}
	\end{figure}

Equation (\ref{ML11}) is the ML series of the GF in $k$-representation. However, the RSE formulated in \Sec{Sec:Formalism} requires a ML form of the GF in the $\omega$-representation. Being treated as a function of frequency $\omega$, the GF has simple poles due to the RSs, at $\omega=\omega_n$ (where $\omega^2_n=k_n^2+p^2$), which are distributed in the complex $\omega$-plane symmetrically with respect to the imaginary axis, see \Fig{Fig:complex-w}. The residues of the GF at these poles are given by
\bea
\lim_{\omega\to\omega_n}g(z,z')(\omega-\omega_n)&=&\left.\frac{\p\omega}{\p k}\right|_{k_n}
\lim_{k\to k_n}g(z,z')(k-k_n)\nonumber\\
&=&\frac{k_n}{\omega_n} R_n(z,z')\,,
\eea
where $R_n(z,z')$ is found earlier, see \Eq{ML11}.

Apart from these poles, the GF $g(z,z')$ is analytic in the complex $k$-plane, as shown above. However, in the complex $\omega$-plane $g(z,z')$ has branch cuts, owing to the link between $k$ and $\omega$,
\be
k=\sqrt{\omega^2-p^2}\,,
\label{k}
\ee
with the branch points at $\omega=\pm p$. Therefore, applying the ML theorem in the frequency plane results instead in
\be
g(z,z')=\sum_n\frac{E_n(z)E_n(z')}{\omega_n(\omega-\omega_n)}+I_1(z,z';\omega)+I_2(z,z';\omega)\,,
\label{ML21}
\ee
where the sum includes only the RSs on the selected Riemann sheet, which in our case include all the WG and FP modes but does not contain anti-WG modes, unlike the series in \Eq{ML11}. Integrals
\be
I_{1,2}(z,z';\omega)=-\frac{1}{2\pi i} \int_{\pm p}^{\pm p-i\infty} \frac{\Delta g(z,z';\omega')}{\omega-\omega'} d\omega'\,,
\ee
describe the contribution of the cuts, which are chosen as vertical straight lines in the complex $\omega$-plane going from the branch points $\pm p $ down to $-i\infty$, see \Fig{Fig:complex-w}. According to~\cite{LobanovPRA17}, this choice of the cuts almost minimizes their contribution to the GF.
The jump of the GF value across the cut is given by the function
\be
\Delta g(z,z';\omega)=g_k(z,z')-g_{-k}(z,z')\,,
\label{delGF}
\ee
where we have added index $k$ for convenience, in order to emphasize the fact that $k$ is changing to $-k$ when going through the cut with an infinitesimal change of $\omega$.

The jump of the GF \Eq{delGF} can be evaluated in the general case by using the explicit form of the GF,
\be
g(z,z')=\frac{E_L(z_<)E_R(z_>)}{E_L\dot{E}_R-\dot{E}_L E_R}\,,
\label{GFexplicit}
\ee
in terms of the ``left'' and ``right'' functions, $E_L(z)$ and $E_R(z)$, respectively. These are solution of the homogeneous wave equation
(\ref{E-equ}) and the left or right outgoing boundary condition. More explicitly, they are given by
\be
E_L(z)=\left\{
\begin{array}{ll}
e^{-ikz} & z<-a \\
B_+ f_+(z)+B_- f_-(z) & |z|<a
\end{array}
\right.
\ee
and
\be
E_R(z)=\left\{
\begin{array}{ll}
C_+ f_+(z)+C_- f_-(z) & |z|<a \\
e^{ikz} & z>a
\end{array}
\right.
\ee
in terms of $f_+(z;\omega)$ and $f_-(z;\omega)$, two arbitrary linearly independent solutions of \Eq{E-equ} within the slab.
While these functions (depending on $\omega$) do not change when changing the sign of $k$, the coefficients $B_\pm$ and $C_\pm$ do modify, leading to non-vanishing contributions to the jump $\Delta g$ of the GF across the cuts. Here in \Eq{GFexplicit}, $z_<=\min(z,z')$ and  $z_>=\max(z,z')$, and the derivative
\be
\dot{E}(z)\equiv \frac{1}{\mu(z)}\p_z E(z)
\label{derivative}
\ee
is introduced for convenience.

Now, choosing the functions $f(z)$ and $g(z)$ in such a way that
\bea
f_+(a)f_-(a)+f_+(-a)f_-(-a)&=&0\,,
\label{f1}
\\
\dot{f}_+(a)\dot{f}_-(a)+\dot{f}_+(-a)\dot{f}_-(-a)&=&0\,,\\
f_+(-a)\dot{f}_+(a)+f_+(a)\dot{f}_+(-a)&=&0\,,\\
f_-(-a)\dot{f}_-(a)+f_-(a)\dot{f}_-(-a)&=&0\,,
\label{f4}
\eea
which can always be fulfilled, for any profile $\mu(z)$,
we obtain, after simple algebra, a convenient form of the cut integrand
\be
-\frac{1}{2\pi i}\Delta g(z,z';\omega)=\sum_{s=\pm}\sigma_s f_s(z)f_s (z')\,,
\ee
where
\be
\sigma_s(\omega)=\frac{1}{\pi}\,\frac{k}{\dot{f}_s^2(a)+\dot{f}_s^2(-a)+k^2[f_s^2(a)+f_s^2(-as)]}\,.
\label{sigma}
\ee
This form allows us to include the contribution of the cuts on equal footing with the RSs, treating the cuts as continua of poles of the GF:
\bea
g(z,z') &=&\sum_{n}
\frac{E_n(z)E_n(z')}{\omega_n(\omega-\omega_n)}
\nonumber\\
&&+\sum_{s=\pm}\sum_{s'=\pm}
\int_{s' p}^{s' p-i\infty} \frac{E_s(z;\omega')E_s(z';\omega')}{\omega'(\omega-\omega')}d\omega'
\nonumber\\
&\equiv&\sumint{n}\frac{E_n(z)E_n(z')}{\omega_n(\omega-\omega_n)}\,,
\label{ML22}
\eea
where
\be
E_s(z;\omega)=\sqrt{\omega\sigma_s(\omega)} f_s(z;\omega)\,.
\label{Es}
\ee
Note that in \Eqsss{sigma}{ML22}{Es}, we have added $\omega$ to the arguments of $E_s$, $\sigma_s$, and $f_s$, in order to emphasize their frequency dependence, earlier omitted for brevity of notations.

\section{Homogeneous slab with constant permittivity and permeability}
\label{App:HS}

	
Consider a dielectric slab in vacuum, having thickness $2a$ and constant permittivity and permeability. Their profiles in space are described by
\bea
\eps(z)&=&1+(\epsilon-1)\Theta(a-|z|)\,,
\label{eps-hom}
\\
\mu(z)&=&1+(\mu-1)\Theta(a-|z|)\,.
\eea
Within the slab ($|z|,\,|z'|\leqslant a$) the GF has the form
\be
g(z,z')=-\frac{\mu}{2iq}\frac{\varphi(z_<)\varphi(-z_>)}{1-\xi^2}\,,
\label{GFanal}
\ee
see \Eq{GFexplicit} in which, due to the mirror symmetry, the left and right solutions are given by the same function, $E_L(z)=E_R(-z)=\varphi(z)$, with
\bea
\varphi(z)&=&e^{iqz}+\xi e^{-iqz}\,,
\label{phi}\\
\xi&=& \frac{1+\eta}{1-\eta} e^{-2iqa}\,,\ \ \ \ \ \eta=\frac{\mu k}{q}\,,
\label{s}
\\
q^2 &=&\epsilon\mu\omega^2-p^2 \,,\ \ \ \ \ \  k^2=\omega^2-p^2\,.
\label{qk}
\eea
Clearly, the GF has poles at $\xi=\pm1$, determining secular equation for the RS frequencies $\omega_n$:
\be
(q_n+\mu k_n)e^{-iq_n a}=(-1)^n(q_n-\mu k_n)e^{iq_n a}\,.
\label{secular}
\ee
The RS wave functions, which are the solutions of \Eq{E-hom}, are given by
\be
E_n(z)=
\begin{cases}
A_n e^{ik_n z} &  z>a\\
B_n(e^{-iq_n z}+(-1)^n e^{-iq_n z}) & |z|\leqslant a\\
(-1)^n A_n e^{-ik_n z} & z< -a
\end{cases}
\label{RSwf}
\ee
with the continuity condition $A_n=B_n(e^{-iq_n a}+(-1)^n e^{-iq_n a})e^{-ik_n a}$. The eigenvalues $k_n$ and $q_n$ are related to the eigenfrequency $\omega_n$ via \Eq{qk}, and the normalization constants $B_n$ found from \Eq{RS-norm} have the following explicit form:
\be
B_n^{-2}=8(-1)^n\left[\epsilon a+\frac{ip^2(\epsilon\mu-1)}{k_n(q_n^2-\mu^2 k_n^2)}\right]\,.
\label{Bn}
\ee

For the cuts of the GF in the complex $\omega$-plane, the functions satisfying Eqs.\,(\ref{f1})--(\ref{f4}) are given by
\be
f_\pm(z)=e^{iqz}\pm e^{-iqz}
\ee
within the slab $|z|\leqslant a$. They possess a definitive parity ($s=\pm 1$) due to the mirror symmetry of the system, and the same form as the RS wave functions \Eq{RSwf}.
According to \Eq{sigma}, the cut density functions are given by
\be
\sigma_\pm(\omega)= \frac{1}{4\pi}\frac{\mu^2 k}{(\mu^2 k^2-q^2)\cos (2qa)\pm (\mu^2 k^2+q^2)}\,.
\ee

\subsection{GF in $k$-representation}

	
Using the Newton-Raphson method, we have solved the secular equation \Eq{secular} and found all the RS wave numbers in a selected frequency range (within a circle of radius $\omega_{\rm max}$ in the complex frequency plane). The RSs include four categories of modes: WG modes, anti-WG modes, FP modes, and a leaky mode (LM). The WG and anti-WG modes are present only if $p\neq0$. We then use the ML expansion \Eq{ML11} in the $k$-representation which includes all types of modes in the summation and compare it with the analytic GF given by \Eq{GFanal}

	\begin{figure}
		\includegraphics*[trim={0cm 0cm 0cm 0cm},clip,width=0.48\textwidth]{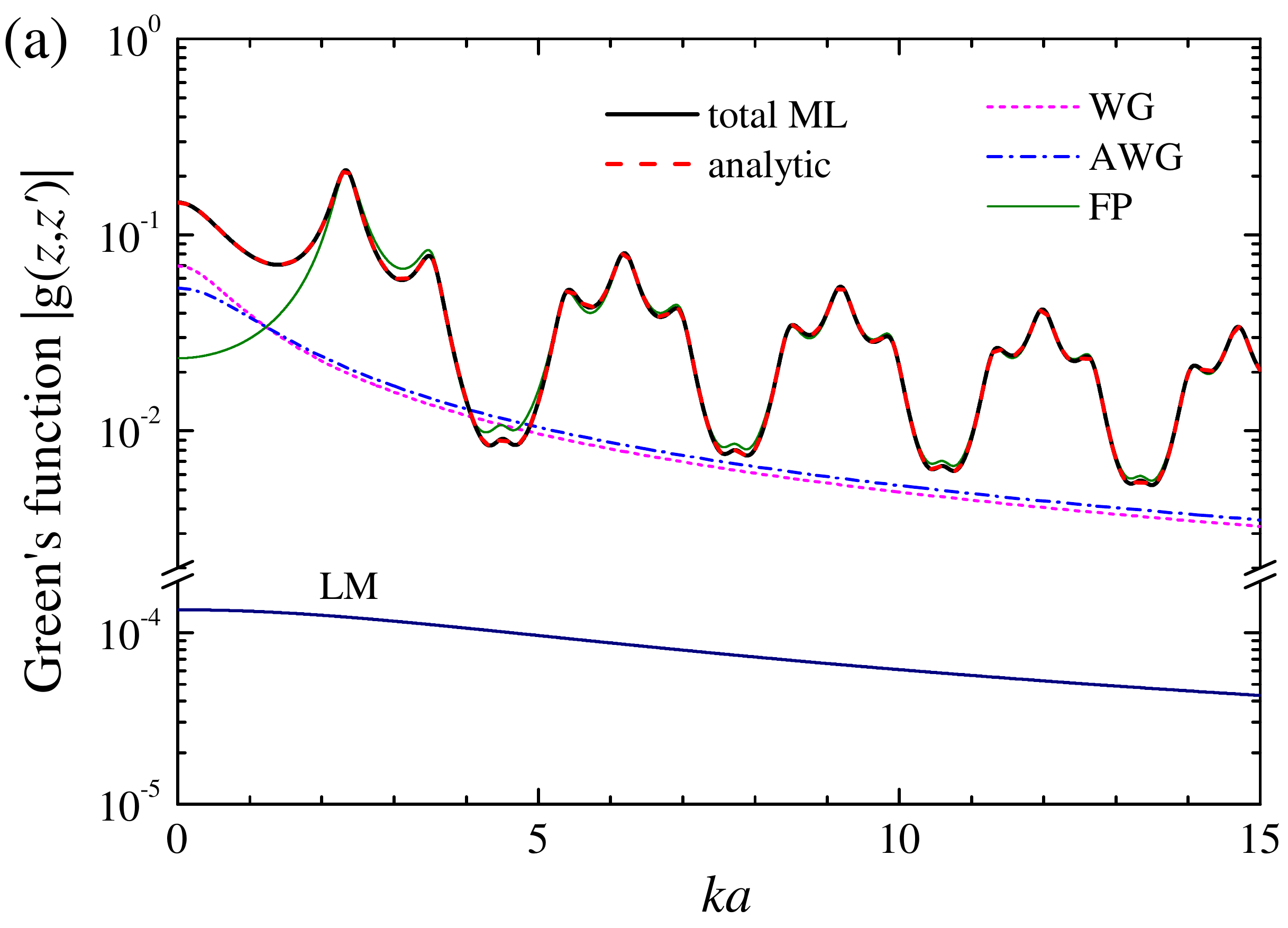}
		\includegraphics*[trim={0cm 0cm 0cm 0cm},clip,width=0.48\textwidth]{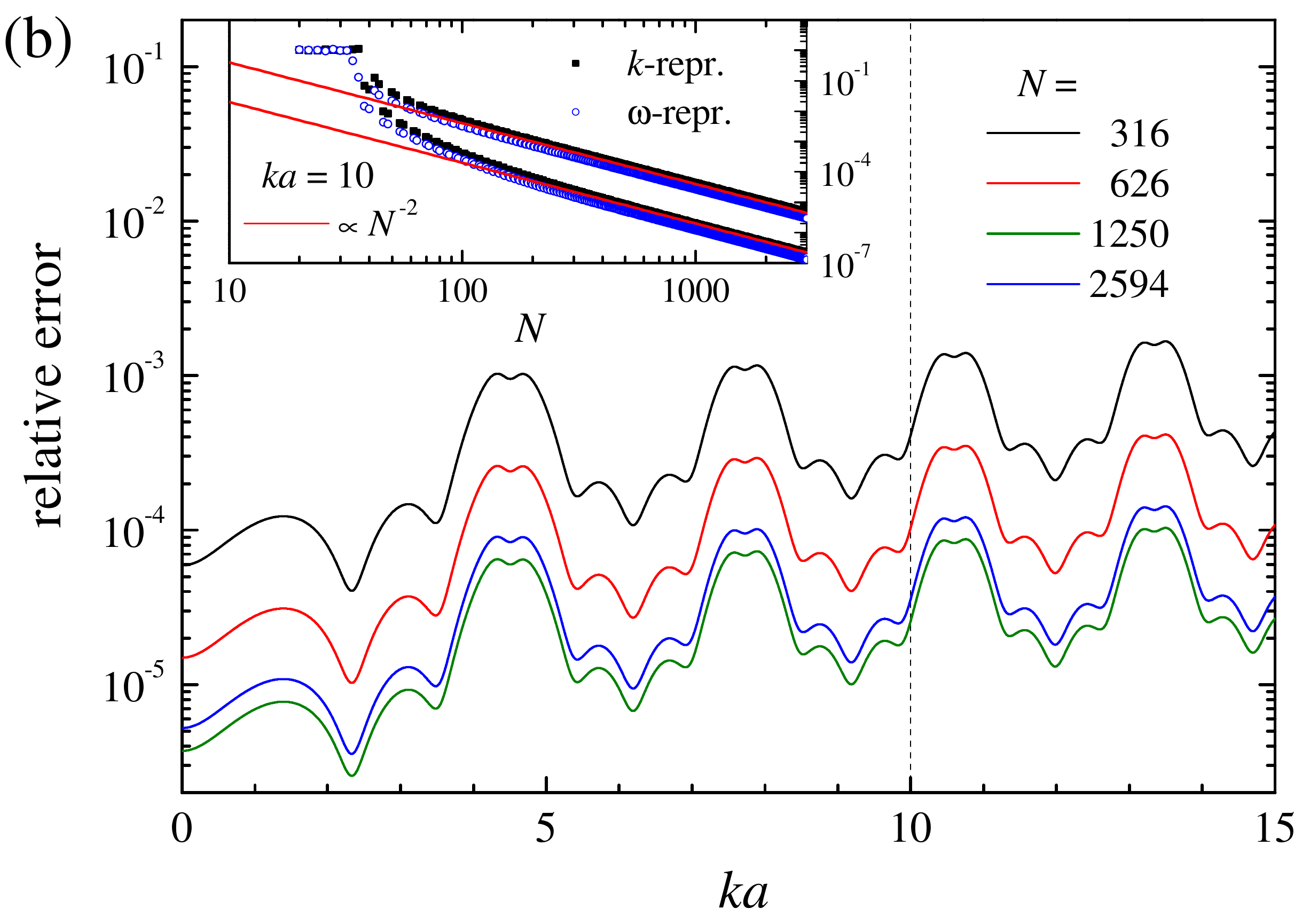}
		\caption{(a) Green's function $g(z,z')$ of a homogeneous dielectric slab in vacuum, with $\epsilon=6$, calculated for $pa=5$, $z=a/2$, and $z'=-a/2$ as a function $k$, the normal component of the wave number in vacuum. Partial contributions of WG, anti-WG (AWG), FP modes and a leaky mode (LM) are shown along with the sum of all the contributions and the analytic values of the GF. (b) Relative error of the GF calculated via \Eq{ML11} as compared to its analytic values  \Eq{GFanal}, for different number of modes $N$ included in the ML series. The inset shows the relative error for $ka=10$ (marked by vertical dashed lines in Figs.\,\ref{Fig:GFk}(b) and \ref{Fig:GFw}(b)) as a function of the basis size $N$. }
\label{Fig:GFk}
	\end{figure}

	\begin{figure}
		\includegraphics*[trim={0cm 0cm 0cm 0cm},clip,width=0.48\textwidth]{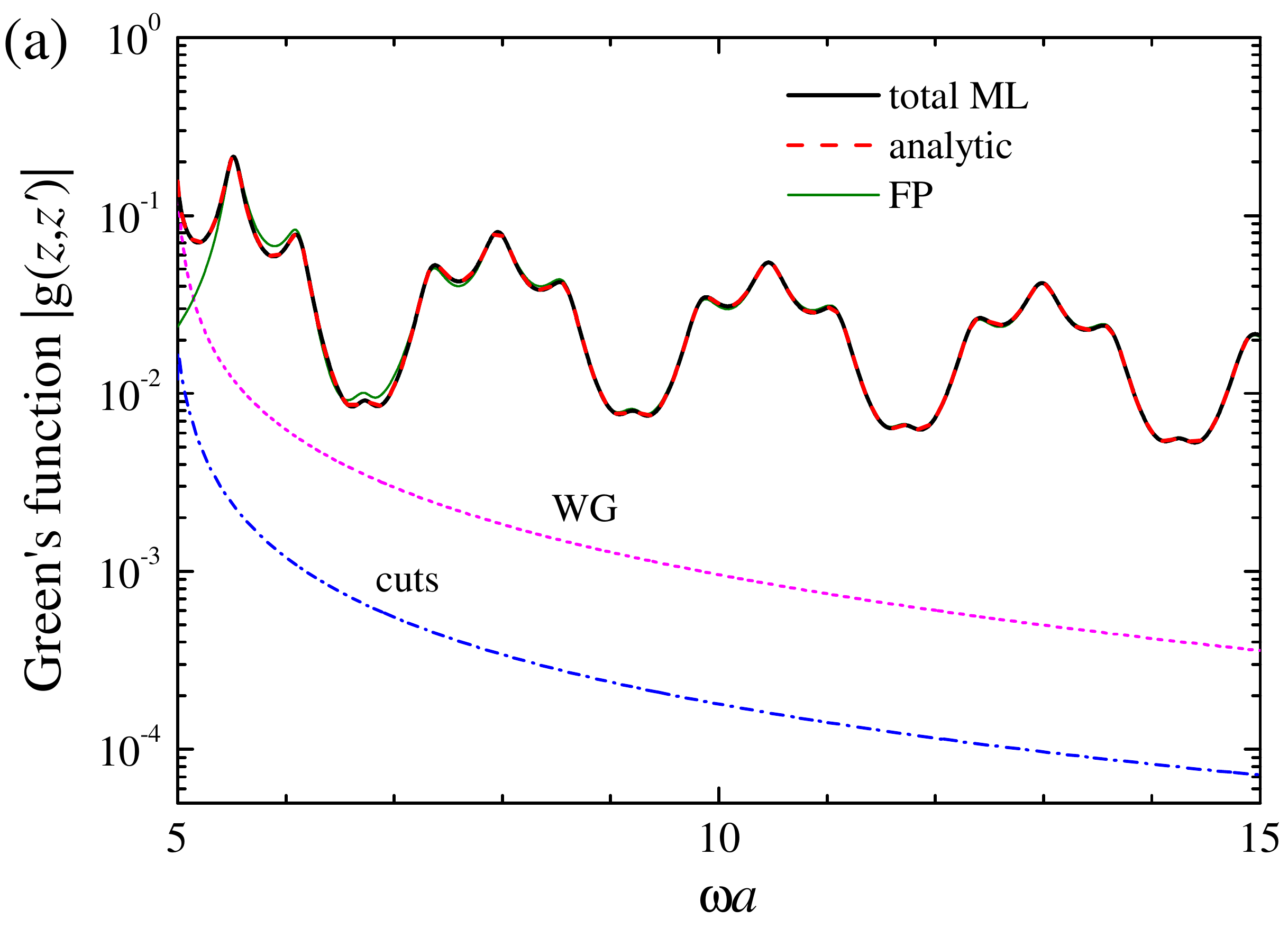}
		\includegraphics*[trim={0cm 0cm 0cm 0cm},clip,width=0.48\textwidth]{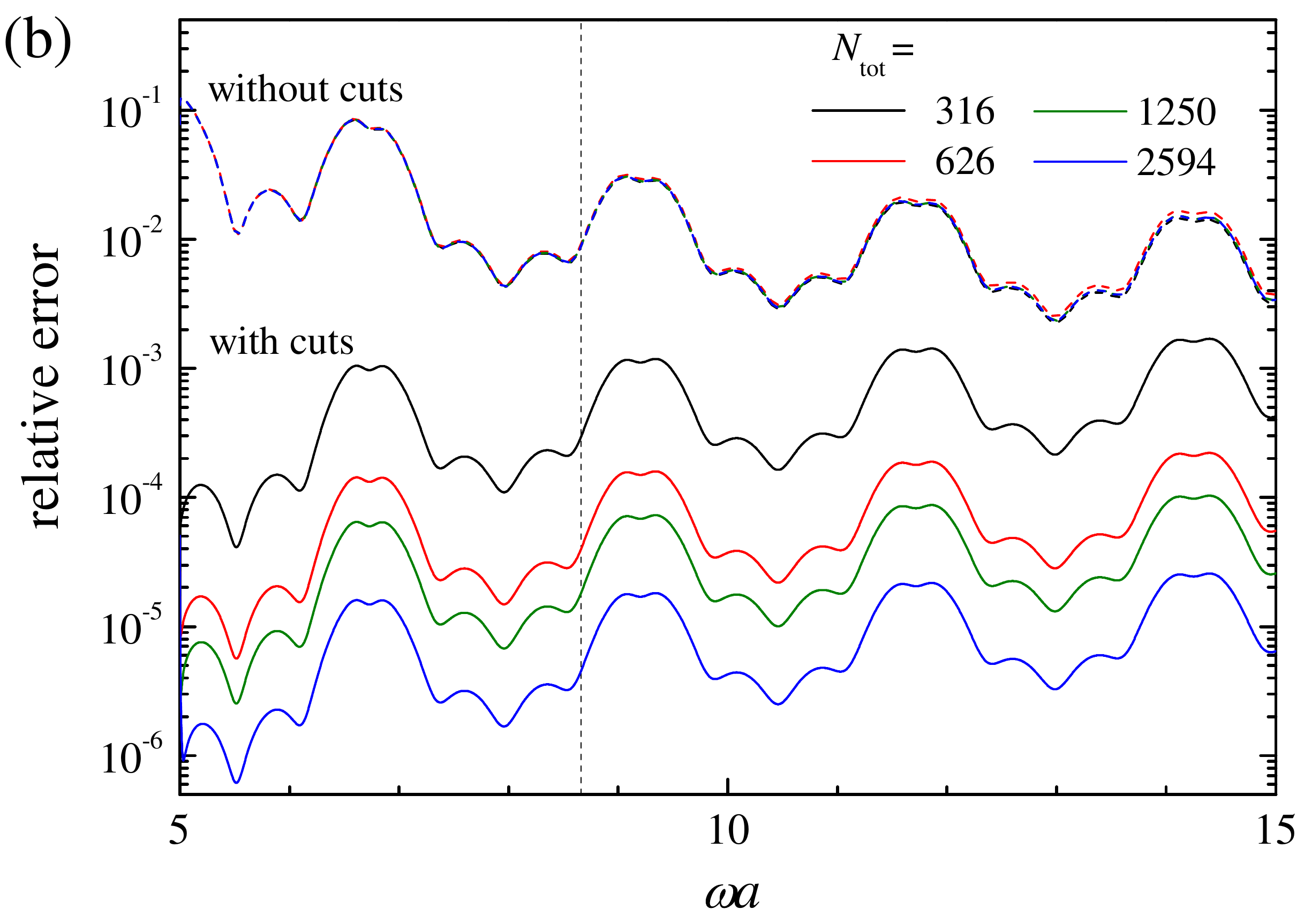}
		\caption{ (a) As \Fig{Fig:GFk}(a) but in the $\omega$-representation, in which the AWG modes and the LM do not contribute to the ML expansion \Eq{ML22}, but there is instead an additional component due to the cuts. (b) Relative error of the GF calculated via \Eq{ML22} as compared to its analytic values \Eq{GFanal}, for different number of modes $N$ included in the ML series, with and without contribution of the cuts.
}
\label{Fig:GFw}
	\end{figure}
	
Results for a slab with $\epsilon=6$  and $\mu=1$ are shown in \Fig{Fig:GFk} for $z=-z'=a/2$. We see that all partial contributions to the GF due to each type of modes is non-vanishing, including that of the LM which  has ${\rm Re}\,\omega_n=0$ and which is discussed in detail in~\cite{ArmitagePRA14,ArmitagePRA18}. Summing up all the contributions to the ML series \Eq{ML11} results in values of the GF approaching its analytic form \Eq{GFanal}. By increasing  $\omega_{\rm max}$, we increase the number of RSs $N$ included in the series \Eq{ML11}, in this way making the ML representation more and more accurate, see \Fig{Fig:GFk}(b). The inset demonstrate the convergence of the ML series to the exact solution, with the error scaling as $1/N^2$.

\subsection{GF in $\omega$-representation}
\label{App:HSw}

The GF can also be represented as a function of frequency. However, the square root in \Eq{k} causes branch cuts at $\omega=\pm p$, which separates the frequency plane into two Riemann sheets with modes split across both sheets. Only the modes found on one of the sheets are taken into account. This is chosen to be the ``physical''  sheet, on which the WG and FP modes are found, while anti-WG modes and the leaky mode turn out to be on the other, unphysical sheet and are thus excluded from the ML expansion \Eq{ML22}. Figure~\ref{Fig:GFw} shows a comparison, for the same parameters as in \Fig{Fig:GFk}, of the ML expansion \Eq{ML22} with the analytic solution \Eq{GFanal}, again showing different contributions, including WG, FP modes, and the cuts. We see that the ML series in the $\omega$-representation again converges to the exact solution, provided that the cut integrals are included. Note also that the convergence is very similar to that in the $k$-representation, as it is clear from the inset in \Fig{Fig:GFk}(b). In fact, in both cases the error alternates between two different boundaries but nevertheless decreases with the basis size as $N^{-2}$.

	\begin{figure}
		\includegraphics*[clip,width=0.48\textwidth]{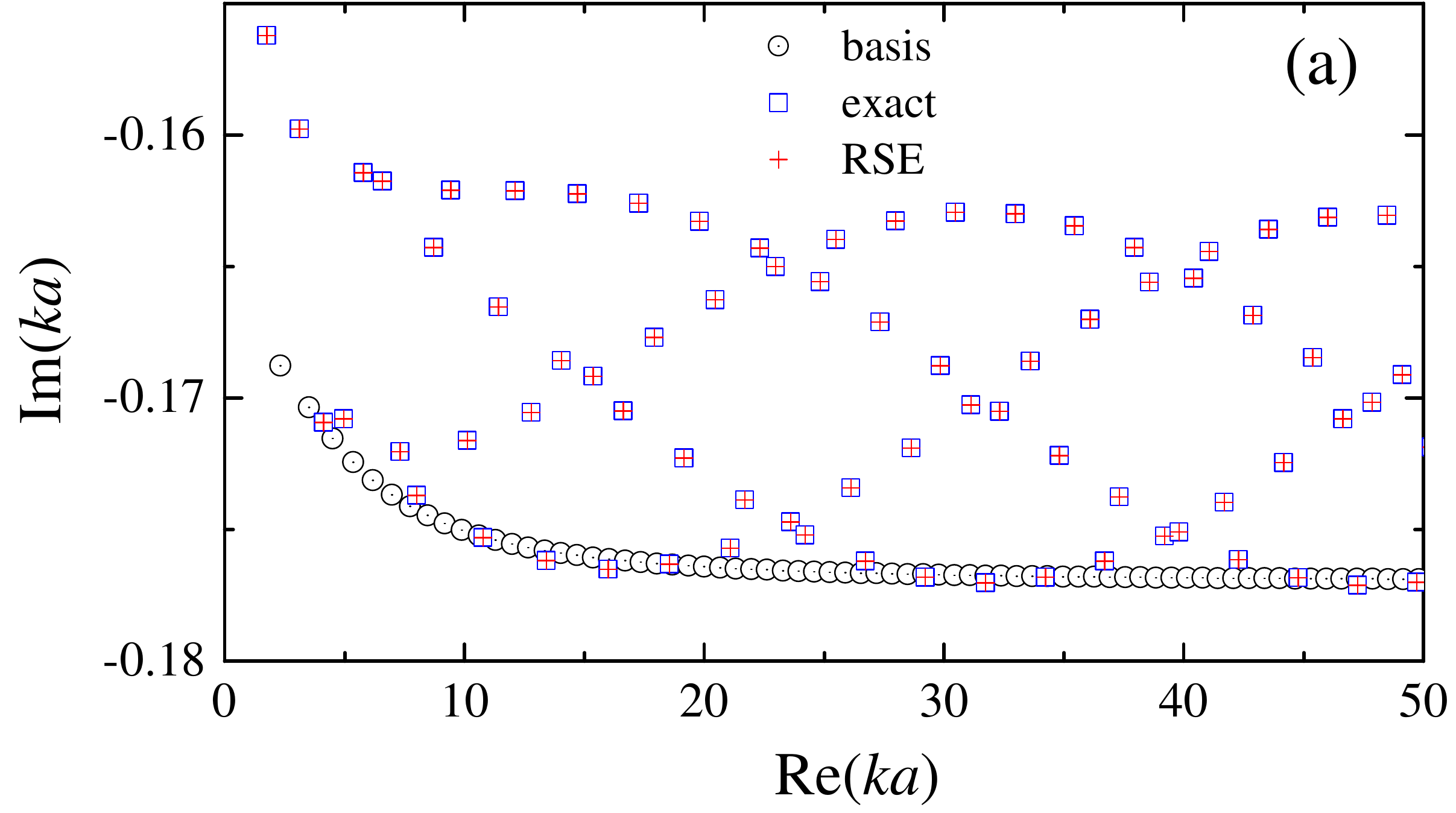}
		\includegraphics*[clip,width=0.48\textwidth]{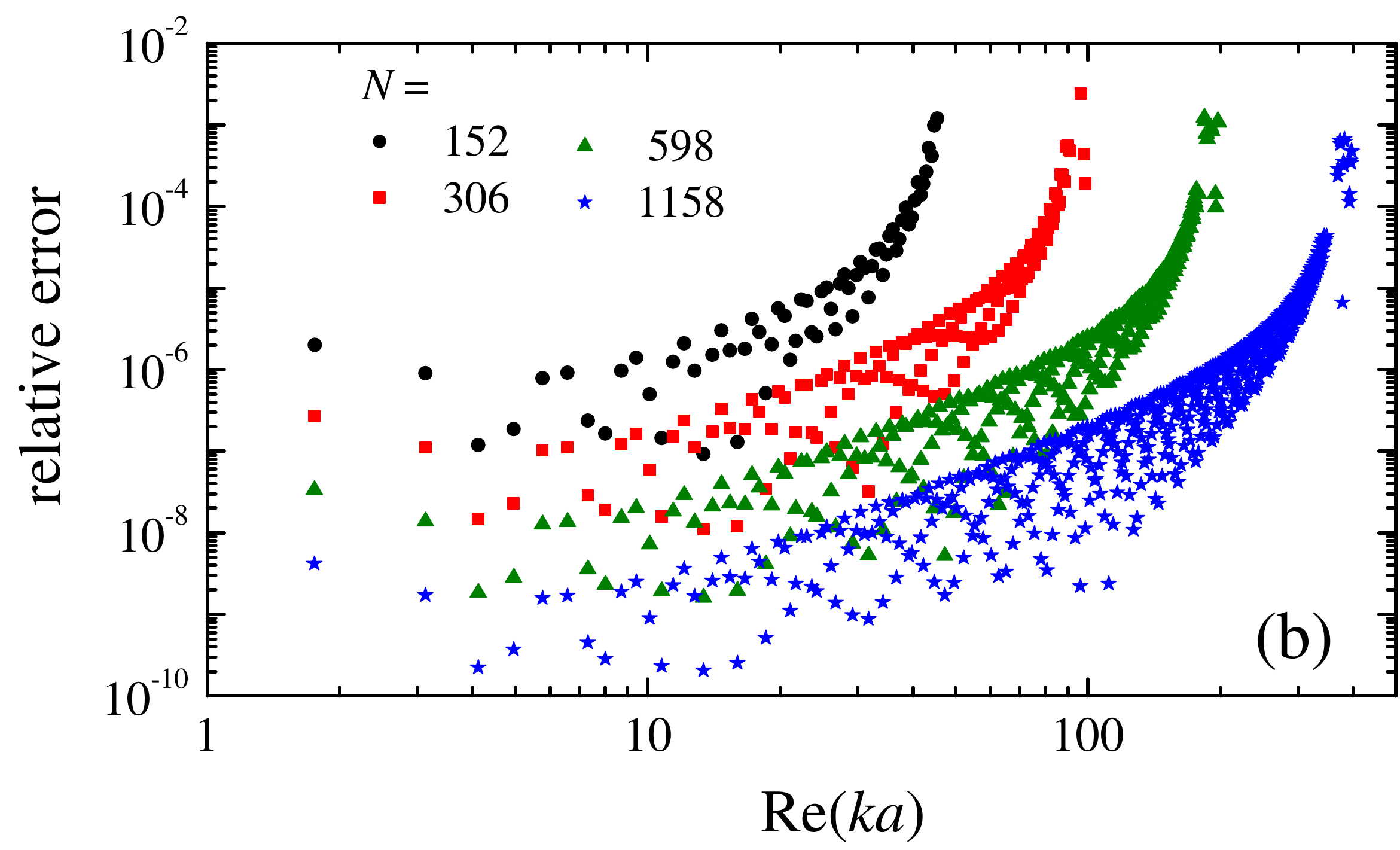}
		\includegraphics*[clip,width=0.48\textwidth]{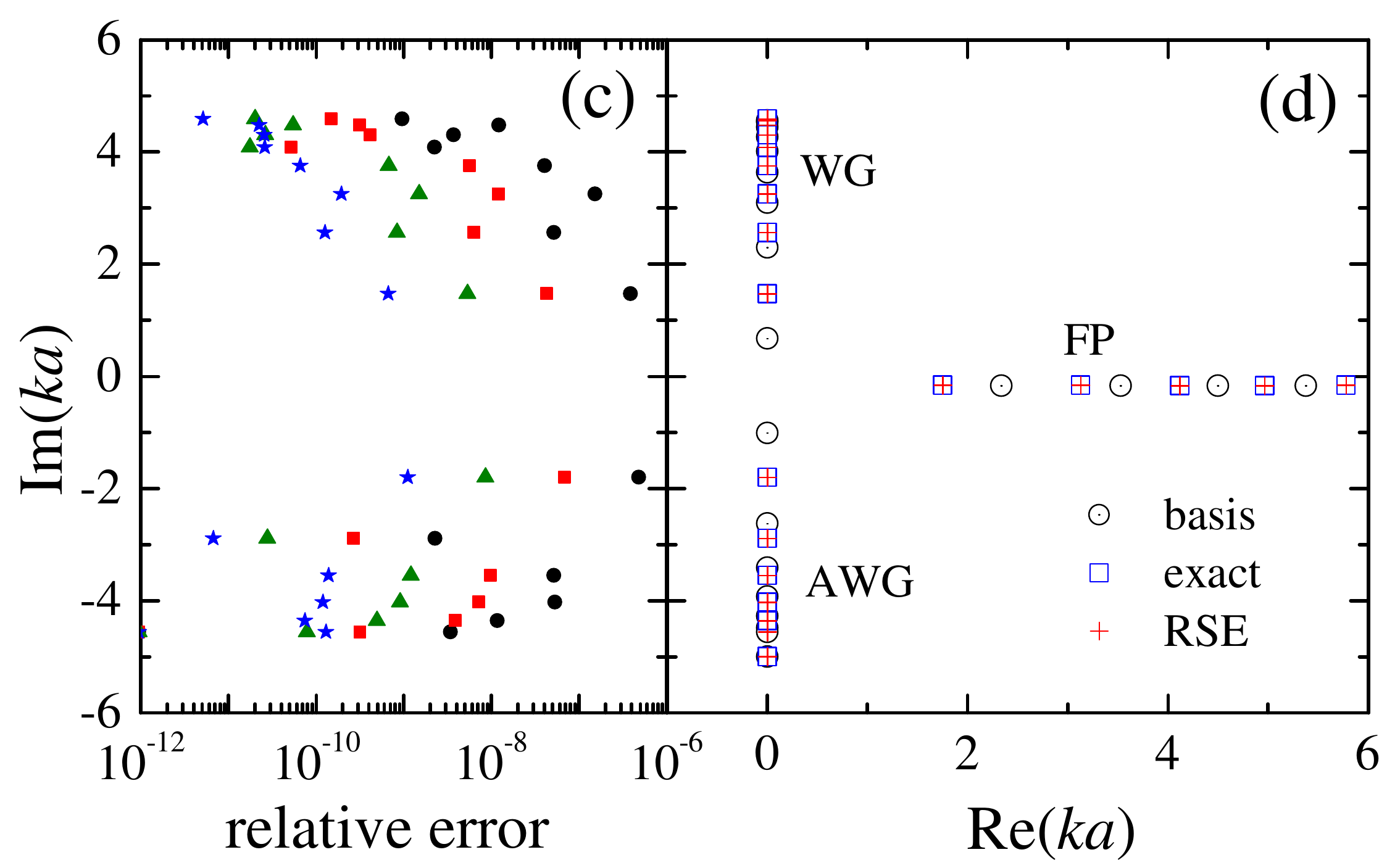}
		\caption{ (a) Wave numbers $k_n$ of the FP modes calculated for $pa=5$ using the RSE in the $k$-representation (red crosses) for the core-shell dielectric slab with $\epsilon=6$, $\Delta\epsilon=1$, and $b=a/2$, along with the exact solution (blue squares) and the basis RSs (black circles with dots).
(b) Relative error for the wave numbers of the FP modes compared to the exact solution, for different basis sizes as given. (c) Relative errors for WG and anti-WG modes, for the same basis sizes as in (b). (d) As (a) but for all WG and anti-WG modes, and the first few FP modes.
}
		\label{k-RSE-homo}
	\end{figure}

\label{App:RSE-w}
	\begin{figure}
		\includegraphics[clip,width=0.48\textwidth]{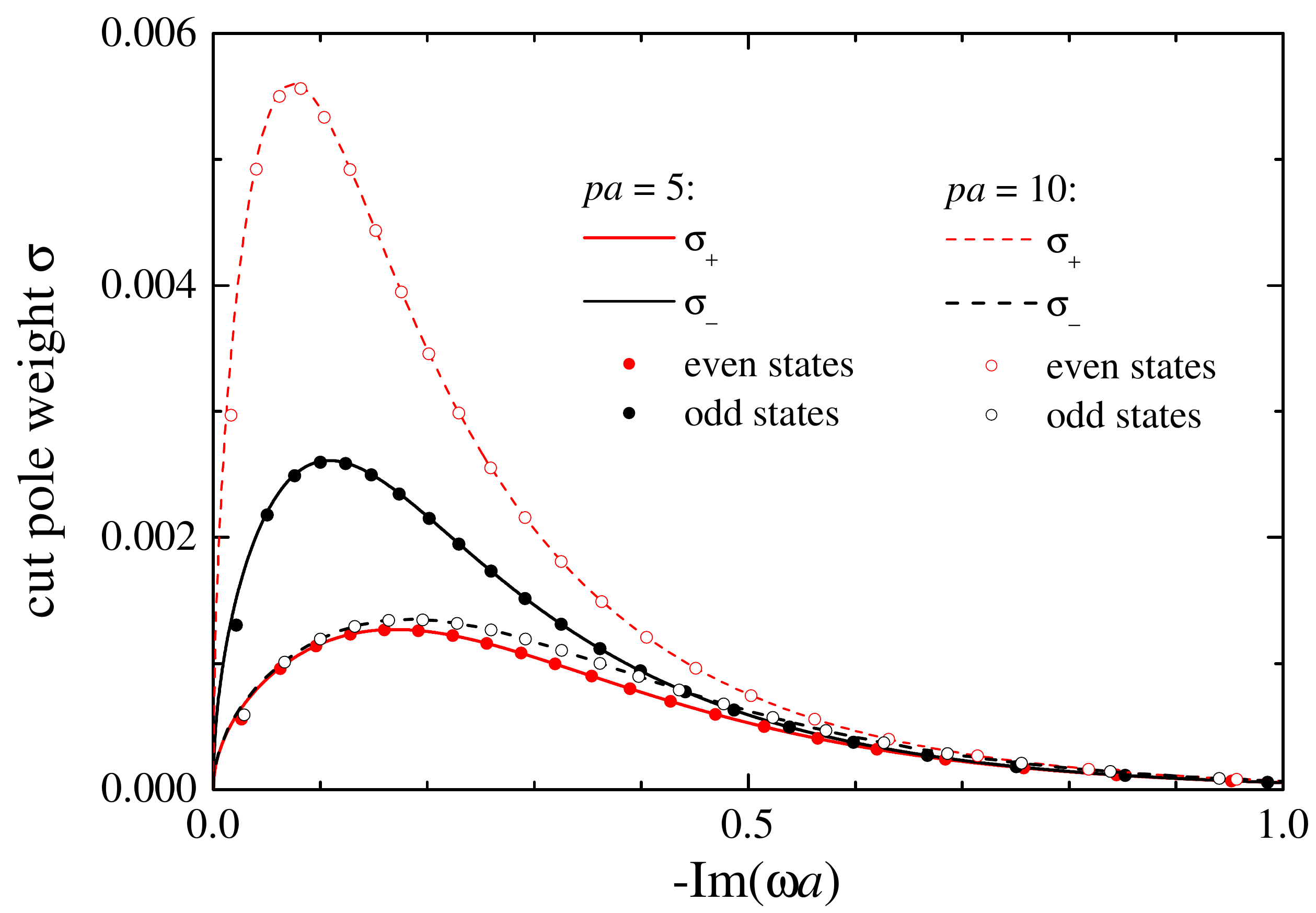}
		\caption{Cut weights $\sigma_\pm$ (solid and dashed lines) as functions of the imaginary part of the frequency along the cut, for different values of $p$ as given, along with the discretized cut pole values $C_\nu/\Delta\omega_\nu$ (circles).}
\label{Fig:Sigma}
	\end{figure}
	\begin{figure}
		\includegraphics[clip,width=0.48\textwidth]{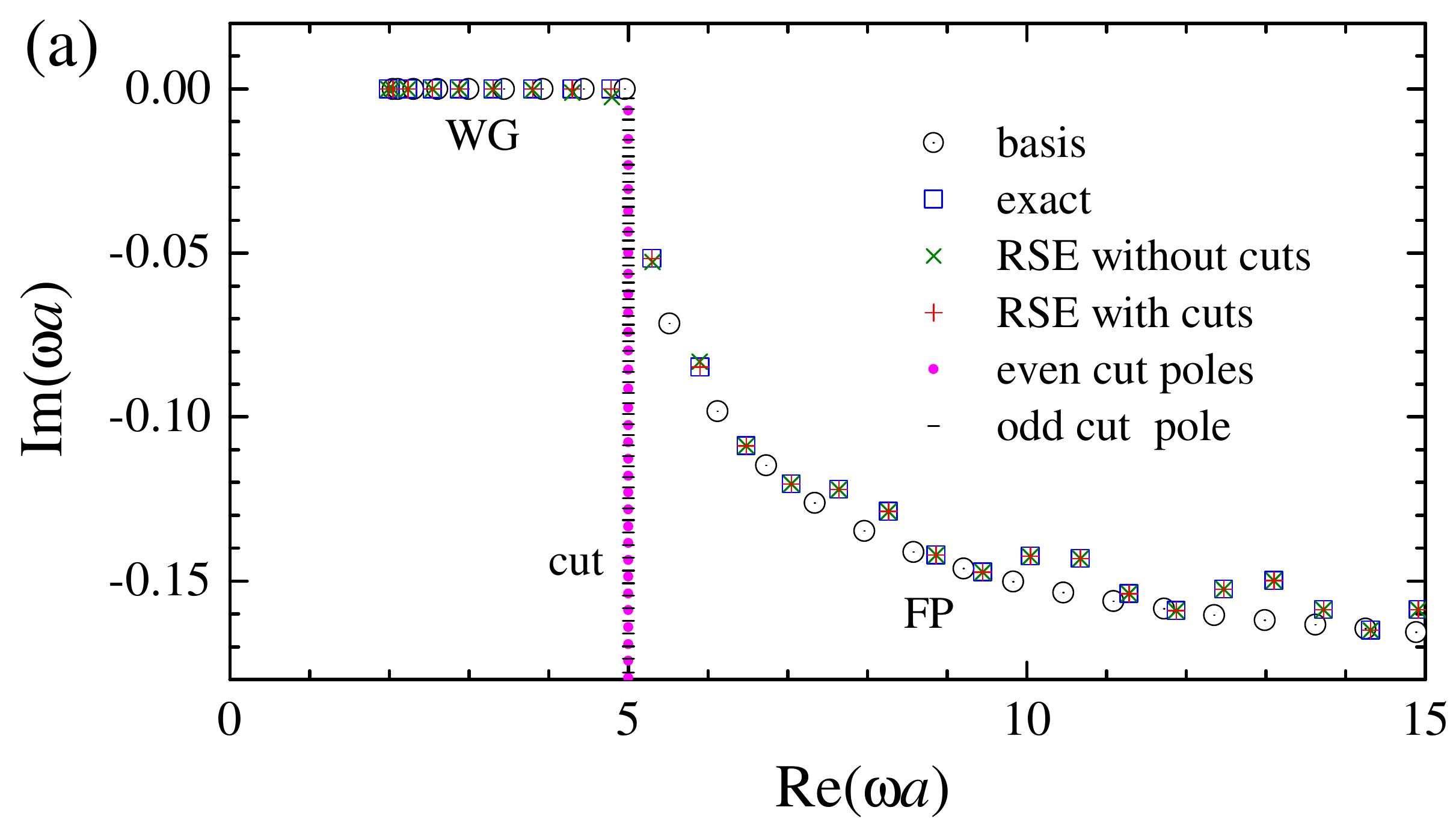}
		\includegraphics[clip,width=0.48\textwidth]{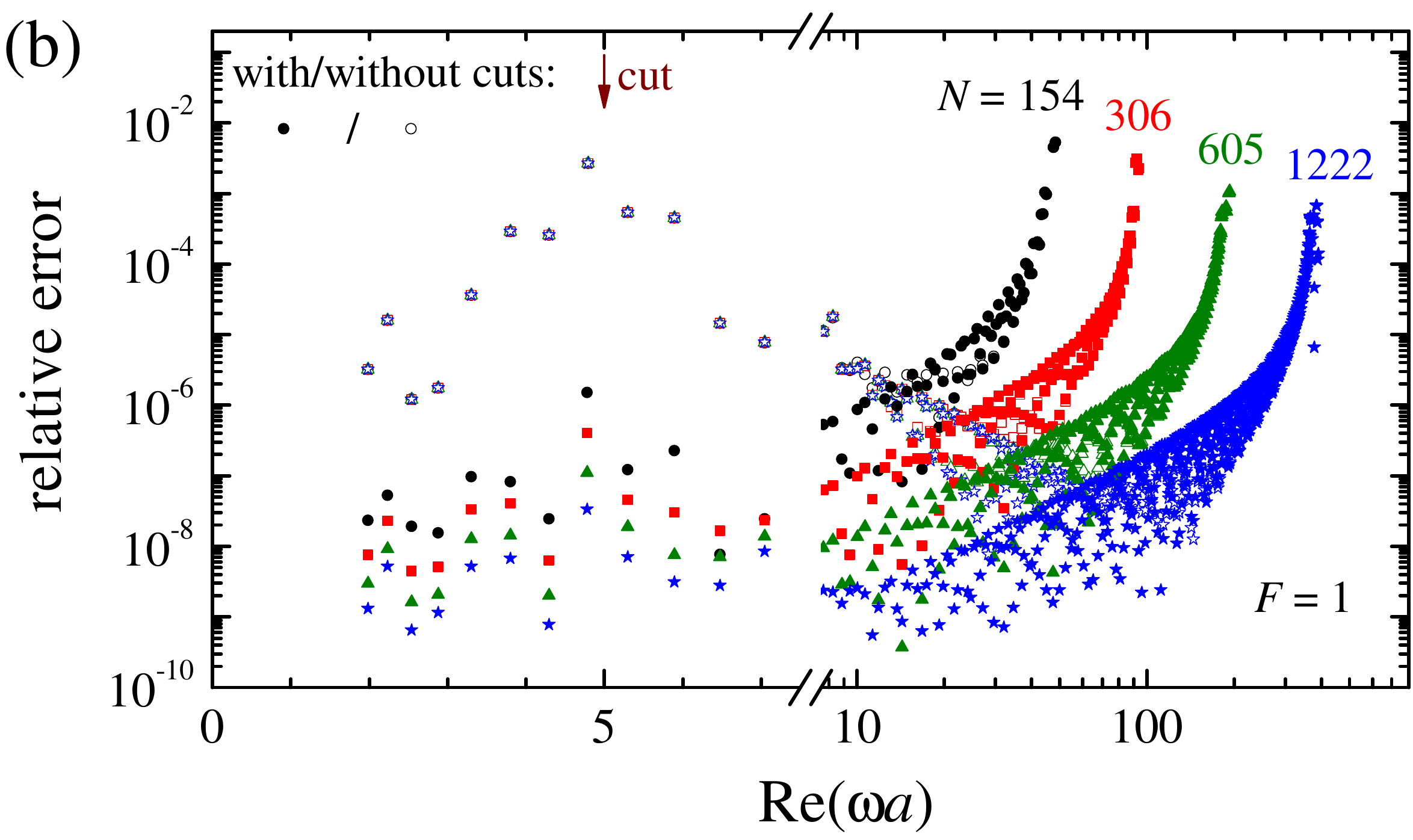}
		\includegraphics[clip,width=0.48\textwidth]{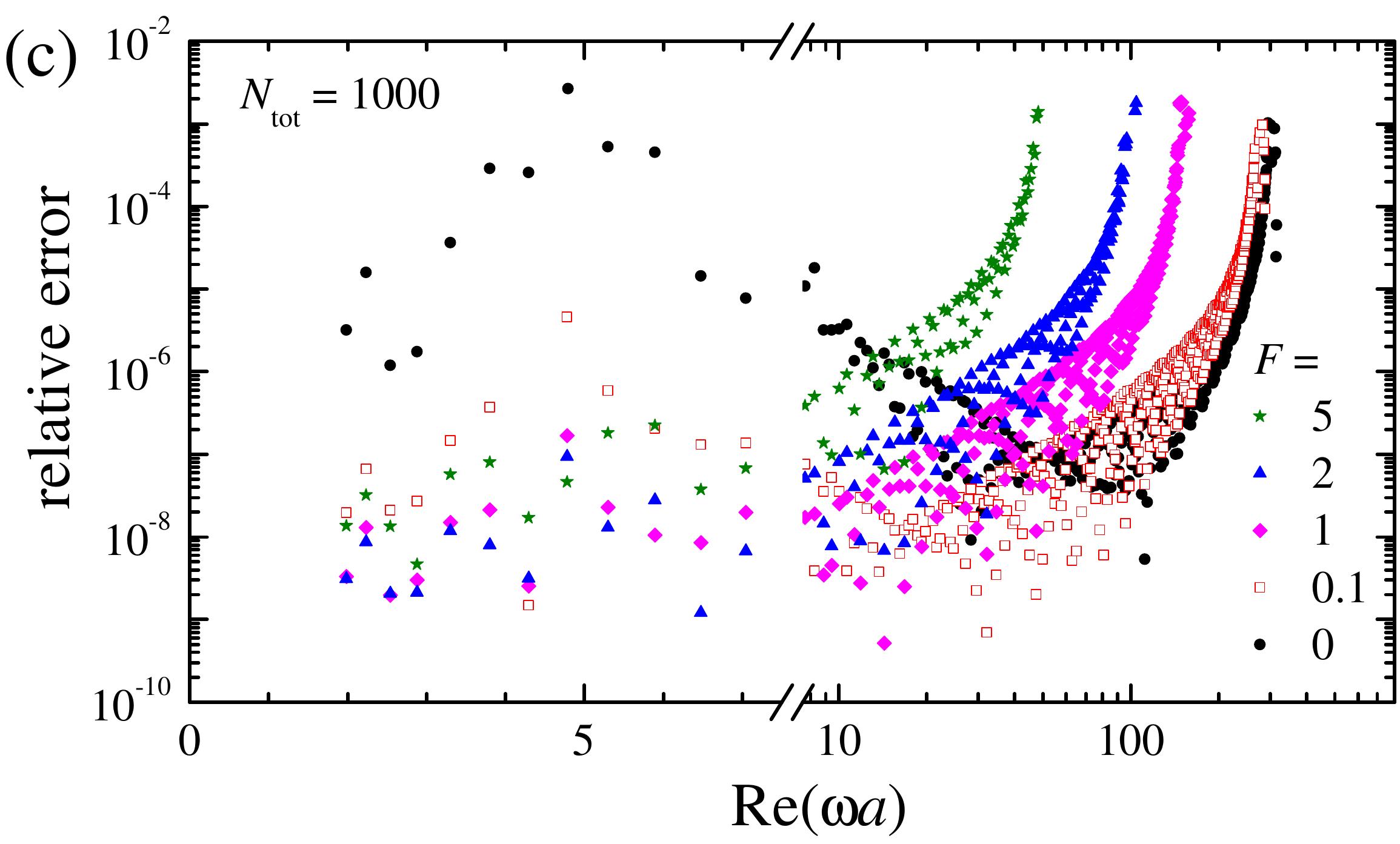}
		\caption{(a) As \Fig{k-RSE-homo} but in frequency representation, also showing even and odd parity cut modes. (b) Relative errors compared to the exact solution \Eq{Secular equation core} of the RS frequencies calculated using the RSE, with and without cut modes in the basis and for different basis sizes as given. (c) Relative error of the RSE results calculated for a fixed basis size $N_{\rm tot}=1000$ and different values of $F$, the ratio of the number of cut modes to the number of RSs in the basis. }
\label{w-RSE-homo}
	\end{figure}
	
\section{RSE for a homogeneous core slab}
\label{App:RSE for slab}

	Here we demonstrate how the RSE in the $k$- and $\omega$-representations is applied to a homogeneous perturbation. This is a special case of the PC-RSE which allows an exact analytic solution, but obviously lacking any periodic modulations. We use $\alpha=\Delta\epsilon$ and $\beta=0$ in \Eq{pert}. The perturbation of the permittivity thus has the form
\be
\Delta\eps(z)=\Delta\epsilon \Theta(b-|z|)\,.
\label{pert2}
\ee
The perturbed system presents a core-shell slab consisting of three homogeneous regions.
The secular equation for this system has the following analytic form:
	\begin{equation}
	\left(\gamma_+\lambda_--\gamma_-\lambda_+\right)e^{iq_1b}\pm\left(\gamma_+\lambda_+-\gamma_-\lambda_-\right)e^{-iq_1b}=0\,,
	\label{Secular equation core}
	\end{equation}
where $\gamma_\pm=(q\pm k)e^{\mp iq(a-b)}$, $\lambda_\pm=q\pm q_1$, and $q_1=\sqrt{\epsilon_1\omega^2-p^2}$ with $\epsilon_1=\epsilon+\Delta\epsilon$. The secular equation (\ref{secular}) for the homogeneous slab  can be restored  by setting $b=a$ and $q=q_1$, or simply $b=0$.

\subsection{RSE in $k$-representation}
\label{App:RSE-k}

In the $k$-representation, the RSE equation for treating planar homogeneous systems is given by Eq.\,(22) of \cite{ArmitagePRA14}, which we write here as
	\begin{equation}
	k\sum_{n'} \left(\delta_{nn'} + {\cal V}_{nn'}\right)c_{n'}=\sum_{n'}\left(k_n\delta_{nn'} +\frac{p^2}{k_n}{\cal V}_{nn'}\right) c_{n'}\,,
	\label{kRSE-homo}
	\end{equation}
with the matrix elements ${\cal V}_{nn'}=V_{nn'}^{00}$ given by \Eq{Vnm} for $g=g'=0$.

Its application to the perturbation given by \Eq{pert2} is shown in \Fig{k-RSE-homo}, in comparison with the exact solution \Eq{Secular equation core} and the basis RSs of the homogenous slab.  The quasi-periodic pattern of the wave numbers of the perturbed RSs seen in \Fig{k-RSE-homo} is caused by the perturbation covering only the middle half of the slab, so that the original homogeneous slab of thickness $2a$ is now split into three subsystems of thickness $a/2$, $a$, and $a/2$, each acting as a resonance cavity. The distance in frequency between the modes is given by a fundamental period of $\pi/(2a)$ as in the basis cavity, but the cavities between $-a$ and $-b$ and between $b$ and $a$ have caused additional quasi-periodicities, one of them having the period of $\pi/(a-b)$, which for $b=a/2$ is four times larger than the fundamental period. Other cavities present in the system also contribute to the rich spectrum of RSs seen in \Fig{k-RSE-homo}.

Looking at the relative error shown in Figs.\,\ref{k-RSE-homo}(b) and (c), we see that the RSE in the $k$-representation quickly converges to the exact solution. The relative error scales as $1/N^3$, which is typical for effective 1D systems, see~\cite{MuljarovEPL10,DoostPRA12,DoostPRA13,DoostPRA14,ArmitagePRA14}.

	\begin{figure}
		\includegraphics[clip,width=0.48\textwidth]{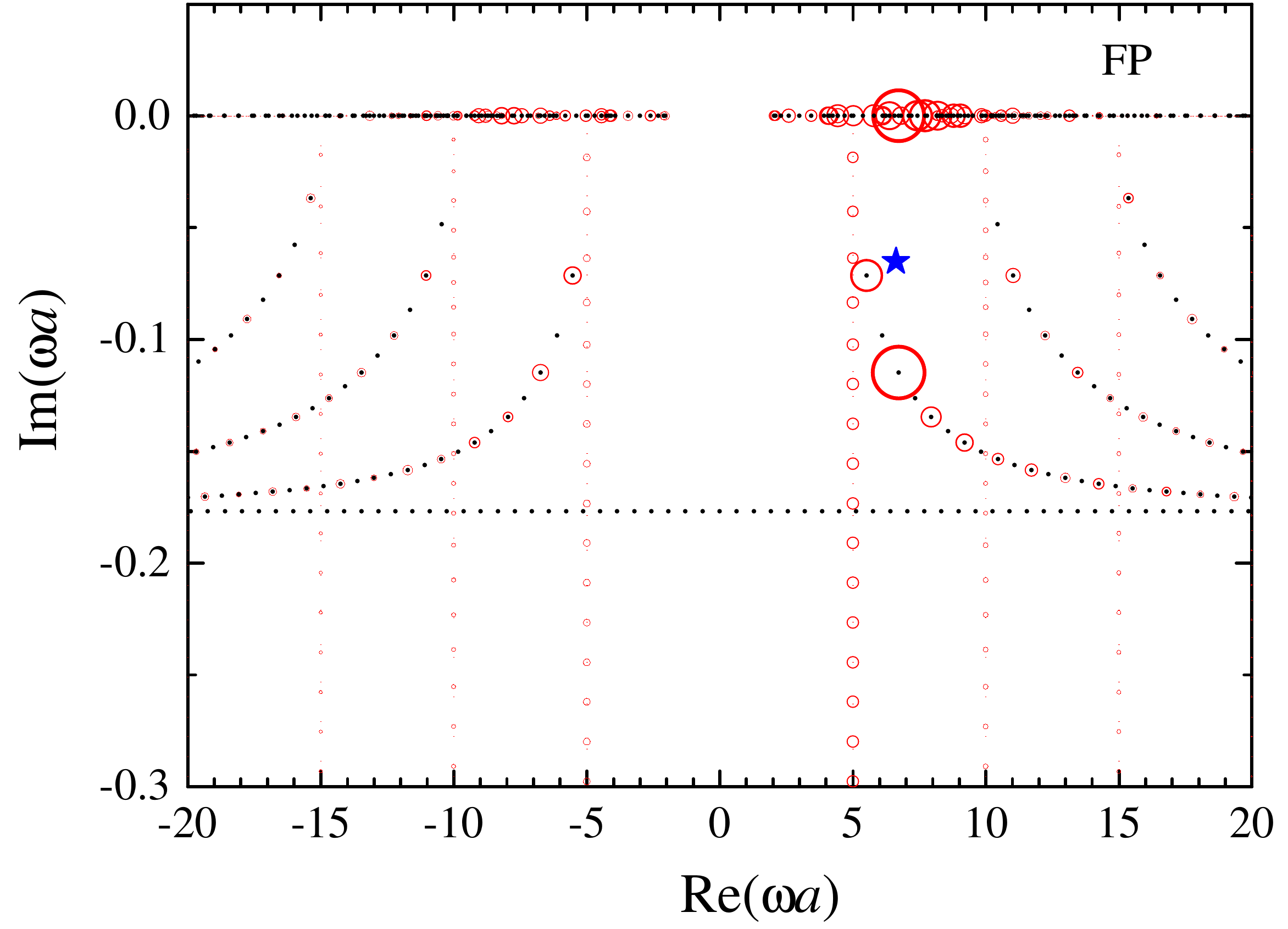}
		\includegraphics[clip,width=0.48\textwidth]{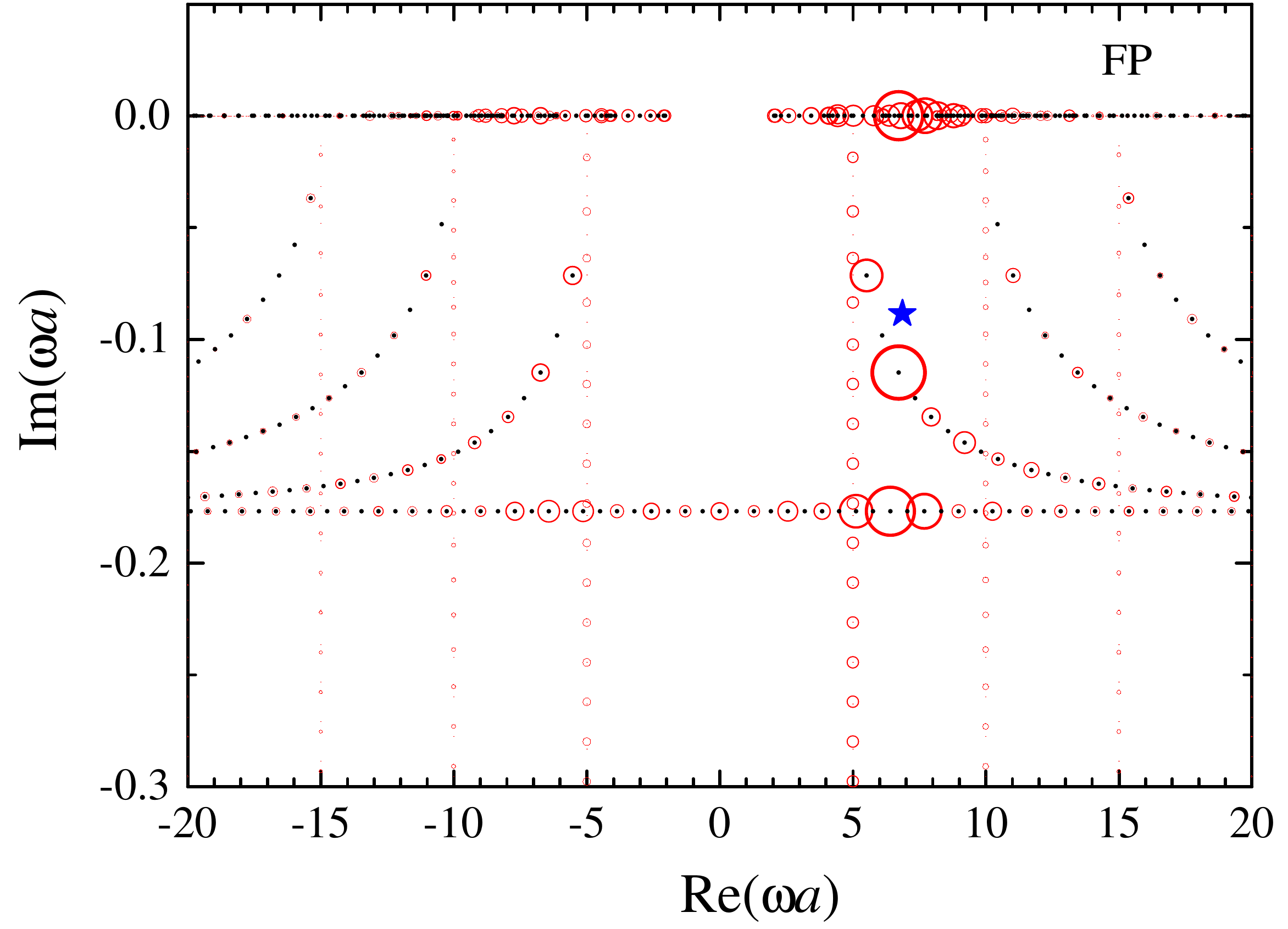}
		\caption{As \Fig{Mode Contributions} but for two RSs originating from a degenerate pair of FP modes.}
		\label{Mode Contributions FP}	
	\end{figure}
	\begin{figure}
		\includegraphics[clip,width=0.48\textwidth]{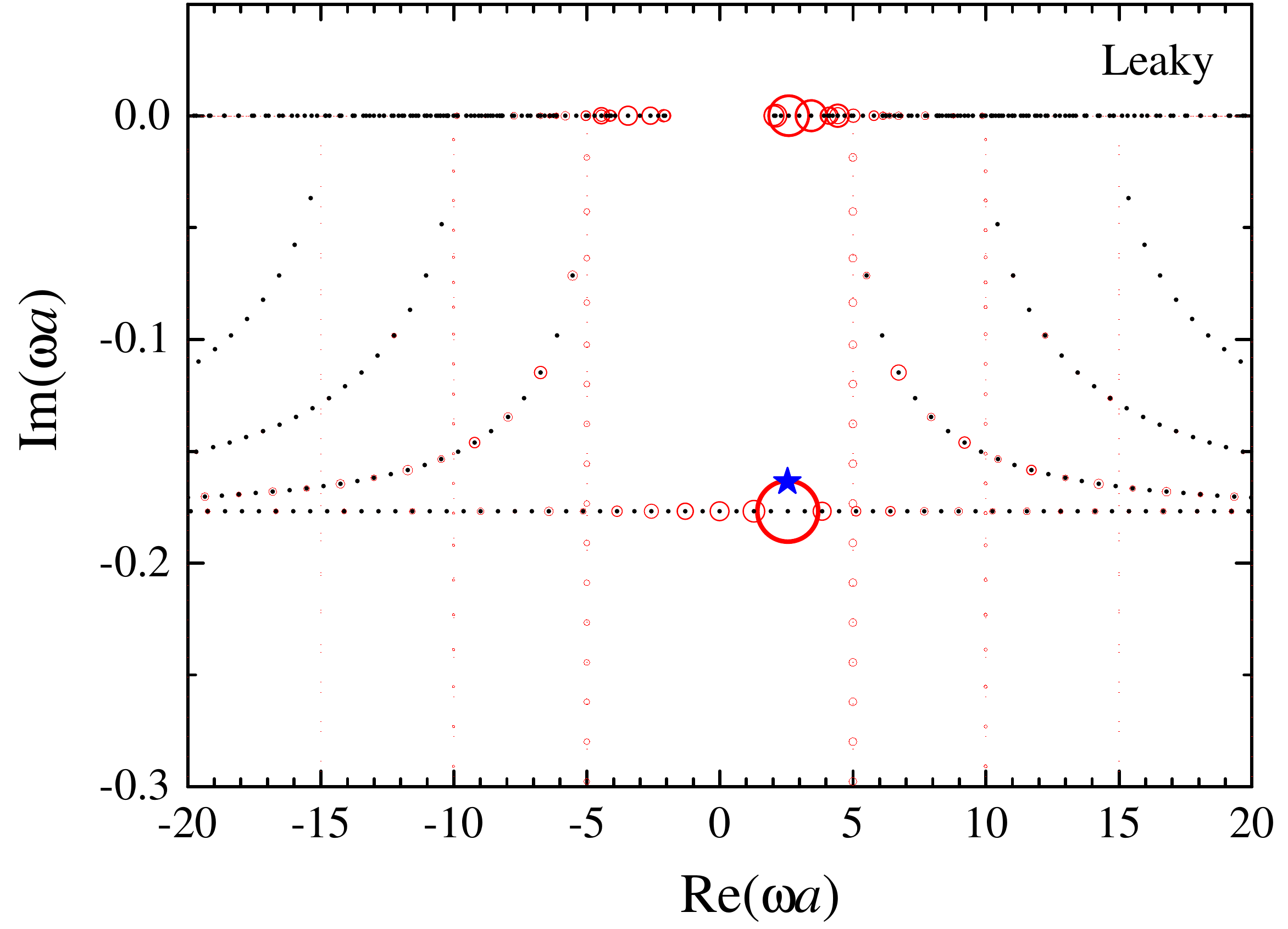}
		\includegraphics[clip,width=0.48\textwidth]{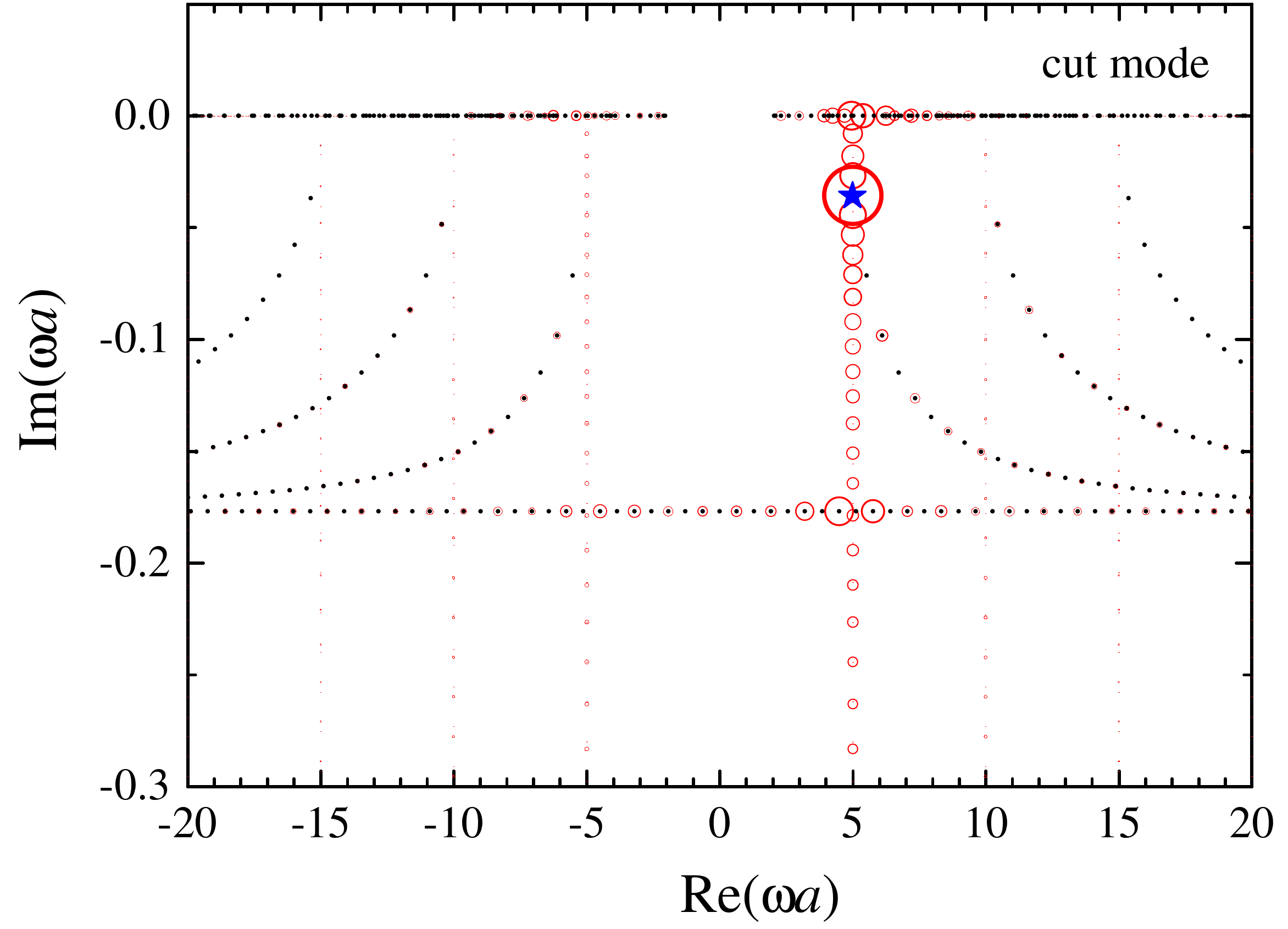}
		\caption{As \Fig{Mode Contributions} but for a RS coming from a non-degenerate $m=0$ leaky mode (top) and for a perturbed cut mode (bottom).}
		\label{Mode Contributions Other}	
	\end{figure}
	\begin{figure}
		\includegraphics[clip,width=0.48\textwidth]{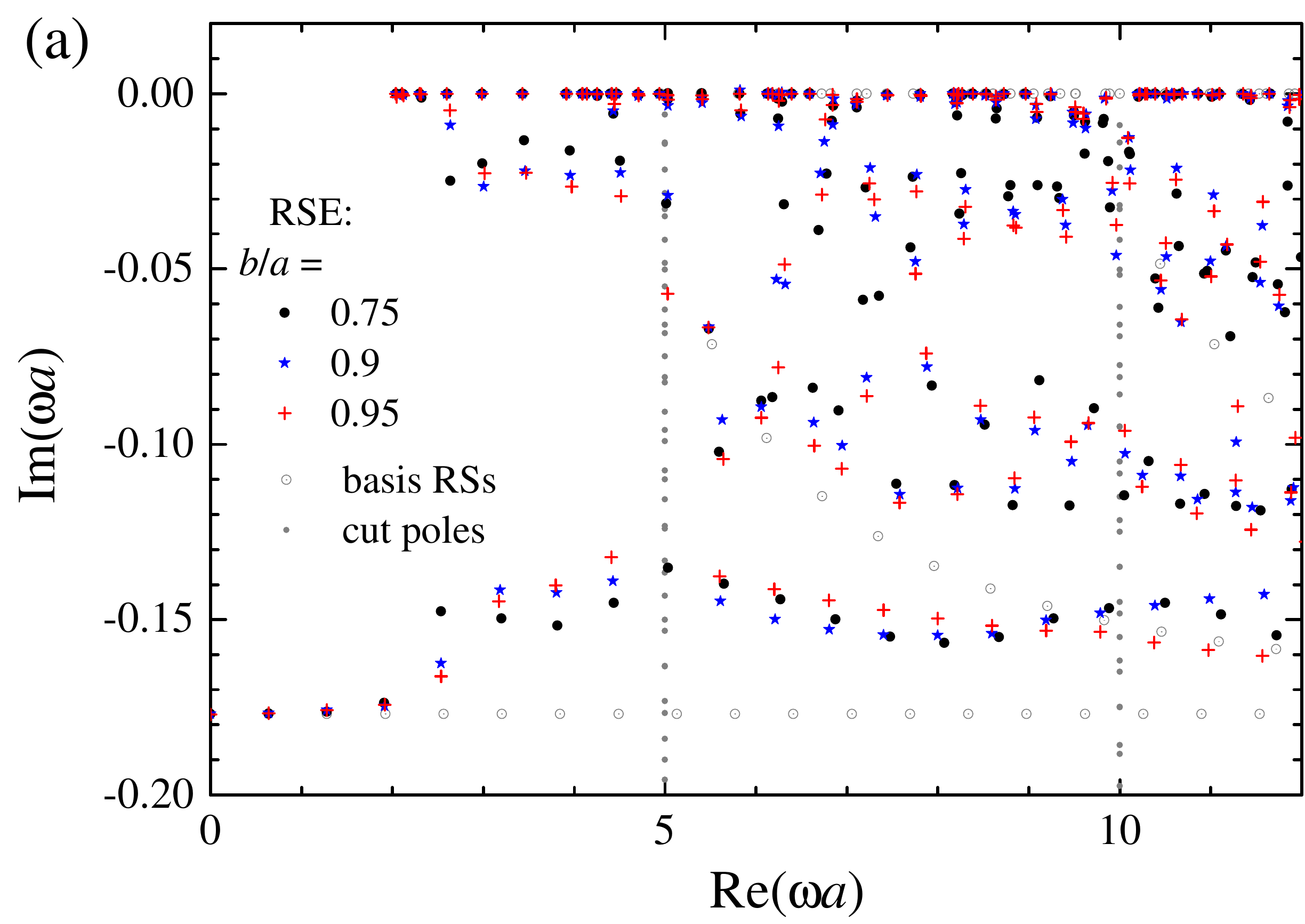}
		\includegraphics[clip,width=0.48\textwidth]{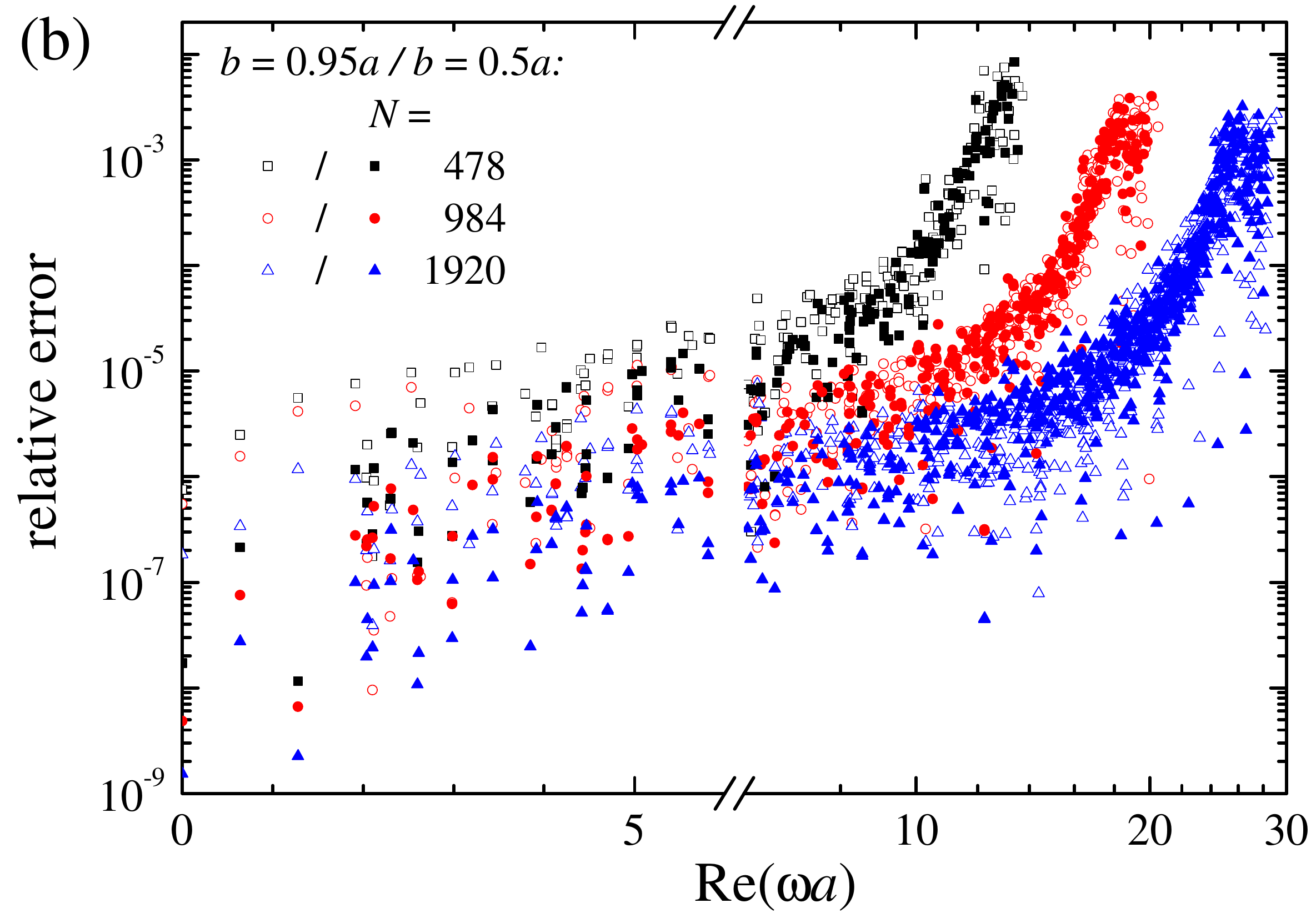}
		\caption{(a) RS frequencies of a PC slab with $\epsilon=6$, $\alpha=0$, $\beta=1$, $d=2\pi/5$, and $p=0$, calculated for $F=1$ and $M=5$ via the PC-RSE for different values of $b$ as given. Unperturbed RSs and cut modes are also shown (black circles with dots and black dots, respectively).
(b) Relative error of the PC-RSE compared to the SMM result, taking the latter as ``exact'',  calculated for different values of $b$ and different basis sizes $N$ as given.
}
		\label{Different b}	
	\end{figure}
	\begin{figure}
		\includegraphics[clip,width=0.48\textwidth]{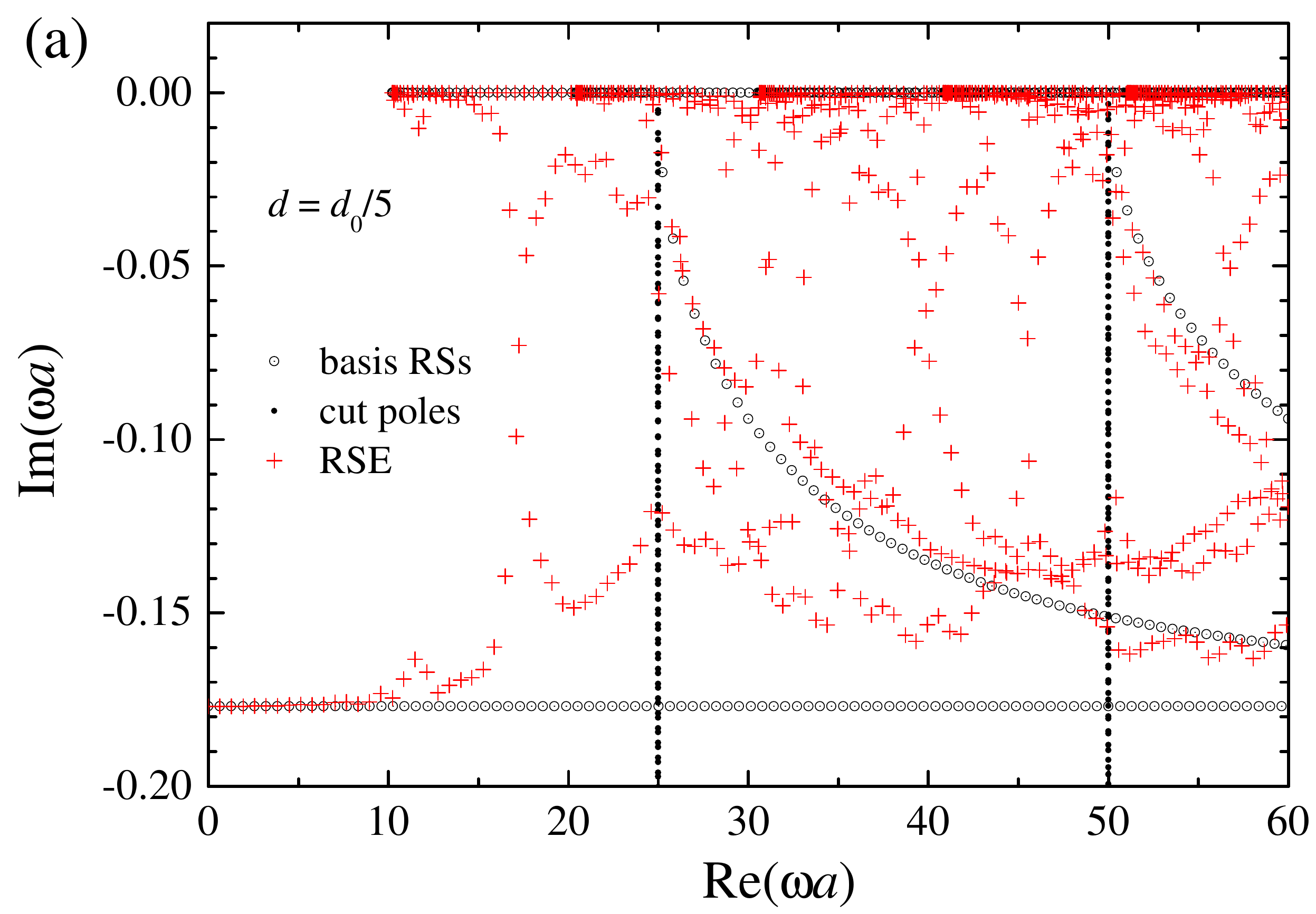}
		\includegraphics[clip,width=0.48\textwidth]{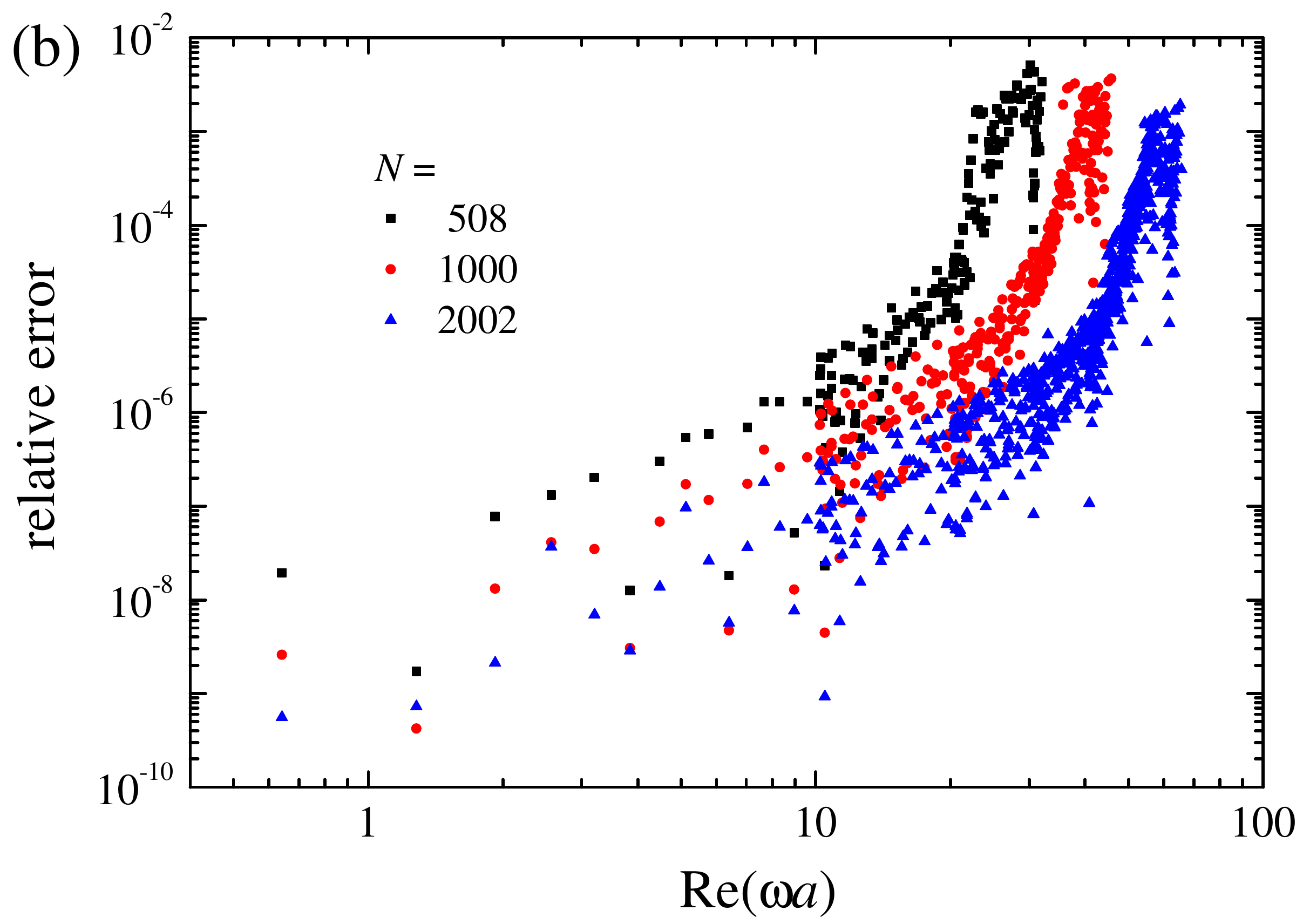}
		\caption{
(a) RS frequencies of a PC slab with $\epsilon=6$, $\alpha=0$, $\beta=1$, $d=2\pi/25$, and $p=0$, calculated for $F=1$ via the PC-RSE (red crosses). Unperturbed RSs and cut modes are also shown (black circles with dots and black dots, respectively).
(b) Relative error of the PC-RSE compared to the RSE with $N\approx4000$, taking the latter as ``exact'', calculated for different basis sizes as given.}
		\label{d/5}	
	\end{figure}	
	\begin{figure}
		\includegraphics[clip,width=0.48\textwidth]{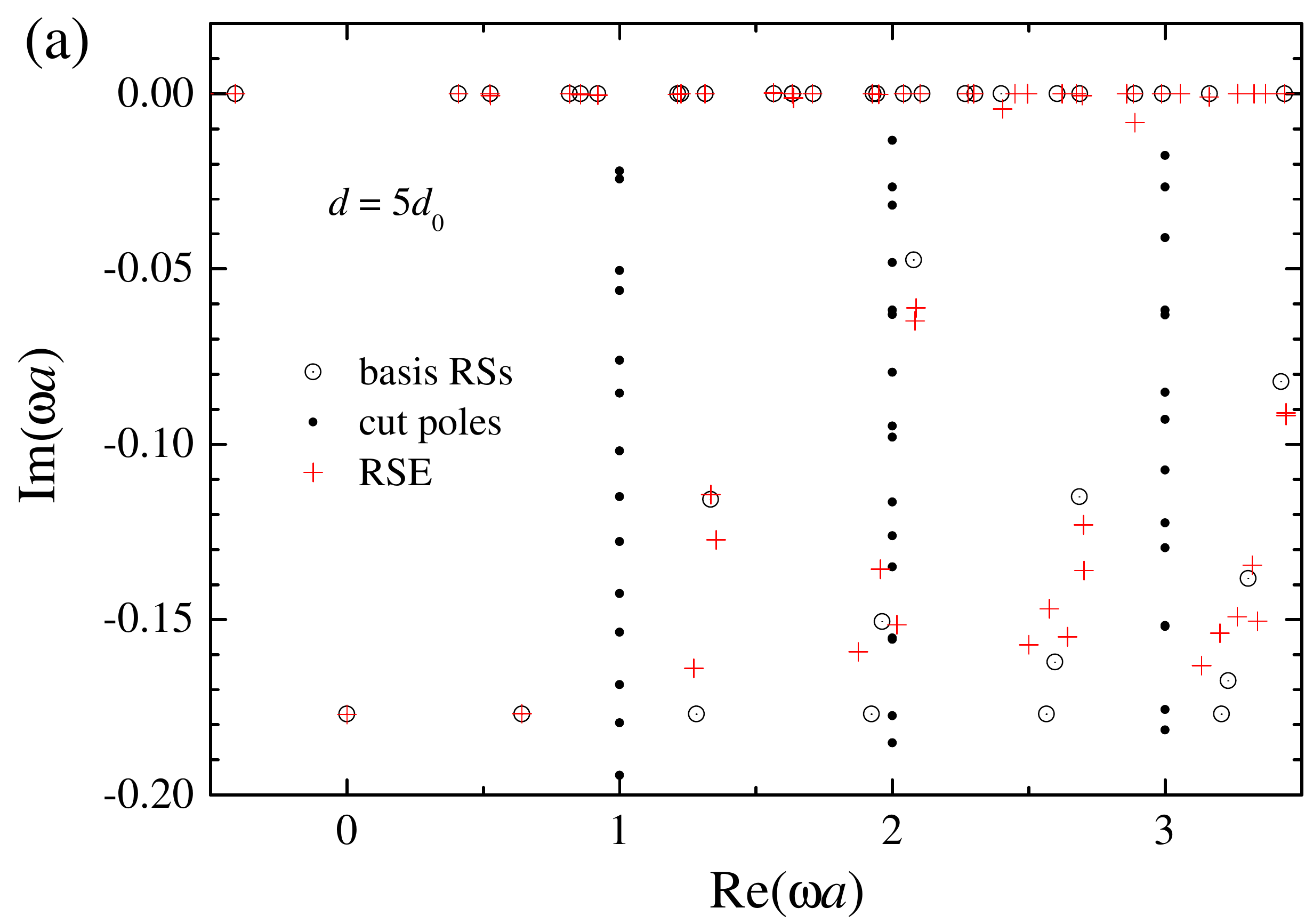}
		\includegraphics[clip,width=0.48\textwidth]{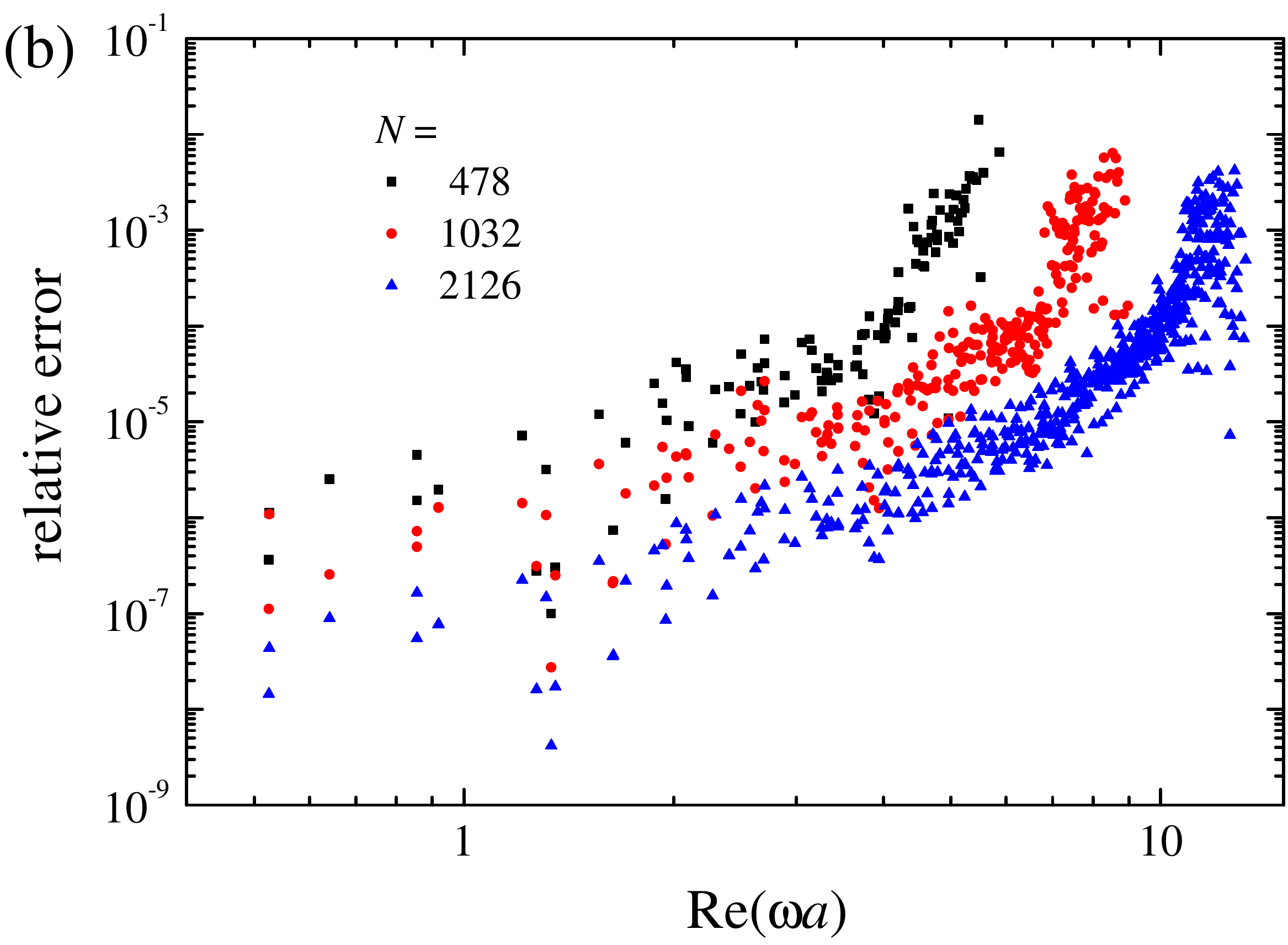}
		\caption{As \Fig{d/5} but for  $d=2\pi$.}
		\label{5d}	
	\end{figure}
\subsection{RSE in $\omega$-representation}
\label{App:RSE-w}

The RSE equation in this case is given by the general formula \Eq{RSE} of the PC-RSE,
but since this is a homogeneous perturbation, there is no mixing of channels, so we use again $g=g'=0$.
Also, \Eq{RSE} includes the contribution of the cuts which need to be discretized, giving rise to cut modes to be used in the RSE on equal footing with the RSs.
	
The discretization of the cuts is done following the procedure described in \cite{DoostPRA13,LobanovPRA17}.
For each parity $s$, the cut with the branch point at $\omega=p$ is divided into $N_c$ intervals bounded by $[\tilde{\omega}_\nu,\tilde{\omega}_{\nu+2}]$, where $\nu$ is even (odd) for $s=+$ ($s=-$),
with a weight given by
\be
W_s=\int_{\tilde{\omega}_\nu}^{\tilde{\omega}_{\nu+2}}\sqrt{|\sigma_s(\omega)|}d\omega\,,
\ee
where $\tilde{\omega}_1=\tilde{\omega}_2=p$ and $\tilde{\omega}_{2N_c+1}=\tilde{\omega}_{2N_c+2}= p-i\infty$
The cut is split into finite intervals in such a way that $W_s$ is the same for each interval (for the given parity $s$). Within each interval, an artificial cut mode is defined at the frequency $\omega_\nu$, given by
\be
\omega_\nu=\frac{1}{C_\nu}\int_{\tilde{\omega}_\nu}^{\tilde{\omega}_{\nu+2}} \sigma_s(\omega)\omega d\omega
\label{omega-nu}
\ee
with $\nu=1,\,2,\dots \,,2N_c$, where $s=(-1)^\nu$ and
\be
C_\nu=\int_{\tilde{\omega}_\nu}^{\tilde{\omega}_{\nu+2}} \sigma_s(\omega)d\omega\,.
\label{Cut weights}
\ee	
Applying the same discretization to the other cut with the branch point at $\omega=-p$ and extending the $\nu$ numbers to negative integers, the ML expansion \Eq{ML22} takes the form
\be
g(z,z') \approx \sum_{\bar{n}}
\frac{E_{\bar{n}}(z)E_{\bar{n}}(z')}{\omega_{\bar{n}}(\omega-\omega_{\bar{n}})}
\label{ML31}
\ee
where
\be
\bar{n}=\begin{cases}
n & {\rm for \ RSs}\\
\nu & {\rm for \ cut\ modes}
\end{cases}
\label{nbar}
\ee
and
\be
E_{\bar{n}}(z)=B_{\bar{n}}(e^{iq_{\bar{n}} z}+(-1)^{\bar{n}} e^{-iq_{\bar{n}}z})\ \ \ \ |z|\leqslant a\,,
\ee
with the normalization constant $B_n$ for the RSs given by \Eq{Bn} and for the cut modes by
\be
B_\nu=\sqrt{\omega_\nu C_\nu}\,,
\label{Bnu}
\ee
where $C_\nu$ is defined in \Eq{Cut weights}.
	Using \Eq{Cut weights}, we can graphically show in \Fig{Fig:Sigma} how these artificial cut modes compensate for the cut, comparing  	$\sigma_\nu(\omega)$ with $C_\nu/\Delta\omega_\nu$, 	where $\Delta\omega_\nu=\tilde{\omega}_{\nu+2}-\tilde{\omega}_\nu$ is the interval of integration.

	With these modes added to the basis, the RSE can now be performed in the $\omega$-representation using \Eq{RSEdis} with the index $g$ dropped:
\be
\omega\sum_{\bar{n}' }\left(
\delta_{\bar{n}\bar{n}'}+ {\cal V}_{\bar{n}\bar{n}'}\right)c_{\bar{n}'}=\omega_{\bar{n}}c_{\bar{n}}\,.
\label{RSE2}
\ee
Figure~\ref{w-RSE-homo}(a) shows the RS frequencies calculated using the RSE equation (\ref{RSE2}), with and without cut modes in the basis. The unperturbed RSs and even- and odd-parity cut modes of the basis are also shown. The distribution of perturbed RSs repeats the oscillatory pattern seen in \Fig{k-RSE-homo}(a) and discussed above. The RSE frequencies match well the analytic values given by \Eq{Secular equation core} even if the cut modes are not taken into account. In fact, in this case the relative error is still rather low, as can be seen in \Fig{w-RSE-homo}(b). Obviously, it is  higher for the modes which are close to cut and is not improving for these modes with increasing $N$, the number of the RSs in the basis. Including the cut modes in the RSE results in a relative error decreasing with $N$ as $1/N^3$, almost uniformly for all the RSs, which is essentially the same as in the RSE used in the $k$-representation.

The total number of modes in the basis is given by $N_{\rm tot}=N+4N_c=(1+F)N$, where we have introduced the factor $F$, the ratio of the number of cut modes to the number of RSs included. In \Fig{w-RSE-homo}(c) we show how the error depends on $F$ for a fixed $N_{\rm tot}$. Higher values of $F$ imply more cut modes included in the basis at the expense of RSs. It is clear that larger values of $F$ give generally lower errors for the RSs close to the cut. However, modes with larger frequencies away from the cut are less accurately determined in this case, as the number of basis RSs $N$ reduces with $F$. We found that the value $F=1$ is close to the optimal one, as all RSs in a wide spectral range have a similar level of errors. We have made a similar study of the relative error in the case of the PC-RSE and found the same optimal value of $F$. Therefore, unless stated differently, the value of $F=1$ is used in all calculations throughout this paper.	

	\section{Mode Contributions}
\label{App:cng}

In addition to \Fig{Mode Contributions} of the main text, showing the expansion coefficients for a BIC-QGM pair, \Figs{Mode Contributions FP}{Mode Contributions Other} of this Appendix show basis mode contributions for three other types of modes: FP, leaky, and cut modes. Similar to the BIC-QGM pair in  \Fig{Mode Contributions}, two perturbed FP RSs in Fig.\,\ref{Mode Contributions FP} originate from a  pair of degenerate unperturbed FP modes, and only one of them has a nonzero contribution of $m=0$ leaky modes. This makes the Q-factor for that mode slightly lower than for the other one, not affected by any $m=0$ modes due to symmetry. Interestingly, both perturbed modes are almost equally strongly influenced by a pair of FP and a pair of WG basis modes matching the perturbed mode frequency.

Like in \Fig{Mode Contributions}, the leaky and cut modes shown in \Fig{Mode Contributions Other} have a dominant contribution of only one basis mode (or one pair of modes), while the contribution of any other mode in the basis does
not exceed a few per mille.

	\section{Other parameters}
\label{App:other}

We finally study the dependence of the RS frequencies calculated via the PC-RSE and their errors on the two structural parameters of the PC slab: the half width of the core layer $b$ and the period of modulation $d$, while keeping $\epsilon=6$, $\alpha=0$, $\beta=1$, and $p=0$ as before.

Increasing $b$ from half-width ($b=a/2$) to the full-width value ($b=a$) does not lead to any significant changes in the spectrum, as one can see in \Fig{Different b}(a). The relative error is however getting sensitive to $b$ as $b\to a$. In fact, the error in \Fig{Different b}(b) shows that the case of $b=0.95a$ can produce up to an order of magnitude higher errors (relative to the SMM) compared to the system with $b=0.5a$. The reason for this increase is related to the ML series changing its convergence properties on the borders of the system, which requires a further study. Presently, it prevents the PC-RSE from being used with exactly $b=a$.
	
Figures \ref{d/5} and \ref{5d} show the RS frequencies and the relative error for the period of modulation $d$, respectively,  5 times smaller and 5 time larger than that used for \Fig{crystal rse}. Such changed of the period  change the spectrum of the RSs dramatically, so for instance, in the first case the number of RSs per cut is much larger than in \Fig{crystal rse} and in the second case -- much smaller. Nevertheless, the PC-RSE is working equally well in all these cases, as we can see from almost the same level of errors.
	

\begin{thebibliography}{10}
		
		\bibitem{LiuAPL10}
		L. Liu, M. Pu, K. Yvind, and J.~M. Hvam, Appl. Phys. Lett. {\bf 96},  051126
		(2010).
		
		\bibitem{McnabOS03}
		S.~J. Mcnab, N. Moll, and Y.~A. Vlasov, Opt. Soc. {\bf 299},  358  (2003).
		
		\bibitem{McGurnPRB00}
		A.~R. Mcgurn, Phys. Rev. B {\bf 61},  13235  (2000).
		
		\bibitem{BayindirAPL00}
		M. Bayindir, B. Temelkuran, and E. Ozbay, Appl. Phys. Lett. {\bf 77},  3902
		(2000).
		
		\bibitem{YablonovitchPRL91}
		E. Yablonovitch, T.~J. Gmitter, and K.~M. Leung, Phys. Rev. Lett. {\bf 67},
		2295  (1991).
		
		\bibitem{WhittakerPRB99}
		D.~M. Whittaker and I.~S. Culshaw, Phys. Rev. B {\bf 60},  2610  (1999).
		
		\bibitem{TikhodeevPRB02}
		S.~G. Tikhodeev {\it et~al.}, Phys. Rev. B {\bf 66},  045102  (2002).
		
		\bibitem{FanPRB02}
		S. Fan and J.~D. Joannopoulos, Phys. Rev. B {\bf 65},  235112  (2002).
		
		\bibitem{ZhouPQE14}
		W. Zhou {\it et~al.}, Prog. Quant. Electr. {\bf 38},  1   (2014).
		
		\bibitem{GamowZP28}
		G. Gamow, Z. Phys. {\bf 51},  204  (1928).
		
		\bibitem{SiegertPR39}
		A.~F.~J. Siegert, Phys. Rev. {\bf 56},  750  (1939).
		
		\bibitem{Weinstein}
		L.~A. Weinstein, {\em Open resonators and open waveguides} (Golden press,
		Boulder, Col., Boulder, 1969).
		
		\bibitem{WoodPM02}
		R.~W. Wood, Phil. Mag. {\bf 4},  396  (1902).
		
		\bibitem{LobanovPRA17}
		S.~V. Lobanov, G. Zoriniants, W. Langbein, and E.~A. Muljarov, Phys. Rev. A
		{\bf 95},  053848  (2017).
		
		\bibitem{GrasOL19}
		A. Gras, W. Yan, and P. Lalanne, Opt. Lett. {\bf 44},  3494  (2019).
		
		\bibitem{Akimov11}
		A.~B. Akimov, N.~A. Gippius, and S.~G. Tikhodeev, JETP Lett. {\bf 93},  427
		(2011).
		
		\bibitem{ArmitagePRA14}
		L.~J. Armitage, M.~B. Doost, W. Langbein, and E.~A. Muljarov, Phys. Rev. A {\bf
			89},  053832  (2014).
		
		\bibitem{GovorovNL10}
		A.~O. Govorov {\it et~al.}, Nano Letters {\bf 10},  1374  (2010).
		
		\bibitem{WeissPRL16}
		T. Weiss {\it et~al.}, Phys. Rev. Lett. {\bf 116},  237401  (2016).
		
		\bibitem{WeissPRB17}
		T. Weiss {\it et~al.}, Phys. Rev. B {\bf 96},  045129  (2017).
		
		\bibitem{VollmerNMe08}
		F. Vollmer and S. Arnold, Nat. Meth. {\bf 5},  591  (2008).
		
		\bibitem{VollmerAPL02}
		F. Vollmer {\it et~al.}, Appl. Phys. Lett. {\bf 80},  4057  (2002).
		
		\bibitem{RosenblitPRA04}
		M. Rosenblit, P. Horak, S. Helsby, and R. Folman, Phys. Rev. A {\bf 70},
		053808  (2004).
		
		\bibitem{FrateschiAPL95}
		N.~C. Frateschi and A.~F.~J. Levi, Appl. Phys. Lett. {\bf 66},  2932  (1995).
		
		\bibitem{LalanneLPR18}
		P. Lalanne {\it et~al.}, Laser Phot. Rev. {\bf 12},  1700113  (2018).
		
		\bibitem{MuljarovEPL10}
		A. Muljarov, W. Langbein, and R. Zimmermann, Europhys Lett. {\bf 92},  50010
		(2010).
		
		\bibitem{DoostPRA14}
		M.~B. Doost, W. Langbein, and E.~A. Muljarov, Phys. Rev. A {\bf 90},  013834
		(2014).
		
		\bibitem{SauvanPRL13}
		C. Sauvan, J.~P. Hugonin, I.~S. Maksymov, and P. Lalanne, Phys. Rev. Lett. {\bf
			110},  237401  (2013).
		
		\bibitem{FloessPRX17}
		D. Floess {\it et~al.}, Phys. Rev. X {\bf 7},  021048  (2017).
		
		\bibitem{DoostPRA12}
		M.~B. Doost, W. Langbein, and E.~A. Muljarov, Phys. Rev. A {\bf 85},  023835
		(2012).
		
		\bibitem{DoostPRA13}
		M.~B. Doost, W. Langbein, and E.~A. Muljarov, Phys. Rev. A {\bf 87},  043827
		(2013).
		
		\bibitem{ArmitagePRA18}
		L.~J. Armitage, M.~B. Doost, W. Langbein, and E.~A. Muljarov, Phys. Rev. A {\bf
			97},  049901  (2018).
		
		\bibitem{MuljarovPRB16}
		E.~A. Muljarov and W. Langbein, Phys. Rev. B {\bf 93},  075417  (2016).
		
		\bibitem{MuljarovOL18}
		E.~A. Muljarov and T. Weiss, Opt. Lett. {\bf 43},  1978  (2018).
		
		\bibitem{Ashcroft}
		N.~W. Ashcroft and N.~D. Mermin, {\em Solid state physics} (Saunders College,
		Philadelphia, 1976), Chap.~8, pp.\ 132--140.
		
		\bibitem{AndreaniPRB06}
		L.~C. Andreani and D. Gerace, Phys. Rev. B {\bf 73},  235114  (2006).
		
		\bibitem{MinkovSR14}
		M. Minkov and V. Savona, Sci. Rep. {\bf 4},  5124  (2014), article.
		
		\bibitem{YanPRB18}
		W. Yan, R. Faggiani, and P. Lalanne, Phys. Rev. B {\bf 97},  205422  (2018).
		
		\bibitem{LobanovPRA18}
		S.~V. Lobanov, W. Langbein, and E.~A. Muljarov, Phys. Rev. A {\bf 98},  033820
		(2018).
		
		\bibitem{WeissPRB18}
		T. Weiss and E.~A. Muljarov, Phys. Rev. B {\bf 98},  085433  (2018).
		
		\bibitem{LalanneOL00}
		P. Lalanne and E. Silberstein, Opt. Lett. {\bf 25},  1092  (2000).
		
		\bibitem{SilbersteinJOSAA01}
		E. Silberstein, P. Lalanne, J.-P. Hugonin, and Q. Cao, J. Opt. Soc. Am. A {\bf
			18},  2865  (2001).
		
		\bibitem{LiJOA03}
		L. Li, Journal of Optics A: Pure and Applied Optics {\bf 5},  345  (2003).
		
		\bibitem{WeissJOA09}
		T. Weiss {\it et~al.}, J. Opt. {\bf 11},  114019  (2009).
		
		\bibitem{GaoSR16}
		X. Gao {\it et~al.}, Sci. Rep. {\bf 6},  31908  (2016).
		
		\bibitem{NeumannPZ29}
		J. von Neumann and E. Wigner, Phys. Z. {\bf 30},  467  (1929).
		
		\bibitem{MarinicaPRL08}
		D.~C. Marinica and A.~G. Borisov, Phys. Rev. Lett. {\bf 100},  183902  (2008).
		
		\bibitem{BulgakovPRB08}
		E.~N. Bulgakov and A.~F. Sadreev, Phys. Rev. B {\bf 78},  075105  (2008).
		
		\bibitem{MoiseyevPRL09}
		N. Moiseyev, Phys. Rev. Lett. {\bf 102},  167404  (2009).
		
		\bibitem{PlotnikPRL11}
		Y. Plotnik {\it et~al.}, Phys. Rev. Lett. {\bf 107},  28  (2011).
		
		\bibitem{Arfken01}
		G.~B. Arfken and H.~J. Weber, {\em {Mathematical Methods for Physicists, 5th
				edition}} (Academic Press, San Diego, 2001), p.\ 448.
		
		\bibitem{BykovJLT13}
		D.~A. Bykov and L.~L. Doskolovich, J. Lightwave Tech. {\bf 31},  793  (2013).
		
		\bibitem{TaghizadehAPL17}
		A. Taghizadeh and I.-S. Chung, Appl. Phys. Lett. {\bf 111},  031114  (2017).
		
		\bibitem{BulgakovPRA19}
		E.~N. Bulgakov and A.~F. Sadreev, Phys. Rev. A {\bf 99},  033851  (2019).
		
		\bibitem{BulgakovPRA14}
		E.~N. Bulgakov and A.~F. Sadreev, Phys. Rev. A {\bf 90},  053801  (2014).
		
		\bibitem{ZhenPRB14}
		B. Zhen, C.~W. Hsu, L. Lu, and A.~D. Stone, Phys. Rev. B {\bf 113},  1  (2014).
		
		\bibitem{HsuNat16}
		C.~W. Hsu {\it et~al.}, Nat. Rev. Mat. {\bf 1},  1  (2016).
		
		\bibitem{BykovPRA19}
		D.~A. Bykov, E.~A. Bezus, and L.~L. Doskolovich, Phys. Rev. A {\bf 99},  063805
		(2019).
		
	\end{thebibliography}
\end{document}